\documentclass[prb,twocolumn,notitlepage,superscriptaddress,nofootinbib]{revtex4-2}
\usepackage{amsmath,amssymb}
\usepackage[hidelinks,colorlinks,linkcolor=blue,
citecolor=blue,urlcolor=blue]{hyperref}
\usepackage{graphicx,siunitx}

\usepackage{bm,color,amsmath,amssymb,mathrsfs,latexsym,graphicx,psfrag,mathtools}
\usepackage{color}
\usepackage[dvipsnames]{xcolor}
\usepackage[hidelinks,colorlinks,linkcolor=blue,
citecolor=blue,urlcolor=blue]{hyperref}
\usepackage{dsfont}

\usepackage{empheq}

\newcommand{\bsl}[1]{\boldsymbol{#1}}

\newcommand{\ket}[1]{|#1 \rangle}

\newcommand{\ii}{\mathrm{i}}

\newcommand{\dsR}{\mathbb{R}}

\newcommand{\U}{\mathrm{U}}

\newcommand{\eqnref}[1]{Eq.\,\eqref{#1}}

\newcommand{\refcite}[1]{Ref.~\onlinecite{#1}}

\newcommand{\mat}[1]{\left(\begin{matrix}#1\end{matrix}\right)}

\newcommand{\eq}[1]{\begin{equation} #1 \end{equation}}

\newcommand{\eqa}[1]{\begin{align}\begin{split} #1 \end{split}\end{align}}

\let\oldAA\AA
\renewcommand{\AA}{\text{\normalfont\oldAA}}

\newcommand{\cc}{\mathcal{K}}

\newcommand{\eV}{\mathrm{eV}}

\newcommand{\mbf}[1]{\mathbf{#1}}
\newcommand{\V}{\mathcal{V}}

\newcommand{\Q}{\mathcal{Q}}

\newcommand{\K}{\text{K}}

\newcommand{\BZ}{\text{1BZ}}

\usepackage{comment}

\newcommand{\Te}{\text{Te}}

\makeatletter 
    
\renewcommand\onecolumngrid{% <<<<<<
\do@columngrid{one}{\@ne}%
\def\set@footnotewidth{\onecolumngrid}% <<<<<<<<<<<<<<<<
\def\footnoterule{\kern-6pt\hrule width 1.5in\kern6pt}%
}

\renewcommand\twocolumngrid{% <<<<<<
        \def\footnoterule{% restore rule
        \dimen@\skip\footins\divide\dimen@\thr@@
        \kern-\dimen@\hrule width.5in\kern\dimen@}
        \do@columngrid{mlt}{\tw@}
}%

\makeatother

\begin{document}

\title{When Could Abelian Fractional Topological Insulators Exist 
in Twisted MoTe$_2$ \\ (and Other Systems) 
}

\author{Yves H. Kwan}\thanks{These authors contributed equally.} 
\affiliation{Princeton Center for Theoretical Science, Princeton University, Princeton, NJ 08544}

\author{Glenn Wagner}\thanks{These authors contributed equally.} 
\affiliation{Department of Physics, University of Zurich, Winterthurerstrasse 190, 8057 Zurich, Switzerland}

\author{Jiabin Yu}\thanks{These authors contributed equally.} 
\affiliation{Department of Physics, Princeton University, Princeton, New Jersey 08544, USA
}

\author{Andrea Kouta Dagnino}\thanks{These authors contributed equally.} 
\affiliation{Department of Physics, University of Zurich, Winterthurerstrasse 190, 8057 Zurich, Switzerland}

\author{Yi Jiang}
\affiliation{Donostia International Physics Center, P. Manuel de Lardizabal 4, 20018 Donostia-San Sebastian, Spain}

\author{Xiaodong Xu}
\affiliation{Department of Materials Science and Engineering University of Washington Seattle Washington 98195 USA}
\affiliation{Department of Physics, University of Washington, Seattle, Washington, 98195, USA
}

\author{B. Andrei Bernevig}
\affiliation{Department of Physics, Princeton University, Princeton, New Jersey 08544, USA
}
\affiliation{Donostia International Physics Center, P. Manuel de Lardizabal 4, 20018 Donostia-San Sebastian, Spain}
\affiliation{IKERBASQUE, Basque Foundation for Science, Bilbao, Spain
}

\author{Titus Neupert}
\affiliation{Department of Physics, University of Zurich, Winterthurerstrasse 190, 8057 Zurich, Switzerland}

\author{Nicolas Regnault}
\affiliation{Laboratoire de Physique de l’Ecole normale superieure, ENS, Universite PSL, CNRS, Sorbonne Universite}
\affiliation{Department of Physics, Princeton University, Princeton, New Jersey 08544, USA
}

\begin{abstract}

Using comprehensive exact diagonalization calculations on $\theta \approx 3.7 ^{\circ}$  twisted bilayer MoTe$_2$ ($t$MoTe$_2$), as well as idealized Landau level models also relevant for lower $\theta$, we extract general principles for engineering fractional topological insulators (FTIs) in realistic situations. First, in a Landau level setup at $\nu=1/3+1/3$, we investigate what features of the interaction destroy an FTI.  For both pseudopotential interactions and realistic screened Coulomb interactions, we find that sufficient suppression of the short-range repulsion is needed for stabilizing an FTI. We then study $\theta \approx 3.7 ^{\circ}$  $t$MoTe$_2$ with realistic band-mixing and anisotropic non-local dielectric screening. Our finite-size calculations only find an FTI phase at $\nu=-4/3$ in the presence of a significant additional short-range attraction $g$ that acts to counter the Coulomb repulsion at short distances. We discuss how further finite-size drifts, dielectric engineering, Landau level character, and band-mixing effects may reduce the required value of $g$ closer towards the experimentally relevant conditions of $t$MoTe$_2$. 
Projective calculations into the $n=1$ Landau level, which resembles the second valence band of $\theta\simeq 2.1^\circ$ $t$MoTe$_2$, do not yield FTIs for any $g$, suggesting that FTIs at low-angle $t$MoTe$_2$ for $\nu=-8/3$ and $-10/3$ may be unlikely. While our study highlights the challenges, at least for the fillings considered, to obtaining an FTI with transport plateaus, even in large-angle $t$MoTe$_2$ where fractional Chern insulators are experimentally established, we also provide potential sample-engineering routes to improve the stability of FTI phases.
\end{abstract}

\maketitle

\textit{Introduction.---} Fractional Chern insulators (FCIs) \cite{neupert,sheng,regnault} are the zero-field analogue of the venerable fractional quantum Hall (FQH) effect that appear in fractionally filled narrow  Chern bands~\cite{sun2011nearly,yang2012arbitrary,tang2011high}. The interactions between the electrons 
lead to a state with fractionalized excitations, a fractionally quantized Hall conductivity and a topological ground state degeneracy. Moir\'e materials are known to host flat topological bands with relatively uniform Berry curvature, making them ideal platforms for realizing FCIs.

Recent experiments on twisted homobilayer MoTe$_2$ ($t\text{MoTe}_2$) have observed direct consequences of an FCI~\cite{cai2023signatures,zeng2023integer}, 
especially a fractionally quantized Hall conductance at filling factors $\nu=-2/3$ and $-3/5$ \cite{Park2023,Xu2023,park2024ferromagnetism,xu2024interplay}.
Although exact diagonalization (ED) calculations without band mixing can yield FCIs at certain fillings~\cite{Li2023,Crepel2023,moralesdurán2023pressureenhanced,wang2023fractional,reddy2023global,Reddy2023}, inclusion of band mixing \cite{Yu2024MFCI0,xu2024maximally,abouelkomsan2023band} is essential for theoretically capturing key aspects of the experimental phenomenology.
Furthermore, quantitative comparison to experiments requires careful consideration of the band structure~\cite{jia2023moire,mao2023lattice,zhang2024polarizationdriven,wang2023topologydft}.
Even more recently, pentalayer graphene with a hexagonal boron nitride (hBN) substrate has been experimentally shown to host a Chern insulator at $\nu=1$~\cite{Han2023,lu2023fractional} and FCIs at multiple fillings $\nu<1$ \cite{lu2023fractional}, which have motivated several recent theoretical studies~\cite{dong2023anomalous,zhou2023fractional,dong2023theory,herzog2023moir,kwan2023moire,guo2023theory}, though the appearance of an FCI is still not reproduced in unbiased calculations that account for multi-band effects.

\begin{figure}\centering\includegraphics[width=1.0\linewidth]{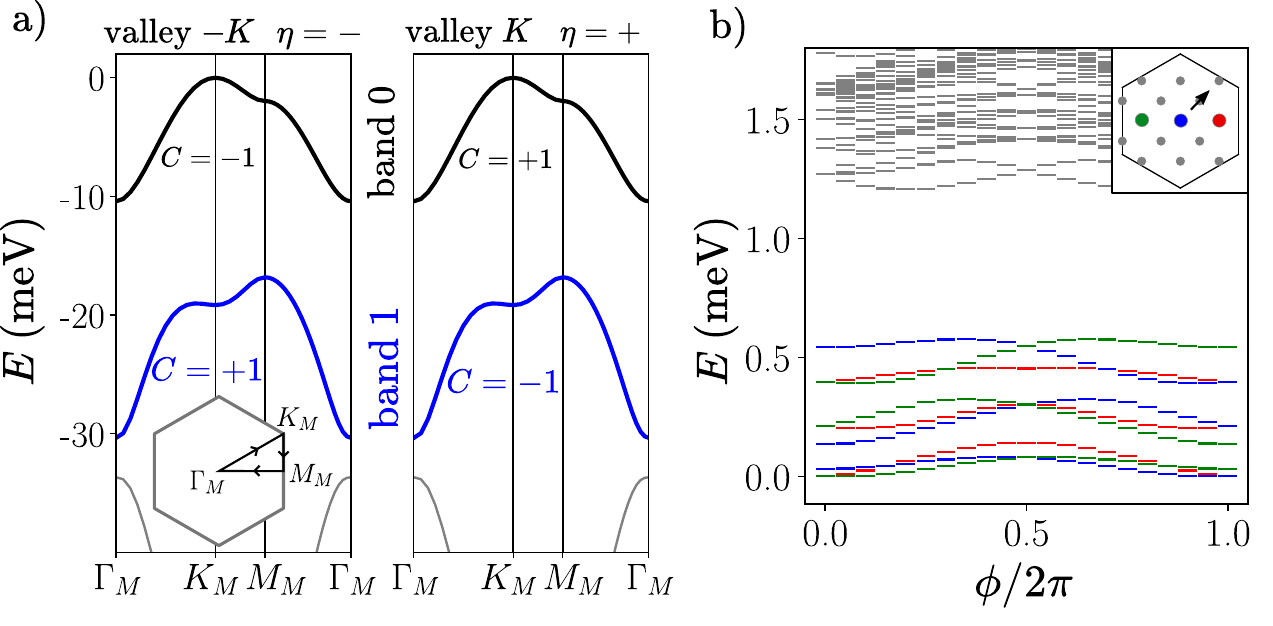}
    \caption{\textbf{$t\text{MoTe}_2$ at $\theta=3.7^\circ$.} \textbf{a)} Continuum model band structure of the first-harmonic model of Ref.~\cite{jia2023moire}. The highest valence bands have non-trivial valley Chern numbers. Inset shows the corresponding path in the moir\'e Brillouin zone (mBZ). \textbf{b)} Many-body spectrum for $\nu=-4/3$, $S_z=0$, and $N^\text{1}_{\text{max}}=0$ as a function of flux $\phi$ threaded along one handle of the torus (inset shows momentum grid of the $N_s=15$ lattice and shift under one flux insertion). Colored markers denote the lowest three states in each of the three momentum sectors (colored in inset) belonging to the FTI ground state manifold. The long-range interaction is computed using the sample configuration in Fig.~\ref{fig:main_tMoTe2_dielectric} with $\epsilon^{\perp}_{\text{MoTe}_2}=10$, $\epsilon^{\parallel}_{\text{MoTe}_2}=21$ and no spacer. In this figure, we add an additional short-range interaction with strength 
    $g=-1400\,\text{meVnm}^2$ (Eq.~\ref{eq:g}). The FTI survives until $g=-1200\,\text{meVnm}^2$ for these parameters.
    }
    \label{fig:main_bandstruct_flux}
\end{figure}

A class of related topologically ordered phases are fractional topological insulators (FTIs), which are distinguished from FCIs in that they are not chiral, but 
time-reversal (TR) symmetric~\cite{bernevigQSHE2006,Levin_Stern,Neupert2011FTI,Stern2015review,Neupert_2015,Levin2012classification}. 
{If spin is a good quantum number, these states exhibit the fractional quantum spin Hall effect (FQSHE). 
Both FTI and FQSHE physics\footnote{Neither spin conservation nor TR symmetry are strictly required to protect the topological order, i.e.~the anyon content, of these phases, in sharp contrast to $Z_2$ topological insulators which require TR symmetry.} represent a substantial departure from FQH physics, and are still experimentally elusive, although there exists a recent report of a potential FQSHE in low-angle $t$MoTe$_2$~\cite{kang2024evidence_nature}.}
Indeed, FTIs have a vanishing Hall response, while also supporting fractionally charged excitations and a topological ground state degeneracy. The simplest FTIs (which we will be concerned with in this work) are adiabatically connected to a direct product of a Laughlin state for spin-up electrons and its time-reversed copy for spin-down electrons. They can be constructed by partially filling two topological bands with spin-locked Chern numbers $C=\pm1$.  
FTIs have been numerically investigated via ED in both lattice toy models~\cite{Neupert2011FTI,Repellin2014FTI,Furukawa2014bose,crepel2024attractive} and lowest Landau level (LLL) models~\cite{Chen2012FQHtorusFTI,Mukherjee2019FQHsphereFTI}. However, no theoretical proposal for an FTI with a material-realistic model has been put forward to date.

The isolated valley Chern bands (Fig.~\ref{fig:main_bandstruct_flux}a) and experimental signatures of FCIs in $t\text{MoTe}_2$ naturally suggest the possibility of realizing a $\nu=-4/3$ FTI in this platform, namely by combining time-reversed copies of the experimentally-observed $\nu=-2/3$ FCI (at twist angle $\theta\approx 3.7^\circ$) in opposite valleys (and hence opposite spins due to spin-valley locking), but several potential obstacles need to be addressed. First, a prerequisite for obtaining FTIs is the absence of magnetic order, while mean-field and single-band ED studies routinely find spin-valley polarization at $\nu=-4/3$, in contradiction with experiments. Reference~\onlinecite{Yu2024MFCI0} shows that band mixing favors states with small magnetization at $\nu=-4/3$, where the ground state still has non-zero spin due to finite-size effects.  Second, an FTI phase needs to be obtained with physical intervalley interactions.
For vanishing intervalley interactions, the existence of the FTI in the spin-unpolarized sector as a product of two decoupled FCIs follows directly from the existence of the latter. However, numerical studies in other idealized models consistently find that FTIs are rapidly destabilized already by a moderate intervalley interaction~\cite{Neupert2011FTI,Repellin2014FTI}, calling into question the feasibility of FTIs in realistic $t$MoTe$_2$ where the dominant interactions are expected to be valley-isotropic.

In this paper, we address these challenges in the twisted TMD homobilayer material platform.
First working in the Landau level setting, we find that the FTI phase exists for valley-isotropic interactions when (i) the effect of non-local screening of the Coulomb interaction is accounted for and (ii) the short-range repulsion is sufficiently softened. We translate these insights to the realistic material setting by performing comprehensive ED calculations at $\nu=-4/3$ $t$MoTe$_2$ around $\theta=3.7^\circ$. We investigate how factors such as screening, short-range interactions, band structure parameters and band mixing influence the magnetism of the system and the stability of the FTI. 
The main conclusion we draw from our $t$MoTe$_2$ calculations is that the FTI requires an additional onsite attraction, whose value is significantly larger than the microscopic contribution derived from electron coupling to monolayer phonons, to counteract the strong Coulomb repulsion at short distances.

\begin{figure}\centering\includegraphics[width=1.0\linewidth]{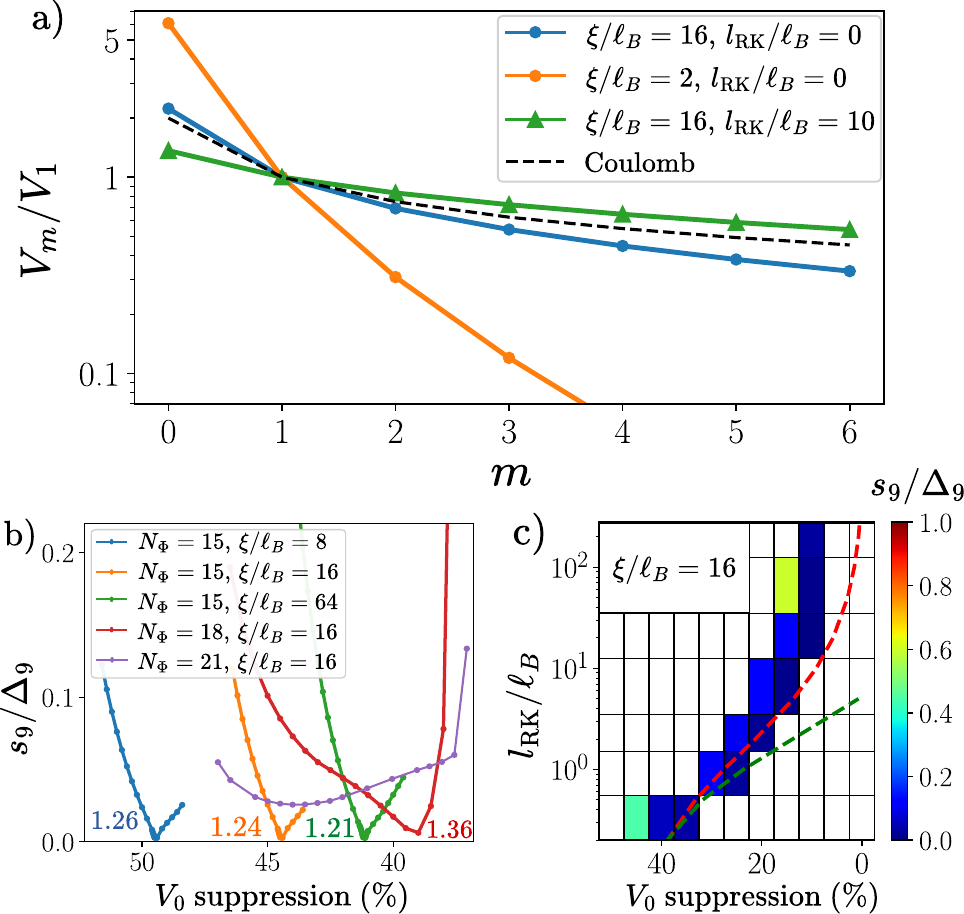}
    \caption{\textbf{TR-symmetric lowest Landau level model on torus geometry at $\nu_+=\nu_-=1/3$.} \textbf{a)} Haldane pseudopotentials $V_m$ for the Coulomb potential with dual-gate distance $\xi$ and Rytova-Keldysh length scale $l_\text{RK}$. Black dashed line indicates the bare Coulomb limit $\xi\rightarrow\infty,\,l_\text{RK}=0$. \textbf{b)} FTI spacing/gap ratio for different $\xi$ and number of flux quanta $N_\Phi$. The horizontal axis shows the percentage suppression of $V_0$. The number next to each curve denotes the value of $V_0/V_1$ corresponding to the minimum spacing/gap ratio. \textbf{c)} FTI spacing/gap ratio (color) as a function of $l_\text{RK}$ and $V_0$ suppression for fixed $\xi/\ell_B=16$ and $N_\Phi=18$. White regions indicate where $\Delta_9<0$ or $s_9/\Delta_9>1$. Red (green) dashed line is contour of constant $V_0/V_1=1.24$ ($V_0/V_2=1.77$).}
    \label{fig:LLL_pseudopotential_phase}
\end{figure}

\begin{figure}\centering\includegraphics[width=1.0\linewidth]{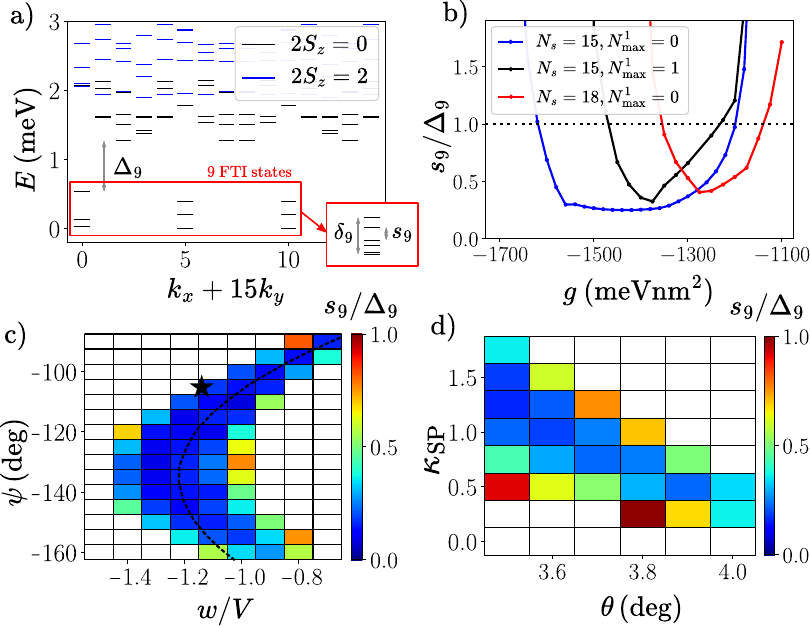}
    \caption{\textbf{ED calculations of FTIs in $t\text{MoTe}_2$ at $\nu=-4/3$.} Unless otherwise stated, calculations are performed at $\theta=3.7^\circ$ with $N^1_\text{max}=0$ for a tilted $N_s=15$ lattice (see Fig.~\ref{fig:main_bandstruct_flux}b inset) in the sample configuration of Fig.~\ref{fig:main_tMoTe2_dielectric}a with $\epsilon^{\perp}_{\text{MoTe}_2}=10$, $\epsilon^{\parallel}_{\text{MoTe}_2}=21$ and no spacer.
    \textbf{a)} Many-body spectrum of FTI phase. Parameters are identical to those of Fig.~\ref{fig:main_bandstruct_flux}b, i.e.~$g=-1400\,\text{meVnm}^2$. Lowest four levels shown for each $S_z$ and momentum sector. Higher spin sectors are above the energy axis limit. The nine states of the FTI manifold, indicated by red box, are separated by a gap $\Delta_9$ to the higher states.  Zoom-in collapses the nine FTI states onto a single column, and indicates how the maximal spacing $s_9$ and spread $\delta_9$ are defined.
    \textbf{b)} Dependence of the FTI spacing/gap ratio as a function of $g$ for different system sizes $N_s$ and occupations $N^\text{1}_{\text{max}}$ of band 1. The ground state is in the $S_z=0$ sector for all data points. 
    \textbf{c)} $s_9/\Delta_9$ (color) as a function of continuum model parameters $\psi$ and $w/V$ for fixed $g=-1400$meVnm$^2$ and $V^2+w^2=25$meV. The black star indicates the continuum model parameters of \refcite{jia2023moire} used in this work. The dashed line tracks the minimum of the Berry curvature fluctuations as a function of $\psi$.
    \textbf{d)} $s_9/\Delta_9$ (color) as a function of bandwidth scaling factor $\kappa_\text{SP}$ and twist angle $\theta$ for fixed $g=-1400$meVnm$^2$.
    }
    \label{fig:main_tMoTe2_ED}
\end{figure}

\textit{LLL model.---} To understand the FTI phase in a system that is free from the complications of dispersion, inhomogeneous band geometry/curvature and band mixing, we first consider a toy model with TR symmetry whose Hilbert space comprises two LLLs arising from opposite magnetic fields $\mbf{B}=\eta B\hat{z}$, where $\eta=+$ ($-$) is the index for valley $K$ ($-K$)~\cite{Chen2012FQHtorusFTI,Mukherjee2019FQHsphereFTI,Bultinck2020mechanism,furukawa2014global,zhang2018composite,kwan2021exciton,kwan2022hierarchy,eugenio2020DMRG,stefanidis2020excitonic,chatterjee2022dmrg,myersonjain2023conjugate,yang2023phase,Wu2024,shi2024excitonic,kwan2024textured}.
In the context of $t$MoTe$_2$ with spin-valley locking, $\eta$ also corresponds to the spin projection along $S_z$. We add density-density interactions $V(q)$ that preserve the valley-$U_\text{v}(1)$ symmetry, and define the Haldane pseudopotentials for relative angular momentum $m$
\begin{equation}\label{eq:main_PP}
    V_m\equiv \int \frac{d^2\mbf{q}}{(2\pi)^2}V(q)L_m(q^2)e^{-q^2},
\end{equation}
where $L_m(x)$ are the Laguerre polynomials and we have set the magnetic length $\ell_B=1$. The even $m$ pseudopotentials do not affect the intravalley physics due to fermion antisymmetry.
We perform ED calculations on the square torus geometry with $N_\Phi$ flux quanta, and fix the filling factors $\nu_\eta=N_\eta/N_\Phi=1/3$, where $N_\eta$ is the number of particles in valley $\eta$. Note that the particle-hole symmetry within the LLL ensures that the results are identical  
for $\nu_\eta=2/3$ up to a global energy shift. 
Due to the periodic boundary conditions (PBCs), the FTI has nine topologically-degenerate ground states that are adiabatically connected to decoupled products of $\nu=1/3$ FQH states in the two valleys, and lie in specific momentum sectors~\cite{Bernevig2012emergent}\footnote{The correct momentum-resolved counting of the FTI degeneracy can be straightforwardly inferred by performing a calculation where the intervalley interaction is artificially switched off.}. 
 We characterize the FTI states through the maximum spacing $s_9$ between adjacent levels in the FTI manifold, irrespective of the momentum, and the gap $\Delta_9$ that separates them from higher energy levels (see Fig.~\ref{fig:main_tMoTe2_ED}a).
A smaller positive spacing/gap ratio $s_9/\Delta_9$ indicates a more robust FTI. In this work, we use $s_9/\Delta_9<1$ as a necessary condition for an FTI. A negative $\Delta_9$ means that the lowest nine states do not have the correct momenta for an FTI. 
More details of the conditions are given in App.~\ref{app:subsec:LLL_setup}.

For a purely \emph{intravalley} $V_1$ interaction, there are nine zero-energy ground states corresponding to products of $\nu=1/3$ Laughlin states with no intervalley correlations. 
In \refcite{Chen2012FQHtorusFTI}, the stability of this model FTI against finite $V_0$ has been investigated. While the FTI still survived at $V_0=V_1$, we emphasize that this does not correspond to a valley-isotropic interaction since the \emph{intervalley} $V_1$ was not considered. To our knowledge, there are no existing studies that have demonstrated an FTI phase for valley-isotropic (i.e., purely density-density) interactions in any fermionic system. Thus, we perform calculations in a restricted model with valley-isotropic $V_0$ and $V_1$ (see Fig.~\ref{fig:isotropic_V0_V1} in App.~\ref{app:subsec:LLL_V0V1}), and find $s_9/\Delta_9<1$ for the FTI in a window $V_0/V_1\simeq 1.18-1.35$ ($1.28-1.80$) for $N_\Phi=15$ ($18$).

With an eye towards treating long-range interaction potentials relevant for moir\'e materials, we now consider the Coulomb potential screened by metallic gates positioned at height $\pm\xi/2$, i.e.~\mbox{$V(q)\propto \frac{\tanh{\frac{q\xi}{2}}}{q}$}. The unscreened Coulomb potential has $V_0/V_1=2$ (Fig.~\ref{fig:LLL_pseudopotential_phase}a, dashed line), which is increased further by gate-screening (blue and orange lines). Since large $V_0/V_1$ may lead to phase-separated or disordered phases~\cite{Chen2012FQHtorusFTI,Mukherjee2019FQHsphereFTI}, we introduce an additional short range attraction $\sim-\delta(\mbf{r})$ which only suppresses the $V_0$ pseudopotential, and additionally helps to penalize ferromagnetism. While the resulting interaction has attractive components, the LLL-projected interaction remains purely repulsive as long as $V_0>0$.

The spacing/gap ratio of the FTI as a function of the percentage suppression of $V_0$ is shown in Fig.~\ref{fig:LLL_pseudopotential_phase}b, where we zoom in to highlight the small values of $s_9/\Delta_9$, indicating very close degeneracy of the FTI ground states. It is clear that the FTI persists for a window around \mbox{$V_0/V_1\simeq 1.3$}, similar to our isotropic $V_0-V_1$ calculation mentioned above, despite the large-$m$ tail of $V_m$, whose decay rate decreases for weaker screening (Fig.~\ref{fig:LLL_pseudopotential_phase}a). 
The FTI is more stable for larger $\xi$ 
in that the required suppression of $V_0$ is reduced, and the best spacing/gap ratio in finite-size numerics is smaller (see Fig.~\ref{fig:xi_V0frac_refined} in App.~\ref{app:subsec:LLL_screened} for $\xi/\ell_B=2,4$ results). The latter is surprising since the valley-polarized $\nu=1/3$ Laughlin state only becomes exact in the $\xi\rightarrow0$ limit where $V_{m>1}/V_1\rightarrow 0$. This implies that the ideal conditions for realizing an FQH/FCI state do not necessarily coincide with those for an FTI. As the system size increases, there is a drift of the FTI phase towards greater $V_0/V_1$ (which implies less needed suppression of $V_0$)  for the system sizes studied. We note that for the largest calculations ($N_\Phi=21$), the $s_9/\Delta_9$ curve has a slightly higher minimum (though still much less than 1) but is significantly broadened. This prevents us from performing a finite-size extrapolation to estimate the stability window of $V_0/V_1$ for the FTI in the thermodynamic limit.  While the global ground state for the unsuppressed  gate-screened interaction \mbox{$V(q)\propto \frac{\tanh{\frac{q\xi}{2}}}{q}$} has finite valley polarization, the valley-unpolarized \mbox{$S_z=\frac{1}{2}(N_+-N_-)=0$} sector is the lowest sector for values of $V_0/V_1$ where the FTI is stabilized (see Fig.~\ref{fig:spin_xi16_combined} in App.~\ref{app:subsec:LLL_screened}).

We now account for the  Rytova-Keldysh (RK)~\cite{rytova2018screened,Keldysh1979} correction to the Coulomb interaction arising from the in-plane polarizability and finite width of few-layer materials~\cite{cudazzo2011dielectric,wang2018colloq,zhao2023probing}. We take the phenomenological form
\begin{equation}\label{eq:main_LLL_RK}
    V(q)\propto\frac{1}{ q(1+l_\text{RK}q)}\tanh{\frac{q\xi}{2}},
\end{equation}
where $l_\text{RK}$ is the RK length scale. Increasing $l_\text{RK}$ softens the short-distance repulsion and reduces $V_0/V_1$, while somewhat enhacing $V_m/V_1$ for $m>1$ 
(see Fig.~\ref{fig:LLL_pseudopotential_phase}a, green line). Hence, less additional suppression of $V_0$ is needed to obtain the FTI, as shown in Fig.~\ref{fig:LLL_pseudopotential_phase}c. The FTI phase roughly follows a locus of constant $V_0/V_1$. Relaxing the FTI condition to only require $\Delta_9>0$ (i.e.~the nine lowest states lie in the correct momentum sectors) does not significantly enlarge the stability region (see Fig.~\ref{fig:RK_V0frac_lambda_xi16_LL0_Nphi18_spinpol} in App.~\ref{secapp:LLL_RK}).

Finally, we also consider higher Landau levels (LLs) as potential hosts for FTIs. As discussed in App.~\ref{subsecapp:otherLLs}, while the non-relativistic $n=1$ LL does not stabilize an FTI for the parameters considered, we find that the \emph{Dirac} fermion $n=1$ LL yields an FTI for a significantly smaller suppression of the $V_0$ pseudopotential compared to the LLL case ($n=0$)\footnote{For the LLL, the form factors are identical for the non-relativistic and Dirac cases.}. 
The reason is that for both types of $n=1$ LLs, the different form factors of the wavefunctions lead to Eq.~\ref{eq:main_LLL_RK} having a smaller $V_0/V_1$ ratio, compared to the LLL case. However, the higher $V_m/V_1$ ratios are sufficiently enhanced for the non-relativistic $n=1$ LL such that the FTI does not survive for valley-isotropic interactions.

\textit{Fractional topological insulators in $t\text{MoTe}_2$.---} We now turn to  modeling $t\text{MoTe}_2$ at filling $\nu=-4/3$, focusing on $\theta=3.7^\circ$ about which  FCIs~\cite{cai2023signatures,zeng2023integer,Xu2023,Park2023} were observed (at fillings $<1$). Unless otherwise stated, we use the single-particle first-harmonic continuum model~\cite{Wu2019} with parameters $m_*=0.6m_e$, $w=-18.8$meV, $V=16.5$meV and $\psi=-105.9^\circ$ taken from \refcite{jia2023moire}. Higher harmonics are only necessary for describing substantially smaller twist angles. Due to spin-orbit coupling, the moir\'e bands are spin-valley locked.

\begin{figure}\centering\includegraphics[width=0.9\linewidth]{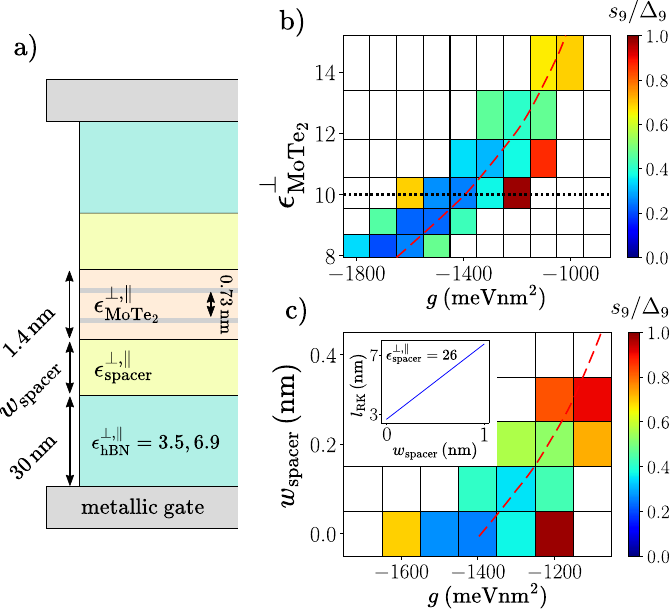}
    \caption{\textbf{Dependence on the dielectric environment.}  1BPV calculations are performed at $\theta=3.7^\circ$ for a tilted $N_s=15$ lattice (see Fig.~\ref{fig:main_bandstruct_flux}b inset).
    \textbf{a)} Schematic cross-section of a $t$MoTe$_2$ device, where we assume mirror symmetry in the vertical direction. The continuum model degrees of freedom are positioned at the Mo planes (grey lines) at $z=\pm 0.365\,\text{nm}$, and embedded in the dielectric environment generated by the device stack. Optional spacer dielectrics (yellow) can be inserted between $t$MoTe$_2$ and the encapsulating hBN substrates. We fix $w_\text{hBN}=30\,\text{nm},w_{\text{MoTe}_2}=1.4\,\text{nm},d_{\text{MoTe}_2}=0.73\,\text{nm}$, $\epsilon^{\perp}_{\text{hBN}}=3.5$ and $\epsilon^{\parallel}_{\text{hBN}}=6.9$.  
    \textbf{b)} $s_9/\Delta_9$ (color) as a function of $\epsilon^\perp_{\text{MoTe}_2}$ and $g$ for fixed $\epsilon^\parallel_{\text{MoTe}_2}/\epsilon^\perp_{\text{MoTe}_2}=2.1$ in the absence of spacers ($w_\text{spacer}=0$). Dotted line indicates $\epsilon^{\perp}_{\text{MoTe}_2}$ used in Fig.~\ref{fig:main_tMoTe2_ED}.  
    \textbf{c)} $s_9/\Delta_9$ (color) as a function of $w_\text{spacer}$ and $g$ for fixed $\epsilon^{\perp}_{\text{MoTe}_2}=10$, $\epsilon^{\parallel}_{\text{MoTe}_2}=21$ and $\epsilon^{\perp,\parallel}_{\text{spacer}}=26$. Inset shows the effective $q\rightarrow 0$ RK parameter $l_\text{RK}$ for the intralayer interaction as a function of $w_\text{spacer}$, which is fitted well by $l_\text{RK}=5.07\,w_\text{spacer}+2.67\,\text{nm}$. Red dashed lines in b,c) indicate constant $V_0/V_1=1.11$ for the intralayer interaction.
    }
    \label{fig:main_tMoTe2_dielectric}
\end{figure}

To capture the screening effects due to the $t$MoTe$_2$ thickness and the dielectric environment, we compute the interaction potential by solving the Poisson equation in a typical dual-gated hBN-encapsulated geometry (see Fig.~\ref{fig:main_tMoTe2_dielectric}a without the spacer slabs). As detailed in App.~\ref{secapp:dielectric}, our treatment of the interaction potential $V_{ll'}(q)$ accounts for the anisotropic dielectric screening from hBN and MoTe$_2$ slabs of widths $w_{\text{hBN}}$ and $w_{\text{MoTe}_2}$, and the dependence on the layer indices $l,l'$ coming from the finite interlayer distance $d_{\text{MoTe}_2}=0.73\,\text{nm}$. Unless stated otherwise, we take  $\epsilon^{\perp}_{\text{MoTe}_2}=10$ and $\epsilon^{\parallel}_{\text{MoTe}_2}=21$ from first principles calculations on bilayer $\text{MoTe}_2$~\cite{laturia2018dielectric}.  At long wavelengths, the inclusion of the MoTe$_2$ screening results in an effective $l_\text{RK}\simeq 2.7\,\text{nm}$ for the $l=l'$ interaction (Fig.~\ref{fig:main_tMoTe2_dielectric}c inset).
Anticipating from our LL results that a suppression of the short-range repulsion will be necessary to stabilize FTIs, we add an additional onsite intralayer interaction 
\begin{equation}
\label{eq:g}
    \delta V_{ll'}(q)=g\delta_{ll'}.
\end{equation} 
Due to fermion antisymmetry, Eq.~\eqref{eq:g} is ineffective between particles in the same valley.

The ED calculations are performed on torus geometries with $N_s$ moir\'e unit cells. In order to access different $N_s$ while maintaining an aspect ratio close to 1, we frequently utilize tilted boundary conditions \cite{lauchli2013hierarchy,Repellin2014FTI} (see Fig.~\ref{fig:main_bandstruct_flux}b inset for mBZ momentum grid for $N_s=15$). 
For large system sizes, computational complexity necessitates projection to the highest valence band (labelled band 0 in Fig.~\ref{fig:main_bandstruct_flux}a) in each valley $\eta=\pm$, i.e.~a one-band-per-valley (1BPV) calculation. However, as emphasized in \refcite{Yu2024MFCI0},  band mixing is important owing to strong interactions and the relatively small gap $\simeq 10\,\text{meV}$ to the next  valence band. To reduce the computational cost, we adopt a truncated Hilbert space approach~\cite{Rezayi2011} in some computations, such that we also allow a maximum number $N^\text{1}_{\text{max}}$ of holes in the next valence band (band 1 in Fig.~\ref{fig:main_bandstruct_flux}a). 
$N^\text{1}_{\text{max}}=N_s$ corresponds to a two-bands-per-valley (2BPV) calculation.

Figure~\ref{fig:main_tMoTe2_ED}a shows a representative 1BPV ($N^\text{1}_{\text{max}}=0$) many-body spectrum in the FTI phase. 
Apart from the small $s_9/\Delta_9$, spectral evolution under flux threading provides additional evidence for an FTI (see Fig.~\ref{fig:main_bandstruct_flux}b). We emphasize that these results are acquired when the long-range part of the interaction is valley-isotropic, which is the physically natural regime since the valleys are related by TR symmetry and hence the corresponding Bloch states occupy the same region in real space.
We find that, similar to the LLL model, an attractive $g<0$ is necessary to stabilize an FTI. 
Additional results, including dependence of FTI stability on valley-anisotropic interactions and displacement field, are provided in App.~\ref{secapp:additional_tMoTe2}.

Figure~\ref{fig:main_tMoTe2_ED}b illustrates that $s_9/\Delta_9<1$ over a range of finite $g$ for $N_s=15$ and $N^1_\text{max}=0$. To connect more quantitatively with the LLL limit, we use the geometric relation~\cite{Liu2015characterization} for the effective magnetic length $\ell_B^*=\sqrt{\frac{A_M}{2\pi}}=2.02\,\text{nm}$ for $\theta=3.7^\circ$, where $A_M$ is the area of the unit cell in $t\text{MoTe}_2$. This allows for evaluation of the pseudopotentials $V_m$ for the intralayer interaction potential, revealing that these FTI states lie in the range $1.03<V_0/V_1<1.18$. This lends support to the notion that, akin to the LLL model, the ratio $V_0/V_1$ is a key indicator for whether an interaction potential admits FTI states. We note that for the potential $V_{ll'}(q)\propto \frac{\tanh{\frac{q\xi}{2}}}{q}$, obtained by setting $w_{\text{MoTe}_2}=d_{\text{MoTe}_2}=0$ and ignoring the hBN anisotropy, we do not find FTI states for \emph{any} $g$, while introduction of a finite $l_\text{RK}$, as in Eq.~\ref{eq:main_LLL_RK}, yields FTIs (see Figs.~\ref{fig:ratio_compiled_sizes} and \ref{fig:ratio_compiled_sizes_l_RK_3} in App.~\ref{subsecapp:effectsparameters}). This implies that the higher pseudopotentials still play a quantitatively significant role, and underscores the importance of detailed analysis of the dielectric screening environment.

Fig.~\ref{fig:main_tMoTe2_ED}b also shows the effects of increasing the system size ($N_s=18$) and band mixing ($N^1_\text{max}=1$). In both cases, we observe a drift of the FTI region towards smaller values of $|g|$, meaning towards the realistic situation. However, current finite-size restrictions prevent an extrapolation to the thermodynamic limit with full band mixing. We have checked that the ground state lies in $S_z=0$ for the parameters shown in Fig.~\ref{fig:main_tMoTe2_ED}b. The system is fully spin-polarized for $g=0$ in 1BPV calculations, and previous work has shown that the ground state in 2BPV calculations is not fully polarized~\cite{Yu2024MFCI0}.
Here, the short-range attraction $g<0$ acts to decrease the repulsive energy between opposite valleys, allowing the possibility for full demagnetization in 1BPV calculations. Increasing $N^\text{1}_{\text{max}}$ also lowers the energy of the $S_z=0$ sector relative to the magnetized sectors~\cite{Yu2024MFCI0} (see Figs.~\ref{fig:remote_band_depol_3x3} and \ref{fig:remote_band_depol_3x4} in App.~\ref{subsecapp:spindepolarization}), ensuring that the FTI phase lies within the non-magnetic regime at $\nu=-4/3$.

In Fig.~\ref{fig:main_tMoTe2_ED}c, we fix $g=-1400$meVnm$^2$ and repeat the calculations for a range of continuum model parameters $\psi,w$ and $V$. The FTI phase roughly tracks where the Berry curvature fluctuations of the lowest valence band are smallest (dashed line), suggesting that the FTI benefits from uniform Berry curvature \cite{Roy}.
For smaller $|w/V|$, the FTI can survive to a significantly lower $|g|$ (see Fig.~\ref{fig:wVrat_g} in App.~\ref{subsecapp:insights}). In Fig.~\ref{fig:main_tMoTe2_ED}d, still keeping $g$ fixed, we instead vary $\theta$ and introduce an artificial factor $\kappa_\text{SP}$ that multiplies the kinetic term. As the twist angle is reduced, the FTI phase attains a lower $s_9/\Delta_9$ and is realized at larger $\kappa_\text{SP}$, including the physical value $\kappa_\text{SP}=1$. 
The FTI phase roughly tracks where the Hartree renormalized bandwidth is minimal (see Fig.~\ref{fig:theta_U_with_Hartree} in App.~\ref{subsecapp:insights}). We caution that the actual continuum model parameters, as fitted to {\it ab initio} data, are expected to vary with $\theta$~\cite{zhang2024polarizationdriven}, but not considerably over the range shown in Fig.~\ref{fig:main_tMoTe2_ED}. Interestingly, 1BPV calculations projected only into band 1 can stabilize FTIs for significantly lower values of $|g|$ (see Fig.~\ref{fig:ED_band1} in App.~\ref{subsecapp:insights}). This suggests that FTIs may potentially be obtained also at higher hole filling factors such as $\nu=-10/3$ by fully hole-occupying band 0 and partially hole-occupying band 1.

\textit{Dielectric screening.---} Changes to the dielectric environment can affect the functional form of the interaction potential and help stabilize an FTI. The 1BPV calculations of Fig.~\ref{fig:main_tMoTe2_dielectric}b demonstrate that the required values of $|g|$ reduce as the relative permittivity of MoTe$_2$ increases. An enhanced $\epsilon^{\perp,\parallel}_{\text{MoTe}_2}$, which could arise due to screening from remote moir\'e bands, acts to confine the electric field lines within the $t$MoTe$_2$ system, which relatively suppresses the short-distance repulsion. Another way to selectively weaken the short-range repulsion is to surround $t$MoTe$_2$ with thin spacers with moderately high\footnote{If $\epsilon_{\text{spacer}}$ is too large, then $V_0/V_1$ is enhanced because the spacer effectively acts as a metallic gate for $\epsilon_{\text{spacer}}\rightarrow\infty$. $V_0/V_1$ can also be substantially reduced with small but unphysical $\epsilon_{\text{spacer}}\ll 1$, which confines the field lines to the $t$MoTe$_2$ (see Fig.~\ref{fig:nonmonotonic} in App.~\ref{subsecapp:V0V1results}).} dielectric constant (Fig.~\ref{fig:main_tMoTe2_dielectric}a), which works to impede the electric field lines from spreading into the encapsulating hBN.
For example with $\epsilon^{\perp,\parallel}_{\text{spacer}}=26$ appropriate for HfO$_2$~\cite{Wilk2001highk}, increasing $w_\text{spacer}$ (Fig.~\ref{fig:main_tMoTe2_dielectric}c) leads to a drift of the FTI region to smaller $|g|$, approximately following a stripe of constant $V_0/V_1$. The inset shows that the spacer increases the effective RK length. Our current calculations find that the spacer needs to be quite narrow $\lesssim 0.4\,\text{nm}$ to preserve the FTI.

\textit{Discussion.---} In our investigation of FTIs we use realistic $t$MoTe$_2$ parameters, with  the only extra ingredient invoked being a short-range attraction $g$. A possible origin for this is electron-phonon coupling, which has been argued to influence the balance between competing phases in moir\'e graphene~\cite{blason2022local,kwan2023electronphonon,shi2024moire}. In App.~\ref{secapp:EPC}, we consider the monolayer phonons, yielding a contribution $g\simeq -20\,\text{meVnm}^2$ which is significantly smaller than the scale $g\sim -1000\,\text{meVnm}^2$ that is needed for the FTI in our calculations. 
While the precise size of this discrepancy can be affected by other sources of short-range interaction, we discuss the possibility that our current finite-size ED studies overestimate the required strength of short-range attraction. 
First, our calculations of the FTI in both $t$MoTe$_2$ and the LLL model exhibit appreciable drifts towards weaker $|g|$ for larger system sizes. The former also displays a similar drift with respect to band mixing (Fig.~\ref{fig:main_tMoTe2_ED}b). It would be interesting to investigate other methods of incorporating additional bands, such as those leveraged to treat Landau level mixing in the FQH setting~\cite{Rezayi2011,Zalatel2015,Herviou2023,Rezayi2013,LLmix1,LLmix2}.  Second, the stability of the FTI varies with twist angle $\theta$ and the band structure parameters (Fig.~\ref{fig:main_tMoTe2_ED}c,d), and our extensive calculations have so far neither optimized these choices, nor considered other filling factors in detail. We note that the continuum model parameters change significantly as a function of $\theta$~\cite{zhang2024polarizationdriven} and for different TMDs. 
Theoretical work has indicated that the second valence band of low-angle $\theta\approx 2.1^\circ$ $t$MoTe$_2$ exhibits similarities with the non-relativistic $n=1$ LL~\cite{xu2024multiple,ahn2024landau,wang2024higher}. Our calculations show that this LL does not yield an FTI, which suggests that an FTI in $\theta\approx 2.1^\circ$ $t$MoTe$_2$ at $\nu=-8/3$ or $-10/3$ is unlikely, at least in the absence of band-mixing. Interestingly, we find that resemblance to the Dirac $n=1$ LL may support an FTI with a smaller $|g|$ compared to the LLL. 
Third, the details of dielectric screening play an important role in shaping the interaction potential, as evidenced in the dependence of the FTI phase on $\epsilon_{\text{MoTe}_2}$ (Fig.~\ref{fig:main_tMoTe2_dielectric}b). As we have briefly discussed, dielectric engineering can help reduce the required $|g|$ (Fig.~\ref{fig:main_tMoTe2_dielectric}c), an approach that should be explored in future work. 
Therefore, we believe that the $t$MoTe$_2$ material platform warrants further attention to determine whether the experimental realization of an FTI is feasible here. We note that our calculations are \emph{unbiased}, and we consider a large number of realistic effects in TMDs.
The fractionalized helical edge modes of the FTI, which could be locally imaged~\cite{ji2024local}, are stable against $S_z$-preserving perturbations, and would yield quantized signatures in (non-local) transport measurements (see App.~\ref{secapp:edgemodes}).

Given the similarities between the results for the FTI in $\theta\approx3.7^\circ$ $t\text{MoTe}_2$ and the LL models, we believe our work has ramifications for obtaining FTIs in other systems. In particular, we expect the pseudopotential ratio $V_0/V_1$ will still play a crucial role in stabilizing FTI phases. We believe density matrix renormalization group studies that can analyze the drift of $|g|$ can reveal whether, as we speculate, the FTI could appear under real-world sample conditions.

\textit{Note added.---} During the preparation of this manuscript, an experimental report~\cite{kang2024observation,kang2024evidence_nature} of a state with a $R_{xx}$ plateau but without a $R_{xy}$ plateau in  $\theta\approx 2.1^\circ$ $t$MoTe$_2$ at $\nu=-3$ appeared; it was suggested, but not yet confirmed, that this could be evidence for the FQSHE. This  was followed by several theoretical works~\cite{zhang2024vortex,maymann2024theory,jian2024minimal,villadiego2024halperin,zhang2024nonabelianvortex,chou2024composite,abouelkomsan2024nonabelian,paperYu} discussing possible related aspects. 
We note that no prior theory has directly addressed the feasibility of FTIs or FQSH states in $t$MoTe$_2$ using microscopic unbiased calculations.
A calculation or demonstration of a fractionalized state at $\theta\approx 2.1^\circ$ $t$MoTe$_2$ at $\nu=-3$ is still missing, and must be able to explain the continuous $R_{xy}$ dependence on density.

\textit{Acknowledgements}.--- We acknowledge helpful discussions with Claudio Chamon, Valentin Crépel, Kin Fai Mak, Christopher Mudry, Shinsei Ryu, Steve Simon, Jie Shan and Sanfeng Wu. Y.H.K is supported by a postdoctoral research fellowship
at the Princeton Center for Theoretical Science.  G.W. acknowledges funding from the University of Zurich postdoc grant FK-23-134. 
J.Y. acknowledges support from the Gordon and Betty Moore Foundation through Grant No. GBMF8685 towards the Princeton theory program.
T.N.\ and A.K.D.\ acknowledge support from the Swiss National Science Foundation through a Consolidator Grant (iTQC, TMCG-2-213805). 
Y.J. is supported by the European Research Council (ERC) under the European Union’s Horizon 2020 research and innovation program (Grant Agreement No. 101020833).
XX acknowledges the support from AFOSR under the award FA9550-21-1-0177. 
B.A.B.’s work was primarily supported by the the Simons Investigator Grant No. 404513, by the Gordon and Betty Moore Foundation through Grant No. GBMF8685 towards the Princeton theory program, Office of Naval Research (ONR Grant No. N00014-20-1-2303), BSF Israel US foundation No. 2018226 and NSF-MERSEC DMR-2011750, Princeton Global Scholar and the European Union’s Horizon 2020 research and innovation program under Grant Agreement No 101017733 and from the European Research Council (ERC), as well  as from the Simons Collaboration on New Frontiers in Superconductivity. N.R. also acknowledges support from the QuantERA II Programme that has received funding from the European Union’s Horizon 2020 research and innovation programme under Grant Agreement No 101017733 and from the European Research Council (ERC) under the European Union’s Horizon 2020 Research and Innovation Programme (Grant Agreement No. 101020833).


%\bibliography{fti}

%\setcounter{figure}{0}
%\let\oldthefigure\thefigure
%\renewcommand{\thefigure}{S\oldthefigure}
%\renewcommand{\thefigure}{S\arabic{figure}}

\setcounter{figure}{0}
\renewcommand{\thefigure}{S\arabic{figure}}
\renewcommand*{\theHfigure}{\thefigure}

\setcounter{table}{0}
\renewcommand{\thetable}{S\arabic{table}}

\newpage
\clearpage

\begin{appendix}
\onecolumngrid
	\begin{center}\textbf{\large --- Appendix ---}

\end{center}

\tableofcontents

\clearpage

\section{Opposite-fields Lowest Landau level model}\label{secapp:LLL}

In this section, we describe in detail a set of ED calculations within a toy model which comprises a pair of lowest Landau levels (LLLs) where the two valleys $\eta=\pm$ experience magnetic fields oriented along opposite directions. Note that we can use the terms `spin' and `valley' interchangeably, since they are locked together in the $t$MoTe$_2$ context owing to strong spin-orbit coupling. In App.~\ref{app:subsec:LLL_setup}, we describe the model and details of the ED implementation. In App.~\ref{app:subsec:LLL_V0V1}, we study the phase diagram within a restricted space of Haldane pseudopotentials $V_m^{\eta\eta'}$ parameterizing the interaction. In App.~\ref{app:subsec:LLL_screened} and \ref{secapp:LLL_RK}, we consider more realistic interactions corresponding to variations of the gate-screened Coulomb interaction. In  App.~\ref{subsecapp:otherLLs}, we discuss how the physics changes for other Landau levels beyond the LLL.

\subsection{Model and methods}\label{app:subsec:LLL_setup}

Consider the single-particle Hamiltonian of  electrons of charge $-e$ confined to a 2D plane and minimally coupled to a valley-dependent vector potential $\mbf{A}^\eta(\mbf{r})$
\begin{equation}
    \hat{H}_0^\eta=\frac{(\mbf{p}+e\mbf{A}^\eta(\mbf{r}))^2}{2m},
\end{equation}
where $\nabla\times\mbf{A}^\eta=\eta(0,0,B)^T$, $B$ is the magnetic field, and $\eta=+$ ($-$) labels electrons in valley $+$ ($-$). $\mbf{p}$ and $\mbf{r}$ are the 2D momentum and position operators, respectively. We define the magnetic length $\ell_B={\frac{h}{eB}}$ which will often be set to unity. The system preserves time-reversal symmetry $\mathcal{T}$ since the magnetic fields in the two valley sectors are equal and opposite, and the corresponding LLs have opposite Chern numbers $C^\eta=\eta$. Furthermore, the Hamiltonian has a $U_\text{v}(1)$ valley conservation symmetry whose generator is $2\hat{S}_z=\hat{N}_+-\hat{N}_-$, where $\hat{N}_\eta$ is the number operator in valley $\eta$. We construct the single-particle Hilbert space by projecting to the lowest Landau levels (LLLs) in the two valley sectors. We refer to this system simply as the LLL model, with the understanding that it always corresponds to the opposite magnetic fields setup described above. Similar setups have been considered previously in~\cite{Chen2012FQHtorusFTI,Mukherjee2019FQHsphereFTI,Bultinck2020mechanism,furukawa2014global,zhang2018composite,kwan2021exciton,kwan2022hierarchy,eugenio2020DMRG,stefanidis2020excitonic,chatterjee2022dmrg,myersonjain2023conjugate,Wu2024,shi2024excitonic,yang2023phase,kwan2024textured}.

To incorporate interactions into the LLL model, we consider $\mathcal{T}$- and $S_z$-preserving valley-dependent interactions of the form (in the continuum)
\begin{equation}
    \hat{H}_\text{int}=\frac{1}{2}\sum_{\eta\eta'}\int \frac{d^2\mbf{q}}{(2\pi)^2}\,V^{\eta\eta'}(q)\rho^{\eta}_{\text{LLL}}(\mbf{q})\rho^{\eta'}_{\text{LLL}}(-\mbf{q})
\end{equation}
where $\rho^{\eta}_{\text{LLL}}(\mbf{q})$ is the LLL-projected density operator. While the bare interaction potential $V^{\eta\eta'}(q)$ is isotropic in real 2D space, we allow for anistropies in valley space. The symmetries require
\begin{equation}\label{app:eq:PP_valleysym}
    V^{++}(q)=V^{--}(q),\quad V^{+-}(q)=V^{-+}(q)
\end{equation}
such that we will often only refer to the $++$ (intravalley) and $+-$ (intervalley) components explicitly. In certain situations, including for some of the $t\text{MoTe}_2$ calculations in this work, we will consider uniformly scaling the intervalley interaction relative to the intravalley interaction
\begin{equation}\label{app:eq:lambda}
    V^{+-}(q)=\lambda V^{++}(q)
\end{equation}
where $\lambda$ is the valley anistropy parameter and $\lambda =1$ its physical value. 

In the LLL model, it is convenient to parameterize the interaction potential in terms of Haldane pseudopotentials 
\begin{equation}\label{appeq:Haldane_PP}
    V^{\eta\eta'}_m\equiv \int \frac{d^2\mbf{q}}{(2\pi)^2}V^{\eta\eta'}(q)L_m(q^2\ell_B^2)e^{-q^2\ell_B^2}
\end{equation}
where $L_m(x)$ is the Laguerre polynomial. $V^{\eta\eta'}_m$ characterizes the interaction energy of two particles with relative angular momentum $m$. Note that $V^{\eta\eta}_m$ for $m$ even does not affect the physics due to fermionic statistics. 

We perform ED calculations on a torus with magnetic periodic boundary conditions, and restrict to square geometries with aspect ratio of 1. In terms of the number of flux quanta $N_\Phi$ and particle numbers $N_+,N_-$, the filling factors are defined by $\nu_+=N_+/N_\Phi$ and $\nu_-=N_-/N_\phi$. The Hamiltonian has a particle-hole symmetry (PHS) that relates $(\nu_+,\nu_-)\leftrightarrow (1-\nu_+,1-\nu_-)$. We will mostly work at filling $\nu_+=\nu_-=1/3$, but owing to PHS, this is analogous to the $\nu=-2/3-2/3$ situation appropriate for $t\text{MoTe}_2$ (if band mixing is neglected). To reduce the computational cost, we exploit the many-body translation symmetries~\cite{Haldane1985manybody,Bernevig2012emergent}  
and diagonalize within many-body momentum sectors labelled by $\mbf{K}=(K_x,K_y)$, where $K_x$ and $K_y$ take values $0,\ldots,N_\Phi-1$. 

To diagnose the presence of FTIs in the finite-size numerics, we consider properties of the many-body spectrum for $S_z=0$. Focusing on the $\nu_+=\nu_-=1/3$ FTI, we first identify the momenta corresponding to the 9 (nearly)-degenerate FTI ground states. These can be determined by considering the limit of two decoupled $\nu_\eta=1/3$ FQH Laughlin states in the two valley sectors. For example for $N_\Phi=15$, this yields a single state for each of $(K_x,K_y)=(5i_x,5i_y)$ with $i_x,i_y=0,1,2$. We define the FTI spread $\delta_9$ as the energy bandwidth of the 9 lowest states with these quantum numbers, i.e.~we postulate that these 9 states form the FTI ground state manifold. We define the FTI spacing $s_9$ as the maximum energy difference between adjacent levels in the FTI manifold, irrespective of momentum. The FTI gap $\Delta_9$ is defined as the energy of the lowest state not in this manifold minus the energy of the highest energy state in this manifold. Note that the gap $\Delta_9$ defined this way can be negative, which certainly rules out the possibility of an FTI phase. Figure~\ref{fig:V0V1_spectra}a illustrates these definitions for a system deep in the FTI regime. 
The condition that the spread is less than the gap (i.e.~$\delta_9<\Delta_9$ for our problem) is commonly used as a condition for the identification of the topological phase in the FCI/FTI literature.
In this work, we take the less stringent condition of the spacing/gap ratio $s_9/\Delta_9$ being less than 1 as a necessary condition for determining an FTI phase. 
The rationale behind this criterion is that to visually separating the 9 low energy states from the higher energy states requires $s_9<\Delta_9$. 
The presence of a path in parameter space where the spacing/gap ratio remains small and connects to the limit of decoupled FQH states demonstrates adiabatic continuity and provides further evidence of an FTI phase. By considering other $S_z$ sectors, we can also determine the presence of a positive valley gap.

\subsection{$V_{0}^{+-}-V_{1}^{++}-V_{1}^{+-}$ phase diagram}\label{app:subsec:LLL_V0V1}

\begin{figure}
    \centering
    \includegraphics[width=0.7\linewidth]{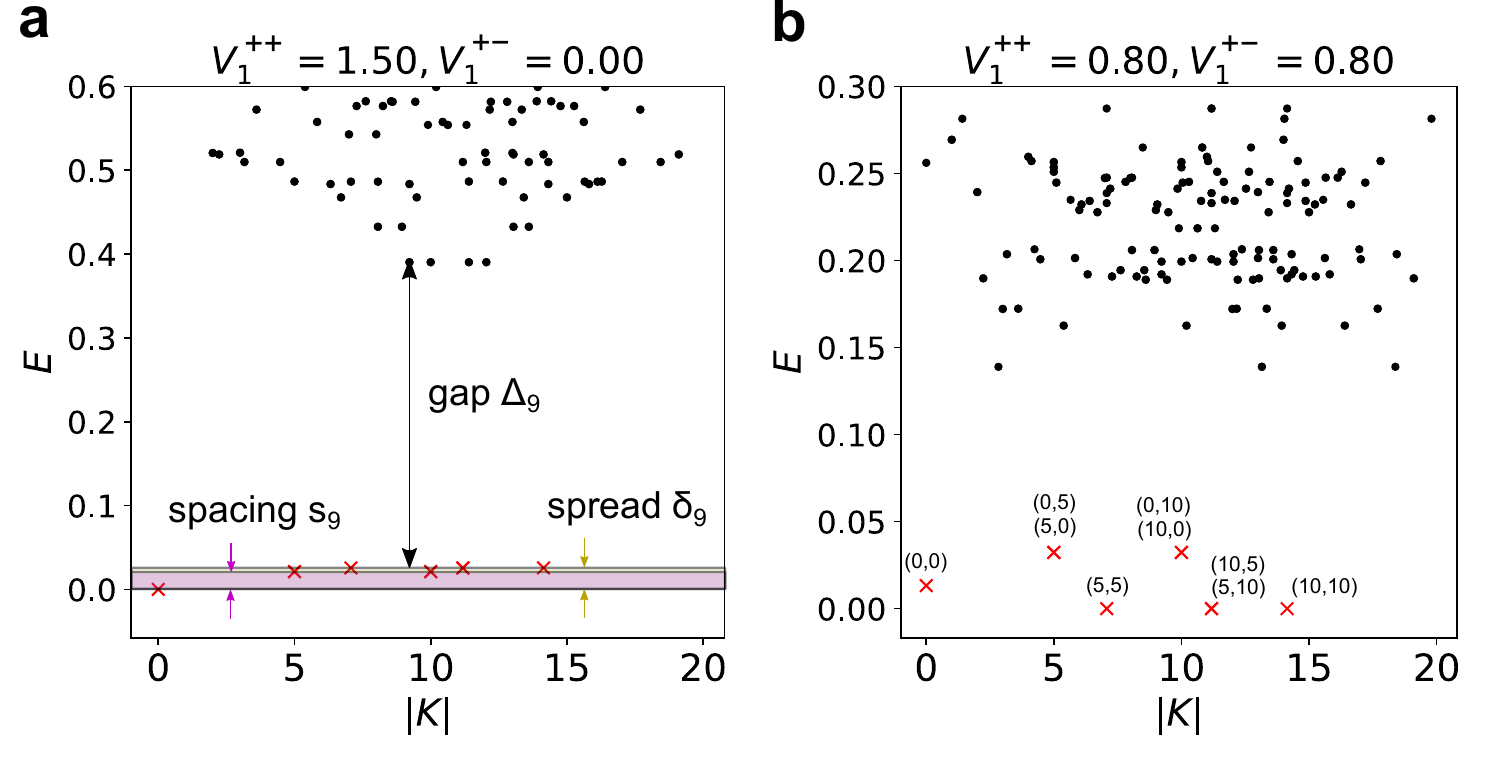}
    \caption{\textbf{Exact diagonalization spectra for FTI phases in the LLL model at $\nu_+=\nu_-=1/3$ for selected $V_{1}^{++}/V_{0}^{+-}$ and $V_{1}^{+-}/V_{0}^{+-}$.} $|K|=\sqrt{K_x^2+K_y^2}$ denotes the magnitude of the many-body momentum. The lowest 4 states (lowest state) are kept for momentum sectors that contain (do not contain) a state in the FTI ground state manifold. The 9 states in the FTI ground state manifold are denoted with red crosses, and their momenta $(K_x,K_y)$ are labelled for b). Note that some ground states are overlapping in the figures, e.g. those with momenta $(K_x,K_y)=(0,5)$ and $(5,0)$. a) illustrates how the FTI 
    spacing $s_9$, spread $\delta_9$  and gap $\Delta_9$ are defined. $V_0^{+-}=1$ for both plots. In this case, $s_9$ is only slightly smaller than $\delta_9$. Square torus geometry with $N_\Phi=15$ flux quanta.}
    \label{fig:V0V1_spectra}
\end{figure}

\begin{figure}[t]
    \centering
    \includegraphics[width=0.8\linewidth]{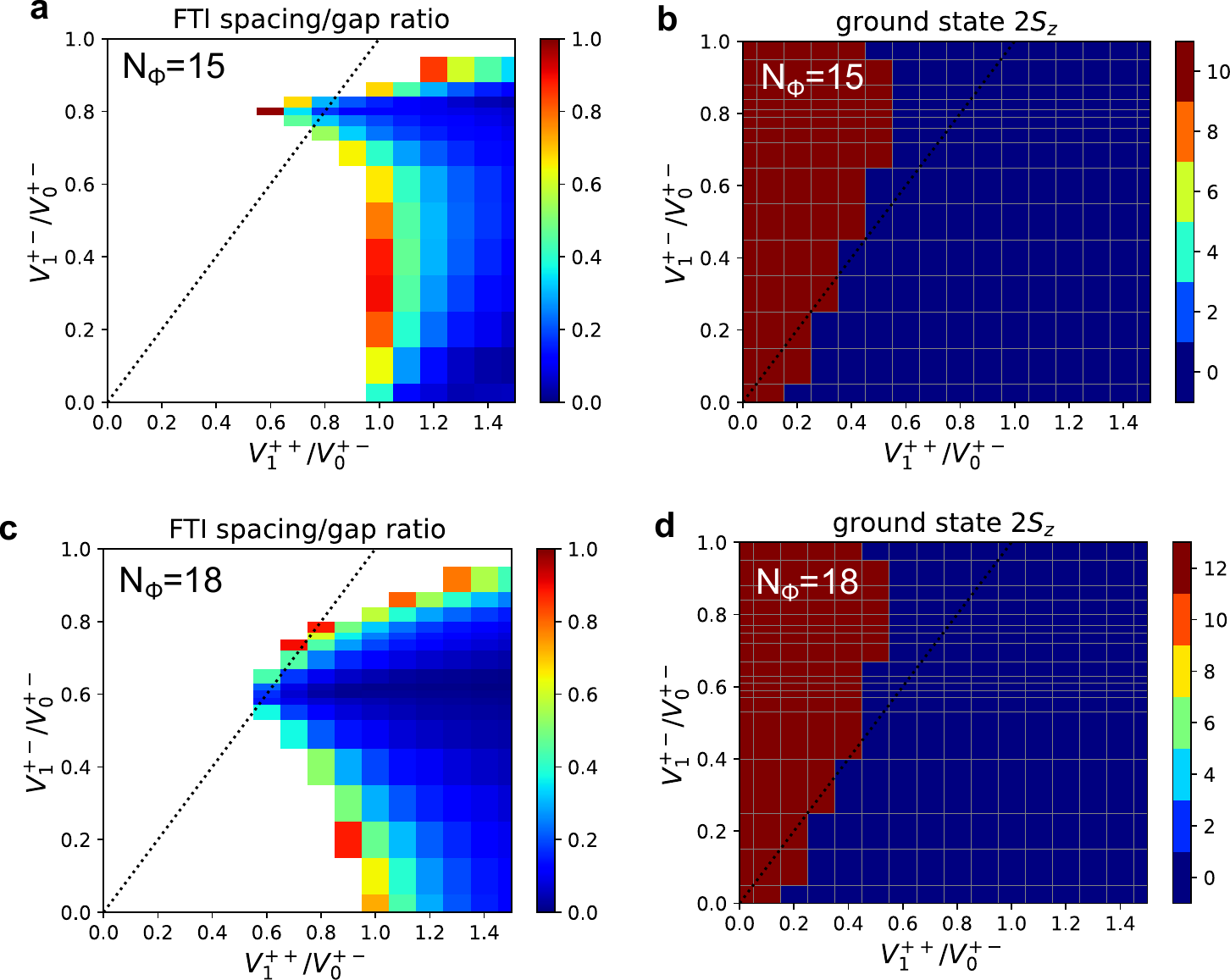}
    \caption{\textbf{FTI spacing/gap ratio and ground state valley polarization as a function of pseudopotentials $V_{1}^{++}/V_{0}^{+-}$ and $V_{1}^{+-}/V_{0}^{+-}$ in the LLL model at $\nu_+=\nu_-=1/3$.} All other pseudopotentials are set to zero. a) See text for definitions of the spacing $s_9$ and gap $\Delta_9$. White regions correspond to where the spacing is greater than the gap, or the gap is negative. Dotted line indicates isotropic interactions $V_{1}^{++}=V_{1}^{+-}.$ $N_\Phi=15$ flux quanta. b) Valley polarization $2S_z=N_+-N_-$ of the ground state across all valley sectors for the same grid of pseudopotentials as in a). c,d) Same as a,b) except for $N_\Phi=18$ flux quanta.  All plots use square torus geometry.}
    \label{fig:V0V1_phase_spin_combined}
\end{figure}

\begin{figure}
    \centering
    \includegraphics[width=0.75\linewidth]{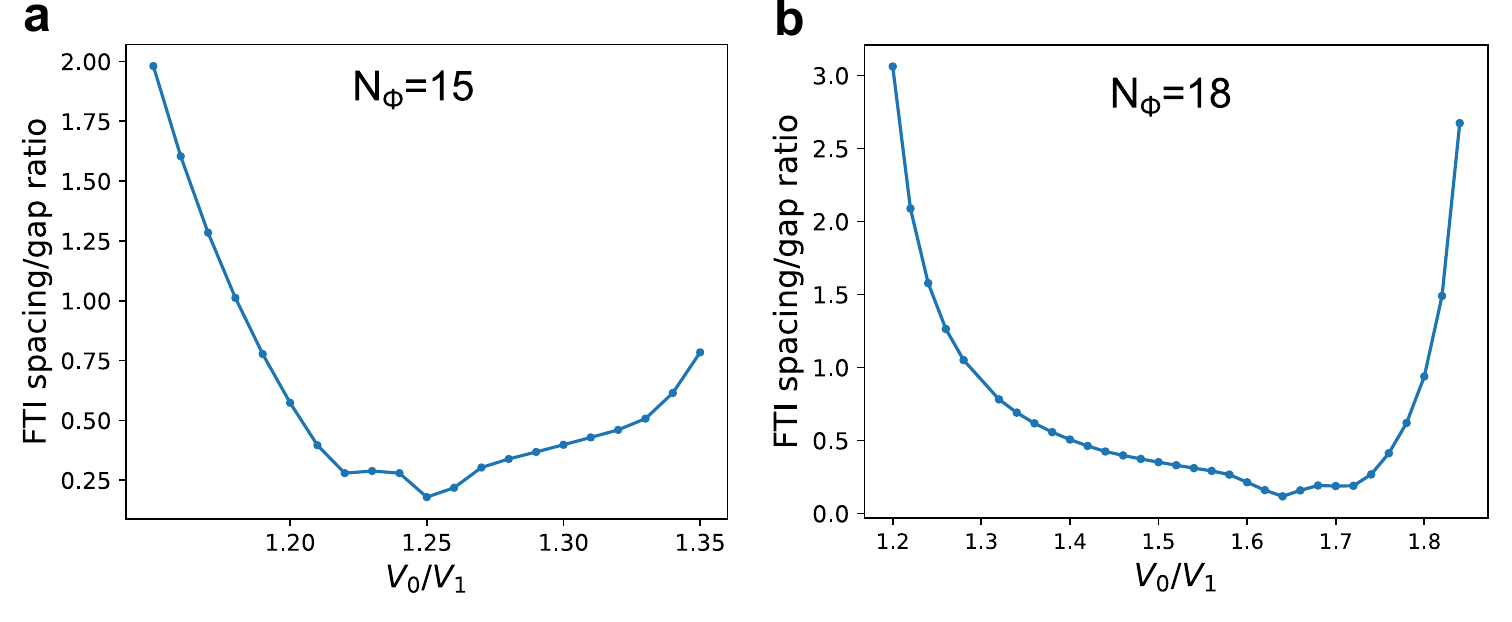}
    \caption{\textbf{FTI spacing/gap ratio as a function of valley-isotropic pseudopotentials $V_{0}^{\eta\eta'}=V_0$ and $V_{1}^{\eta\eta'}=V_1$ in the LLL model at $\nu_+=\nu_-=1/3$.} All other pseudopotentials are set to zero. Square torus geometry with a) $N_\Phi=15$, and b) $18$ flux quanta.}
    \label{fig:isotropic_V0_V1}
\end{figure}

A model $\nu_+=\nu_-=1/3$ FTI state can be trivially obtained by including only a single non-vanishing pseudopotential $V_1^{++}$ (recall from Eq.~\ref{app:eq:PP_valleysym} that $V_1^{--}=V_1^{++}$ due to the valley symmetries). This corresponds to exact $\nu=1/3$ Laughlin states in each valley sector, which are decoupled due to the absence of intervalley interactions. The stability of the FTI against including intervalley interactions was partially addressed by Ref.~\onlinecite{Chen2012FQHtorusFTI}, which carried out ED calculations for the LLL model on the torus for $N_\Phi=15$. In particular, they investigated the phase diagram in the $S_z=0$ sector as a function of the ratio $V_0^{+-}/V_1^{++}$, with all other pseudopotentials, including $V_1^{+-}$, set to zero. For large and repulsive $V_0^{+-}/V_1^{++}$, Ref.~\onlinecite{Chen2012FQHtorusFTI} found that the system was gapless and phase separated into regions of opposite valley polarization. Interestingly, for $V_0^{+-}/V_1^{++}=1$, Ref.~\onlinecite{Chen2012FQHtorusFTI} found that the system remained in the FTI phase based on computing wavefunction overlaps with the model wavefunction corresponding to decoupled Laughlin states. However, we emphasize that $V_0^{+-}/V_1^{++}=1$ does not correspond to a valley-isotropic interaction, since the intervalley interaction is ultra short-range $\sim\delta(\mbf{r})$, while the intravalley interaction has a longer range $\sim\nabla^2\delta(\mbf{r})$. In contrast, the interaction potential relevant for moir\'e materials such as $t\text{MoTe}_2$ is expected to be isotropic in valley space. To the best of our knowledge, there are no prior theoretical studies that have demonstrated an FTI phase for valley-isotropic interactions in a fermionic system, whether for LLs or lattice systems.

In order to address this gap, we first consider an expanded parameter space consisting of pseudopotentials $V_0^{+-},V_1^{++},V_1^{+-}$ in the LLL model. Without loss of generality, we set $V_0^{+-}=1$. As mentioned above, $V_0^{++}$ does not need to be considered since only the odd pseudopotentials are relevant for the intravalley interaction owing to fermionic statistics. Crucially, we include parameter sets which lie along the valley-isotropic line $V_1^{++}=V_1^{+-}$. The results for the FTI spacing/gap ratio for $N_\Phi=15$ and $S_z=0$ are presented in Fig.~\ref{fig:V0V1_phase_spin_combined}a, where parameters that yield ground states that are consistent with our criteria for an FTI phase (see Sec.~\ref{app:subsec:LLL_setup}) are colored. $V_1^{++}\rightarrow\infty$ on the horizontal axis corresponds to the idealized limit of decoupled FQH states and vanishing spacing/gap ratio. In agreement with Ref.~\onlinecite{Chen2012FQHtorusFTI}, we find that the FTI survives to $V_1^{++}=V_0^{+-}$, though it does not persist for much larger $V_0^{+-}$. It is clear that decreasing $V_1^{++}$ generally is deleterious for the FTI. 

Curiously, we find non-monotonic behavior of the spacing/gap ratio as a function of $V^{+-}_1$ in the FTI region at $V_1^{++}\simeq V_0^{+-}$. This suggests that a finite $V_1^{+-}$ can actually be beneficial for the FTI phase, which runs counter to the expectation that intervalley interactions are harmful to the FTI. Strikingly, we observe a small window where an FTI is stabilized with valley-isotropic interactions (see dotted line in Fig.~\ref{fig:V0V1_phase_spin_combined}a).  We find that this occurs for $V_1^{++}=V_1^{+-}\simeq 0.8 V_0^{+-}$, and the spacing/gap ratio remains small along a parameter path connected to the ideal decoupled FQH limit. The many-body spectrum is plotted in Fig.~\ref{fig:V0V1_spectra}b, and exhibits a spacing/gap ratio $\simeq0.3$. 

Figure~\ref{fig:V0V1_phase_spin_combined}b shows the valley polarization $S_z$ of the ground state across all valley sectors (note that the maximum $S_z=N_+-N_-=\frac{2N_\Phi}{3}$ for $\nu_++\nu_-=\frac{2}{3}$). For small $V_1^{++}$, the ground state lies in the fully magnetized sector. However, for the parameters where the lowest states in the $S_z=0$ sector form an FTI, we find that the ground state of the system is indeed non-magnetized. 

Figure~\ref{fig:V0V1_phase_spin_combined}c shows analogous results for the $S_z=0$ phase diagram for $N_\Phi=18$. Compared to the $N_\Phi=15$ calculations, the window of FTI stability along the isotropic line is wider, and shifts to higher $V_0/V_1$. As demonstrated in Fig.~\ref{fig:V0V1_phase_spin_combined}d, the ground state of the system is non-magnetized for parameters where we find an FTI.

In Fig.~\ref{fig:isotropic_V0_V1}, we show additional results at $S_z=0$ along the valley-isotropic line $V_1^{\eta\eta'}=V_1$ for $N_\Phi=15,18$ and $S_z=0$. For the larger system size, the FTI stability window increases, and the minimum spacing/gap ratio decreases.

\subsection{Gate-screened Coulomb interactions}\label{app:subsec:LLL_screened}

\begin{figure}
    \centering
    \includegraphics[width=0.6\linewidth]{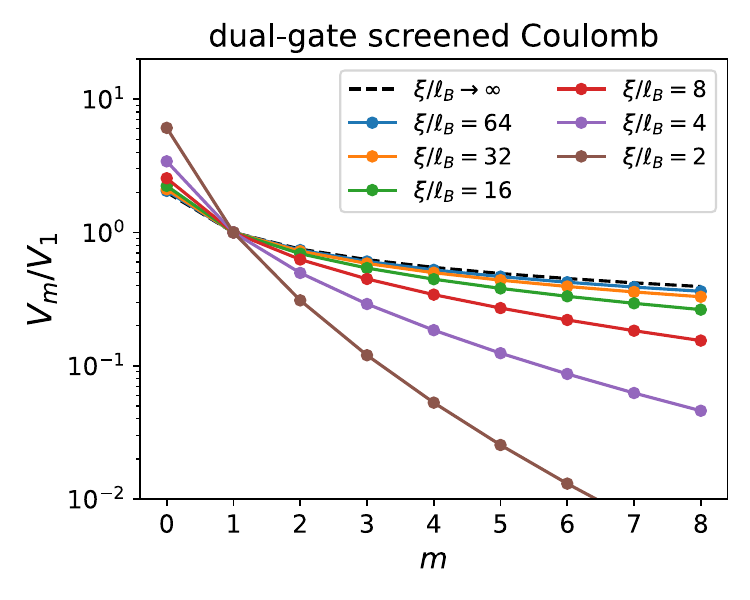}
    \caption{\textbf{LLL pseudopotentials $V_m$ of the dual-gate screened Coulomb interaction (Eq.~\ref{eq:dual_gate}) for different screening lengths $\xi/\ell_B$.} Dashed line corresponds to the unscreened Coulomb limit.}
    \label{fig:double_gated_pseudopotentials}
\end{figure}

Having demonstrated the existence of FTIs for valley-isotropic interactions within a restricted set of allowed pseudopotentials $V_0,V_1$, we now turn to more realistic interactions that are relevant for moir\'e systems. We consider the Coulomb potential screened by two metallic gates with separation $\xi$, and positioned symmetrically on either side of the 2D system
\begin{equation}\label{eq:dual_gate}
    V^{\eta\eta'}(q)=V_{\xi}(q)=\frac{e^2}{2\epsilon_0\epsilon q}\tanh{\frac{q\xi}{2}},
\end{equation}
where $\epsilon$ is the relative permittivity (whose value is irrelevant for the LLL model which lacks band dispersion). The corresponding pseudopotentials are plotted in Fig.~\ref{fig:double_gated_pseudopotentials}. In the absence of screening ($\xi/\ell_B\rightarrow\infty$), the Coulomb pseudopotentials $V_m\sim \binom{2m}{m}4^{-m}$ decay algebraically $\sim m^{-1/2}$ at large $m$. For finite $\xi$, they decay exponentially with sufficiently large $m$.  The limit of small $\xi/\ell_B$ is well approximated by the $(V_{0}^{+-},V_{1}^{++},V_{1}^{+-})$ setup in Sec.~\ref{app:subsec:LLL_V0V1} with $V_{1}^{++}=V_{1}^{+-}$ and large $V_{0}^{+-}/V_{1}^{++}$. 

\begin{figure}
    \centering  \includegraphics[width=0.8\linewidth]{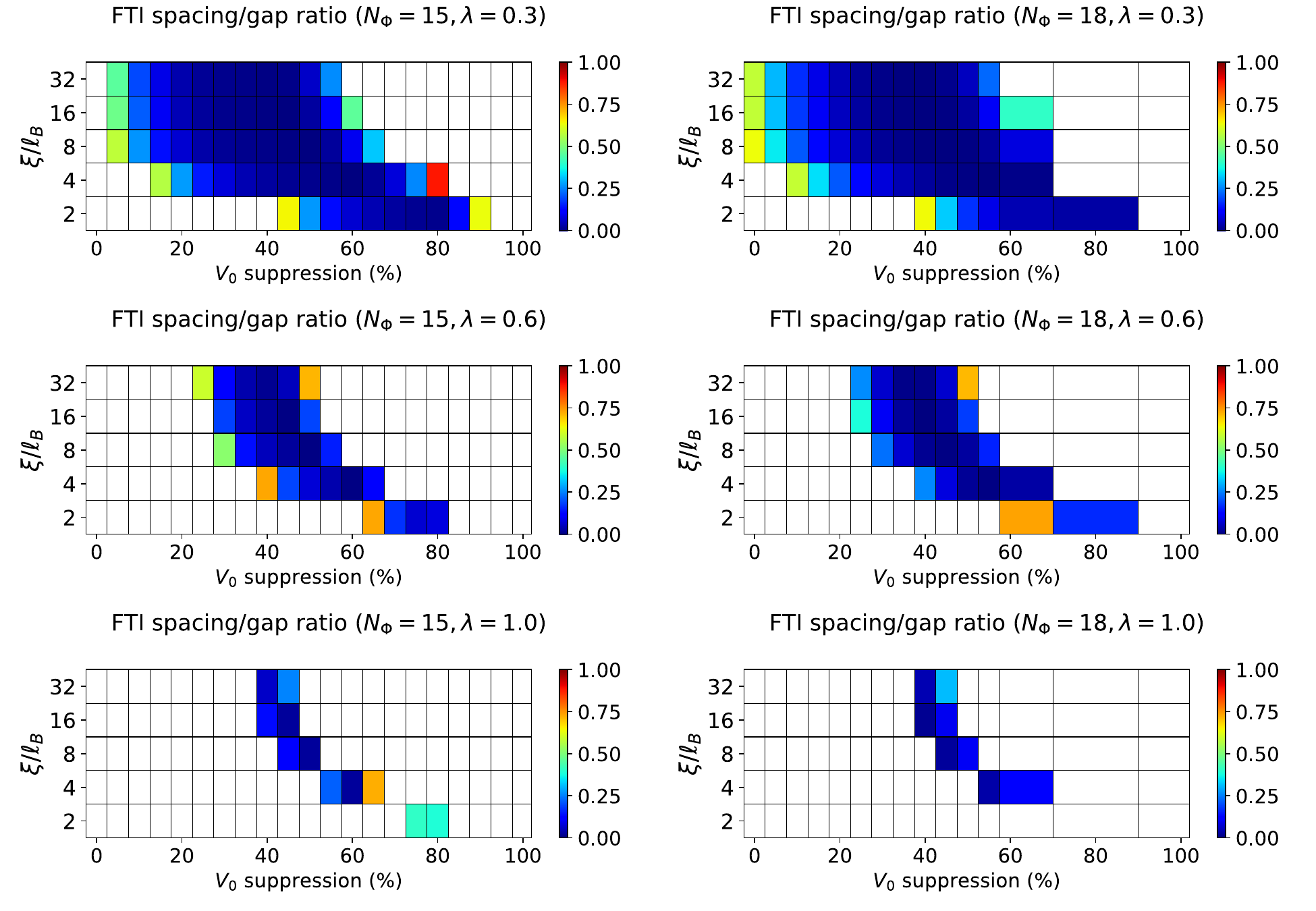}
    \caption{\textbf{FTI spacing/gap ratio as a function of $V_0$ suppression and dual-gate Coulomb screening length $\xi/\ell_B$ for different interaction anisotropy parameters $\lambda$ in the LLL model at $\nu_+=\nu_-=1/3$.} White regions correspond to where the spacing is greater than the gap, or the gap is negative. Square torus geometry with $N_\Phi=15,18$ flux quanta.}
    \label{fig:xi_V0frac_lambda}
\end{figure}

Reference~\onlinecite{Mukherjee2019FQHsphereFTI} has previously performed ED numerics on the LLL model on the sphere using the unscreened ($\xi/\ell_B\rightarrow\infty$) limit of Eq.~\ref{eq:dual_gate}, and found a compressible state at $\nu_+=\nu_-=1/3$ that is likely disordered. For the opposite limit $\xi/\ell_B\rightarrow0$, the system is phase-separated according to the discussion of Sec.~\ref{app:subsec:LLL_V0V1} since $V_0/V_1$ is large. Therefore, to expand the space of parameters and increase the chances of finding a stable FTI phase, we introduce an additional onsite attraction $\propto -\delta(\mbf{r})$, which is equivalent to subtracting a constant from $V_\xi(q)$ in Eq.~\ref{eq:dual_gate}, and only affects the $m=0$ pseudopotential $V_0$. In the following calculations, we parameterize the strength of the onsite attraction by the resulting percentage suppression of $V_0$ compared to using only the screened Coulomb interaction. 

\begin{figure}
    \centering
    \includegraphics[width=0.65\linewidth]{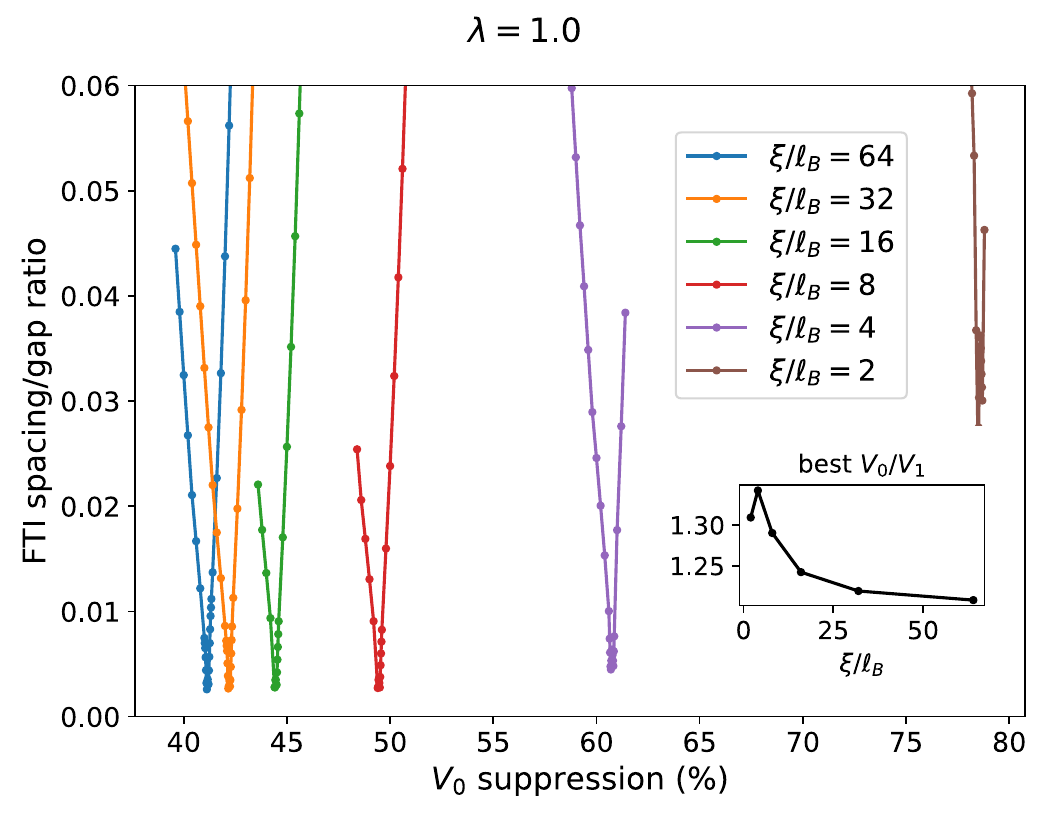}
    \caption{\textbf{FTI spacing/gap ratio as a function of $V_0$ suppression for different dual-gate Coulomb screening lengths $\xi/\ell_B$ in the LLL model at $\nu_+=\nu_-=1/3$.}  A $V_0^{+-}$ suppression of $0\%$ corresponds to the original unaltered dual-gate screened Coulomb potential. The inset shows the effective (suppressed) value of $V_0/V_1$ for the FTI with the lowest spacing/gap ratio, as a function of $\xi/\ell_B$. The interaction potential is isotropic in valley space ($\lambda=1$), such that $V_m^{++}=V_m^{+-}$. Square torus geometry with $N_\Phi=15$ flux quanta.}
    \label{fig:xi_V0frac_refined}
\end{figure}

Figure~\ref{fig:xi_V0frac_lambda} shows the FTI spacing/gap ratio for $N_\Phi=15,18$ as a function of the screening length $\xi/\ell_B$ and the suppression of $V_0$, for different values of the valley anisotropy $\lambda$ [Eq.~\eqref{app:eq:lambda}]. For $\lambda=0$ (not shown) corresponding to decoupled valley sectors, the gap is positive and the spacing is exactly zero (due to center-of-mass degeneracy) for all $\xi$ and $V_0$. For small $\lambda=0.3$, the region of FTI stability persists for a broad range of $V_0$ suppression. This shrinks significantly as $\lambda$ is increased, but even for valley-isotropic interactions ($\lambda=1$), we find that the FTI phase survives in a sliver of the phase diagram. We observe a slight drift towards less $V_0$ suppression for the FTI phase as $N_\Phi$ increases.

\begin{figure}\centering\includegraphics[width=0.5\linewidth]{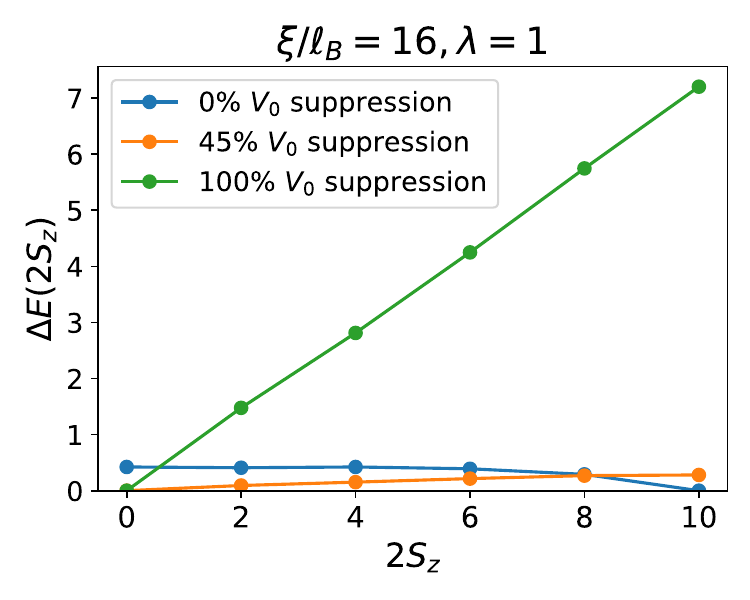}
    \caption{\textbf{Ground state energy across different magnetization sectors in the LLL model with dual-gate screening length $\xi/\ell_B=16$ at $\nu_++\nu_-=2/3$.}  $2S_z=N_+-N_-$ is the valley imbalance, and $\Delta E(2S_z)$ gives the ground state energy in valley sector $S_z$ measured relative to the minimum energy across all $S_z$. For $45\%$ $V_0$ suppression (orange), the system realizes an FTI phase in the $S_z=0$ sector, and the lowest excitation about the FTI ground state manifold remains in the $S_z=0$ sector. Square torus geometry with $N_\Phi=15$ flux quanta.}
    \label{fig:spin_xi16_combined}
\end{figure}

Figure~\ref{fig:xi_V0frac_refined} examines the region of FTI stability for $\lambda=1$ and $N_\Phi=15$ in more detail. The required suppression of $V_0$ decreases for larger $\xi$, mainly because the original $V_0/V_1$ for the gate-screened interaction becomes smaller as the gates are separated from each other (Fig.~\ref{fig:double_gated_pseudopotentials}), and tends to $V_0/V_1=2$ in the unscreened Coulomb limit. Indeed, the inset of Fig.~\ref{fig:xi_V0frac_refined} shows that the ideal value of suppressed $V_0/V_1$ remains relatively constant $\simeq 1.2-1.35$ across a wide range of screening lengths.  We also find the FTI is more robust for larger $\xi$, as evidenced by the minimum spacing/gap ratio $\lesssim 0.03$, compared with $\simeq 0.25$ for the effective $\xi\rightarrow 0$ limit in App.~\ref{app:subsec:LLL_V0V1}. Thus, it appears a longer screening length is preferable for obtaining FTIs because less suppression of $V_0$ is necessary, and the resulting states have a better spacing/gap ratio. This conclusion is somewhat surprising because if we consider a single decoupled valley sector, the FQH state at $\nu=1/3$ is closer to the ideal Laughlin state for shorter $\xi/\ell_B$ since the pseudopotentials decay more rapidly (though for the pure Coulomb limit $\xi\rightarrow\infty$, the Laughlin state still faithfully captures the low-energy physics there). This suggests, again, that the conditions for realizing an ideal FQH state do not strictly coincide with those for stabilizing an FTI for $\lambda=1$. Corresponding data for $N_\Phi=18$, $\xi/\ell_B=16$ are shown in Fig.~\ref{fig:LLL_pseudopotential_phase}b of the main text. 

Finally, we investigate the dependence of the ground state energy on valley polarization $S_z=\frac{1}{2}(N_+-N_-)$ at fixed total filling $\nu=2/3$. Figure~\ref{fig:spin_xi16_combined} shows that for $\xi/\ell_B=16$, using the unaltered gate-screened interaction leads to the global ground state (which is an FQH state) being in the fully valley-polarized sector. We expect a suppression of $V_0$ to favor valley depolarization, as the short-range repulsion between opposite-valley fermions is reduced. For $45\%$ $V_0$ suppression, corresponding to an FTI in the $S_z=0$ sector, we indeed find that the global ground state is valley unpolarized. Furthermore, the lowest energy state above the FTI manifold is also in the $S_z=0$ subspace. These findings are unchanged for the range of $\xi/\ell_B=2-64$ investigated.

\subsection{Rytova-Keldysh correction}\label{secapp:LLL_RK}

\begin{figure}
    \centering
    \includegraphics[width=0.5\linewidth]{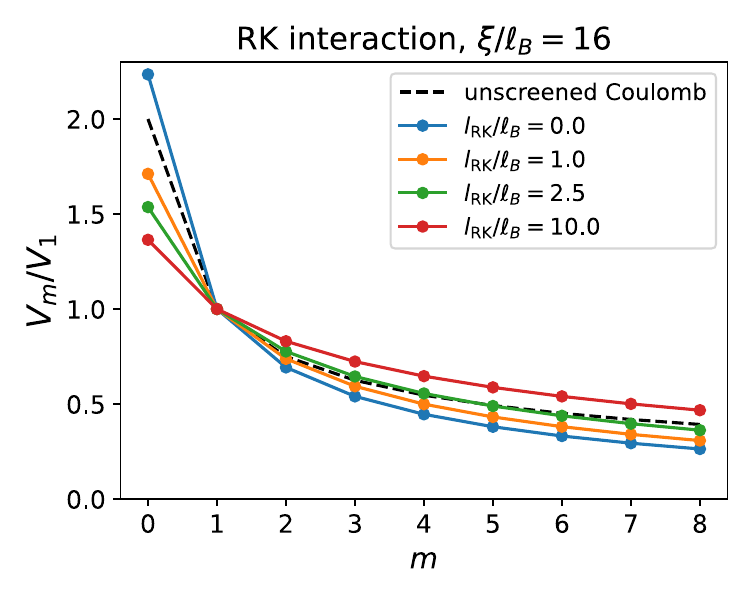}
    \caption{\textbf{LLL pseudopotentials $V_m$ of the dual-gate screened Coulomb interaction with RK corrections (Eq.~\ref{eq:LLL_RK}) for $\xi/\ell_B=16$.} Dashed line corresponds to the unscreened Coulomb limit ($\xi\rightarrow\infty,l_\text{RK}=0$).}
    \label{fig:double_gated_pseudopotentials_RK_xi16}
\end{figure}

\begin{figure}
    \centering  \includegraphics[width=0.5\linewidth]{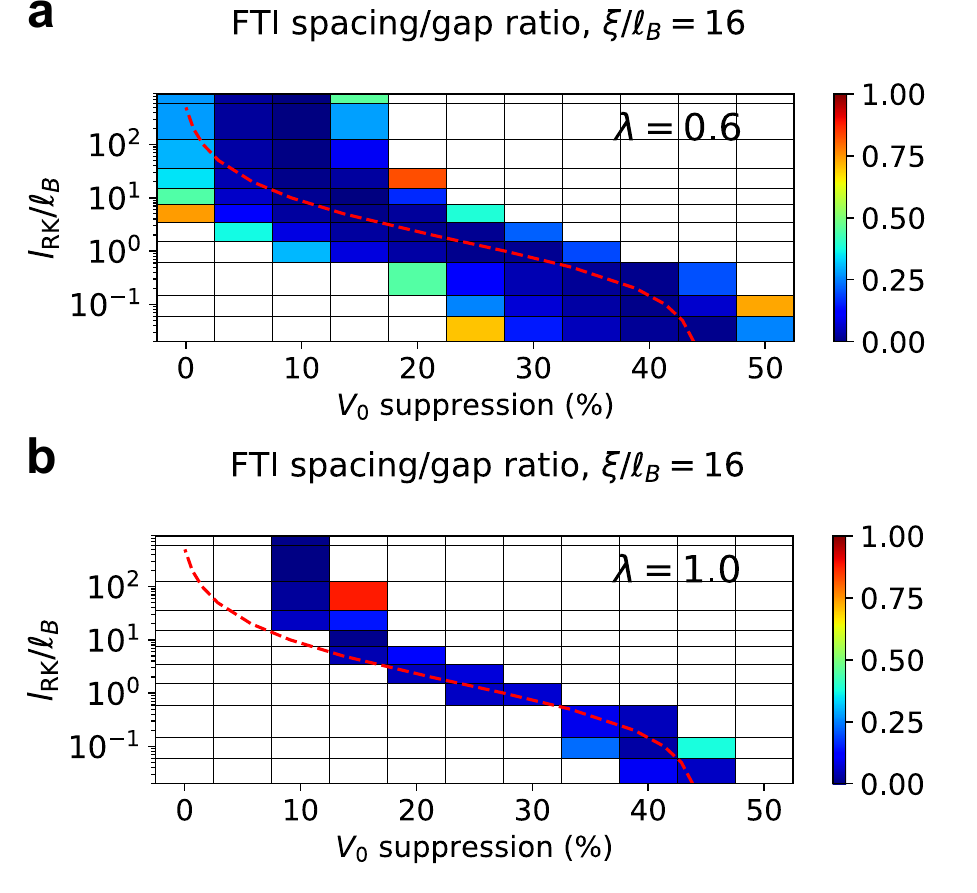}
    \caption{\textbf{FTI spacing/gap ratio as a function of $V_0$ suppression and RK length $l_\text{RK}$ for different interaction anisotropy parameters $\lambda$ in the LLL model at $\nu_+=\nu_-=1/3$.} Gate screening length $\xi/\ell_B=16$, and $\lambda=0.6,1.0$ in \textbf{a)} and \textbf{b)} respectively. White regions correspond to where the spacing is greater than the gap, or the gap is negative. Red dashed line is contour of constant $V_0/V_1=1.24$. Square torus geometry with $N_\Phi=15$ flux quanta.}
    \label{fig:RK_V0frac_lambda_xi16}
\end{figure}

Appendix~\ref{app:subsec:LLL_screened} demonstrated that for long-range gate-screened Coulomb interactions $V_\xi(q)$, the FTI can be stabilized if there is an additional short-range attraction that suppresses the $V_0$ pseudopotential relative to the higher $V_m$. In this section, we consider a Rytova-Keldysh (RK) correction to the long-range interaction that accounts for the in-plane polarizability, which is affected by the sample thickness, in the systems of interest. We use the following interaction potential
\begin{equation}\label{eq:LLL_RK}
    V_{\xi,l_\text{RK}}(q)=\frac{e^2}{2\epsilon_0\epsilon q(1+l_\text{RK}q)}\tanh{\frac{q\xi}{2}}
\end{equation}
where $l_\text{RK}$ is a length scale which will be treated as a phenomenological tuning parameter in the LLL model. Setting $l_\text{RK}=0$ recovers Eq.~\eqref{eq:dual_gate}, while a large $l_\text{RK}$ weakens the interaction at short length scales, effectively reducing $V_0$. The interaction at large distances remains dominated by the screening from the metallic gates. Figure~\ref{fig:double_gated_pseudopotentials_RK_xi16} illustrates the suppression of $V_0/V_1$, as well as the slower decay of higher pseudopotentials with $m$, when $l_\text{RK}$ is finite. We have checked that for $\nu_+=1/3,\nu_-=0$, the many-body gap above the FQH ground states at $l_\text{RK}$ does not vanish as $l_\text{RK}$ is increased.

\begin{figure}
    \centering  \includegraphics[width=0.9\linewidth]{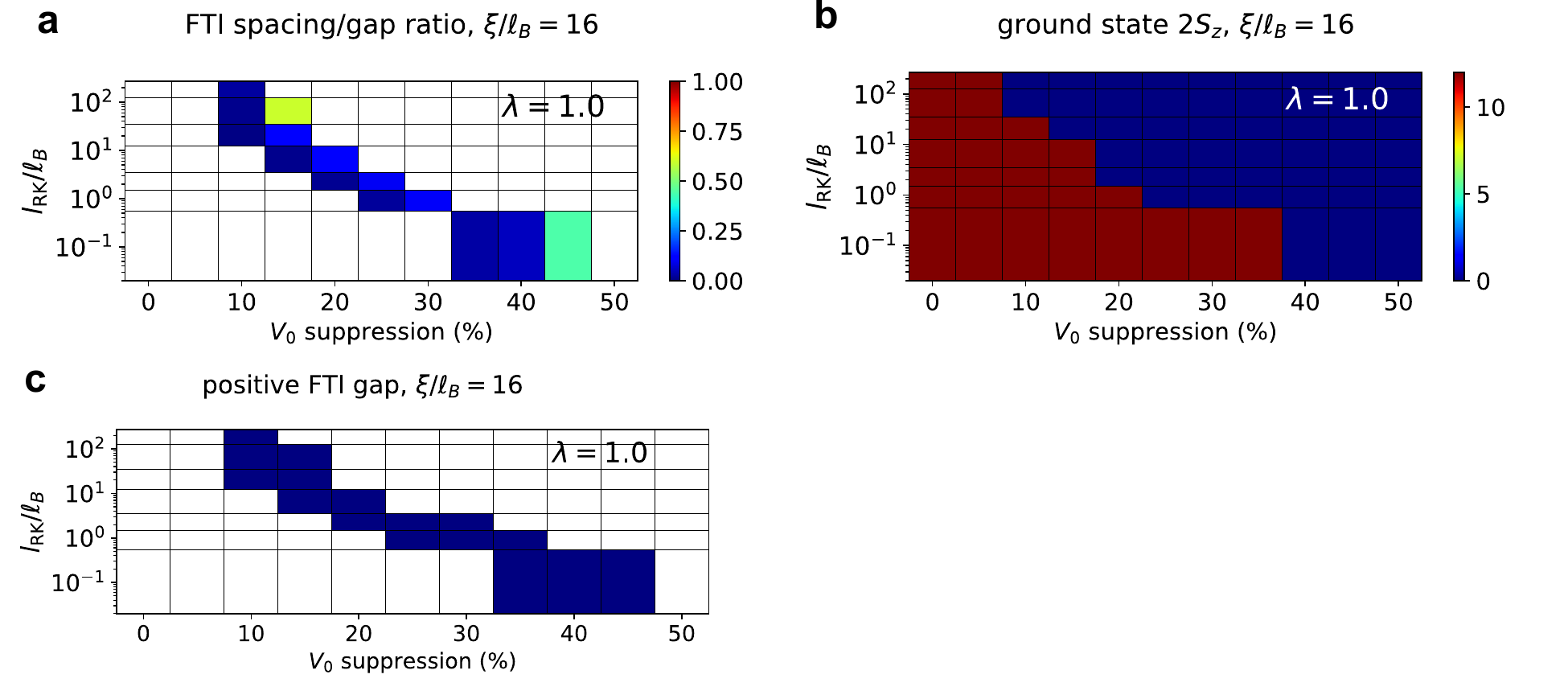}
    \caption{\textbf{FTI spacing/gap ratio, ground state valley polarization, and presence of positive FTI gap as a function of $V_0$ suppression and RK length $l_\text{RK}$ for $\lambda=1.0$ in the LLL model at $\nu_+=\nu_-=1/3$.} Gate screening length $\xi/\ell_B=16$. \textbf{a)} FTI spacing/gap ratio for the unpolarized $S_z=0$ sector. White regions correspond to where the spacing is greater than the gap, or the gap is negative. The data plotted here is identical to Fig.~\ref{fig:LLL_pseudopotential_phase}c of the main text. \textbf{b)} Ground state valley polarization $2S_z$. \textbf{c)} Blue regions indicate where FTI gap $\Delta_9>0$, i.e.~the nine lowest energy states lie in the correct momentum sectors for an FTI. Square torus geometry with $N_\Phi=18$ flux quanta.}
    \label{fig:RK_V0frac_lambda_xi16_LL0_Nphi18_spinpol}
\end{figure}

Figure~\ref{fig:RK_V0frac_lambda_xi16} shows the FTI spacing/gap ratio as a function of the RK length $l_\text{RK}$ and the suppression of $V_0$, for $N_\Phi=15$, gate distance $\xi/\ell_B=16$ and different values of the valley anisotropy $\lambda$ [Eq.~\eqref{app:eq:lambda}]. As $l_\text{RK}$ increases, the requisite percentage suppression of $V_0$ needed to stabilize the FTI decreases. Up to moderate $l_\text{RK}$, the parameters that minimize the spacing/gap ratio correspond roughly to a constant value of $V_0/V_1\simeq 1.24$, though there are deviations for larger $l_\text{RK}$. 

In Fig.~\ref{fig:RK_V0frac_lambda_xi16_LL0_Nphi18_spinpol}a we show the corresponding $\lambda=1.0$ results for $N_\Phi=18$ (the data is identical to Fig.~\ref{fig:LLL_pseudopotential_phase}c of the main text). Furthermore, in Fig.~\ref{fig:RK_V0frac_lambda_xi16_LL0_Nphi18_spinpol}b, we plot the valley polarization $2S_z$ of the global ground state across all magnetization sectors. There is a transition between the fully valley-polarized ($2Sz=N_++N_-$) phase and the unpolarized ($2S_z=0$) phase, which is close to and sometimes overlaps the FTI stability region for $S_z=0$. The system tends towards polarization for smaller $l_\text{RK}$ and $V_0$ suppression. We find that most of the parameters for which we find an FTI in the $S_z=0$ sector belong to the unpolarized phase. In Fig.~\ref{fig:RK_V0frac_lambda_xi16_LL0_Nphi18_spinpol}c, we show the region of parameter space where the FTI gap $\Delta_9>0$ is positive, which indicates that the lowest nine energy states have the correct momenta for an FTI. Comparing to Fig.~\ref{fig:RK_V0frac_lambda_xi16_LL0_Nphi18_spinpol}a, we observe that this region is not significantly larger than the region where $s_9/\Delta_9<1$, suggesting that the FTI stability region is primarily determined by having a positive $\Delta_9$. 

\subsection{Other Landau levels}\label{subsecapp:otherLLs}

\begin{figure}
    \centering
    \includegraphics[width=0.7\linewidth]{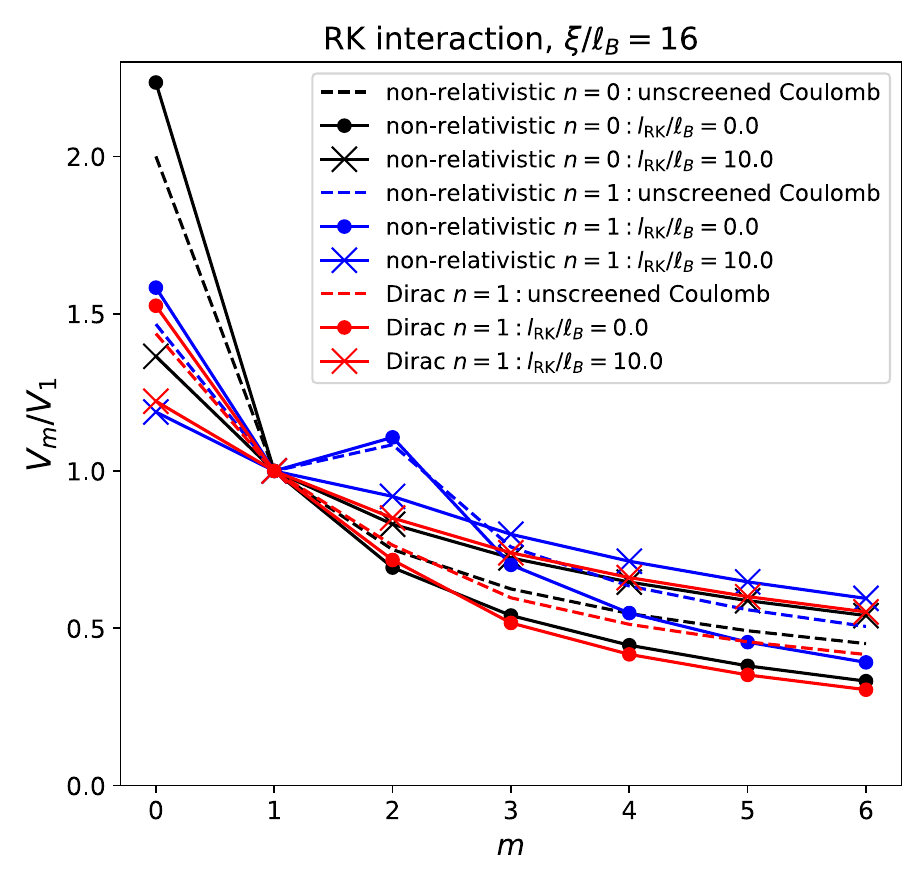}
    \caption{\textbf{Pseudopotentials $V_m$ of the dual-gate screened Coulomb interaction with RK corrections (Eq.~\ref{eq:LLL_RK}) for $\xi/\ell_B=16$ and different Landau levels. } The non-relativistic $n=1$ LL is shown in blue, while the Dirac $n=1$ LL is shown in red. The results for the LLL, i.e.~the $n=0$ LL (the pseudopotentials are identical for the non-relativistic and relativistic cases), are shown in black (see also Fig.~\ref{fig:double_gated_pseudopotentials_RK_xi16}). Dashed lines correspond to the unscreened Coulomb limit ($\xi\rightarrow\infty,l_\text{RK}=0$).}
    \label{fig:double_gated_pseudopotentials_RK_xi16_n}
\end{figure}

So far, we have only investigated the LLL, which we can also refer to as the `non-relativistic'  $n=0$ Landau level (LL). The term `non-relativistic' refers to the fact that the LLs are computed for a parabolic dispersion. We can also project instead into higher LLs with LL index $n>0$. For a given interaction potential $V(q)$, the corresponding Haldane potentials in the non-relativistic  $n$'th LL are
\begin{equation}\label{appeq:Haldane_PP_n}
    V^{\text{non-rel}}_{n,m}\equiv \int \frac{d^2\mbf{q}}{(2\pi)^2}V(q)\left[L_n\left(\frac{q^2\ell_B^2}{2}\right)\right]^2L_m(q^2\ell_B^2)e^{-q^2\ell_B^2},
\end{equation}
which generalize the $n=0$ case of Eq.~\eqref{appeq:Haldane_PP}. Note that we have made the valley indices $\eta,\eta'$ implicit for notational clarity, though just as in Eq.~\eqref{app:eq:lambda}, we can consider scaling the intervalley interaction relative to the intravalley interaction by a factor $\lambda$. The additional Laguerre polynomial factors $L_n$ reflect the different form factors of the $n$'th harmonic oscillator states in the LL problem.  

We can also consider the `relativistic' case corresponding to a Dirac Hamiltonian, which would be appropriate for e.g.~one valley of graphene where the Dirac Hamiltonian acts in sublattice space~\cite{Goerbig2006}. In the presence of a magnetic field, we obtain so-called Dirac LLs. The Dirac $n=0$ LL consists of the $0$'th harmonic oscillator state localized on one sublattice. However, all higher Dirac $n>0$ LLs consist of an equal magnitude superposition of an $n$'th harmonic oscillator state in one sublattice and an $(n-1)$'th harmonic oscillator state in the other sublattice. Assuming that the interaction is density-density in sublattice space, we can define the relativistic Haldane pseudopotentials
\begin{equation}\label{appeq:Haldane_PP_n}
    V^{\text{rel}}_{n>0,m}\equiv \int \frac{d^2\mbf{q}}{(2\pi)^2}V(q)\left[\frac{1}{2}L_n\left(\frac{q^2\ell_B^2}{2}\right)+\frac{1}{2}L_{n-1}\left(\frac{q^2\ell_B^2}{2}\right)\right]^2L_m(q^2\ell_B^2)e^{-q^2\ell_B^2},
\end{equation}
with $V^{\text{rel}}_{0,m}=V^{\text{non-rel}}_{0,m}$.

Figure~\ref{fig:double_gated_pseudopotentials_RK_xi16_n} compares the pseudopotential ratios $V_m/V_1$ for the LLL (where the non-relativistic and Dirac cases are identical), the non-relativistic $n=1$ LL, and the Dirac $n=1$ LL. For the same interaction potential $V(q)$, we find that the $n=1$ LLs have a significantly suppressed $V_0/V_1$ ratio compared to the LLL.  Based on the ED results presented in this appendix so far, this suggests that a comparatively smaller suppression of the $V_0$ pseudopotential would be needed to stabilize the FTI. In the presence of RK corrections with $l_{\text{RK}}/\ell_B=10$, the $V_0/V_1$ ratios of the $n=1$ LLs are $\simeq 1.2$. However, the $n=1$ LLs differ significantly in their pseudopotential ratios for $m>1$. For the (gate-screened) Coulomb interaction, the non-relativistic $n=1$ LL has sizable $V_2/V_1>1$. Furthermore, with the RK correction, $V_m/V_1$ for $m>1$ are all moderately enhanced compared to the LLL case. On the other hand, the pseudopotential ratios $V_m/V_1$ for $m>1$ in the Dirac $n=1$ LL are all similar to the those in the LLL. Motivated by our observations of the LL dependence of the pseudopotentials, in the following we perform ED calculations at $\nu_+=\nu_-=1/3$ analogous to the LLL model, except that we project instead into the non-relativistic or Dirac $n=1$ LL.

\begin{figure}
    \centering  \includegraphics[width=0.5\linewidth]{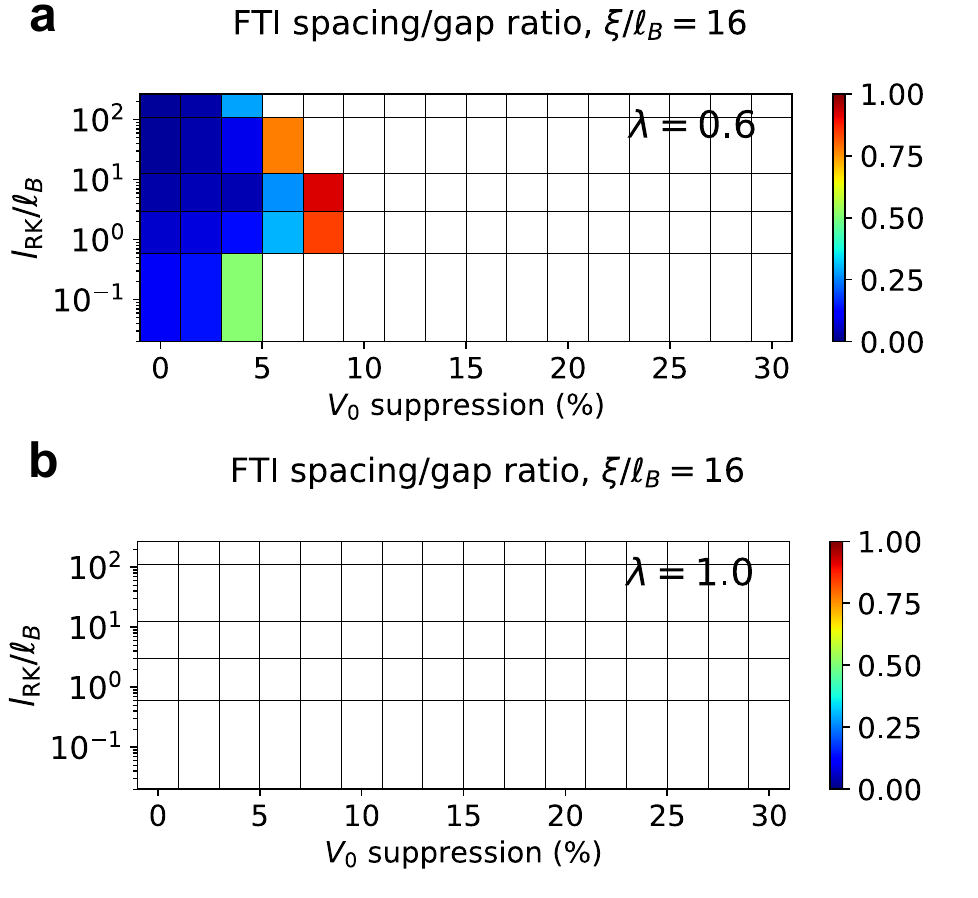}
    \caption{\textbf{FTI spacing/gap ratio as a function of $V_0$ suppression and RK length $l_\text{RK}$ for different interaction anisotropy parameters $\lambda$ in the non-relativistic $n=1$ LL at $\nu_+=\nu_-=1/3$.} Gate screening length $\xi/\ell_B=16$, and $\lambda=0.6,1.0$ in \textbf{a)} and \textbf{b)} respectively. White regions correspond to where the spacing is greater than the gap, or the gap is negative. Square torus geometry with $N_\Phi=15$ flux quanta.}
    \label{fig:RK_V0frac_lambda_xi16_LL1}
\end{figure}

\begin{figure}
    \centering  \includegraphics[width=0.9\linewidth]{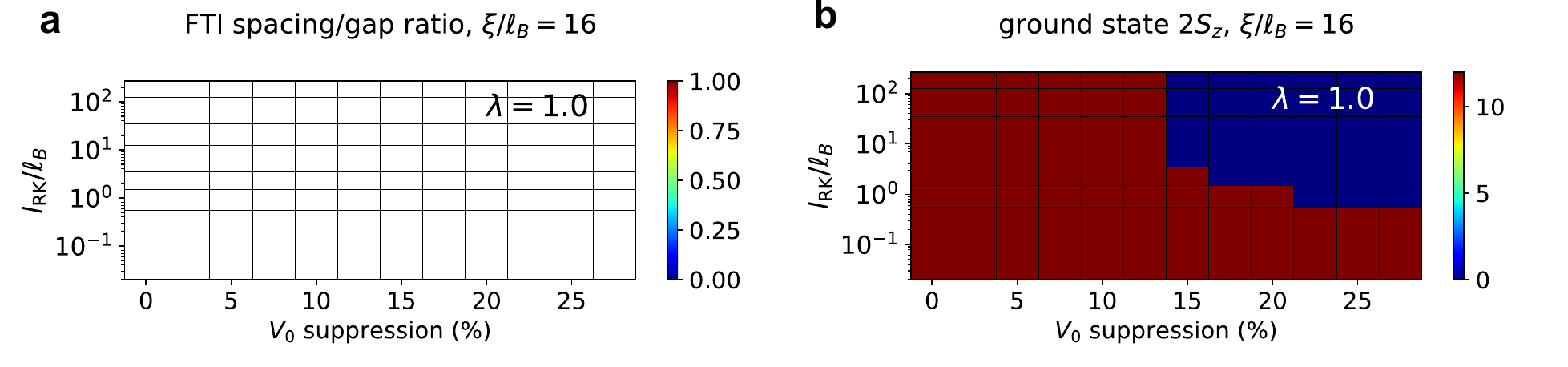}
    \caption{\textbf{FTI spacing/gap ratio and ground state valley polarization as a function of $V_0$ suppression and RK length $l_\text{RK}$ for $\lambda=1.0$ in the non-relativistic $n=1$ LL at $\nu_+=\nu_-=1/3$.} Gate screening length $\xi/\ell_B=16$. \textbf{a)} FTI spacing/gap ratio for the unpolarized $S_z=0$ sector. White regions correspond to where the spacing is greater than the gap, or the gap is negative. \textbf{b)} Ground state valley polarization $2S_z$. Square torus geometry with $N_\Phi=18$ flux quanta.}
    \label{fig:RK_V0frac_lambda_xi16_LL1_Nphi18_spinpol}
\end{figure}

In Fig.~\ref{fig:RK_V0frac_lambda_xi16_LL1}, we show the FTI spacing/gap ratio as a function of the RK length $l_\text{RK}$ and the suppression of $V_0$, for $N_\Phi=15$, gate distance $\xi/\ell_B=16$ and different values of the valley anisotropy $\lambda$ [Eq.~\eqref{app:eq:lambda}] for the non-relativistic $n=1$ LL. For $\lambda=0.6$, we observe that the FTI phase is clustered around zero or small percentage suppression of $V_0$. This is to be contrasted with the LLL case (Fig.~\ref{fig:RK_V0frac_lambda_xi16}), where at small $l_{\text{RK}}$, a sizable suppression $\gtrsim 25\%$ is required to stabilize the FTI. However for the non-relativistic $n=1$ LL, we do not find any FTIs for $\lambda=1.0$. This suggests that the behavior of the pseudopotentials for $m>1$ prevents the existence of the FTI, even if we tune $V_0/V_1$. In Fig.~\ref{fig:RK_V0frac_lambda_xi16_LL1_Nphi18_spinpol}, we show the FTI spacing/gap ratio and ground state valley polarization for $\lambda=1.0$ and $N_\Phi=18$. We note that, like in the LLL (Fig.~\ref{fig:RK_V0frac_lambda_xi16_LL0_Nphi18_spinpol}b), there is a phase boundary between the valley-polarized and unpolarized phases. However, here there is no FTI region in the $S_z=0$ sector anywhere.

\begin{figure}
    \centering  \includegraphics[width=0.9\linewidth]{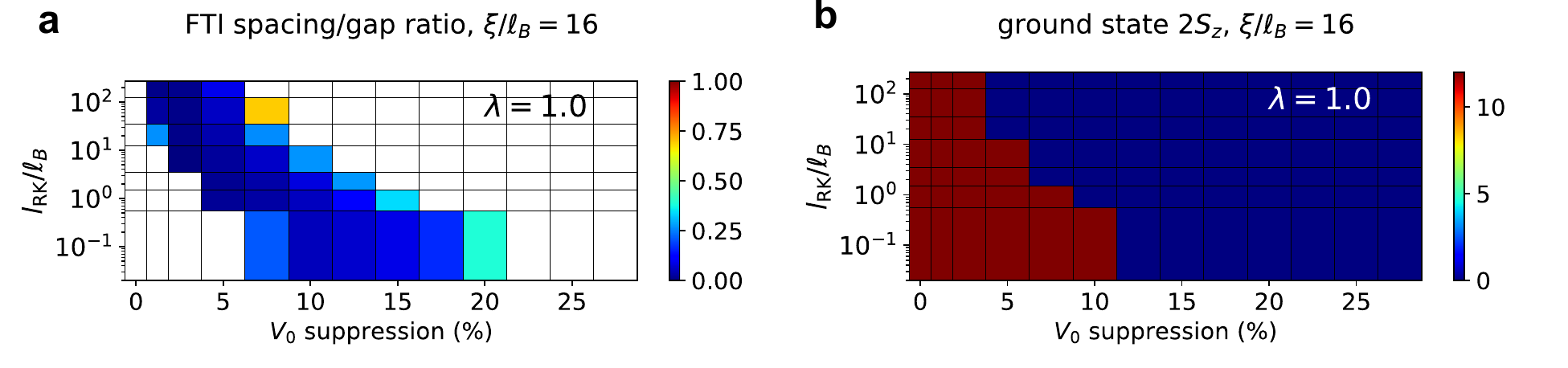}
    \caption{\textbf{FTI spacing/gap ratio and ground state valley polarization as a function of $V_0$ suppression and RK length $l_\text{RK}$ for $\lambda=1.0$ in the Dirac $n=1$ LL at $\nu_+=\nu_-=1/3$.} Gate screening length $\xi/\ell_B=16$. \textbf{a)} FTI spacing/gap ratio for the unpolarized $S_z=0$ sector. White regions correspond to where the spacing is greater than the gap, or the gap is negative. \textbf{b)} Ground state valley polarization $2S_z$. Square torus geometry with $N_\Phi=18$ flux quanta.}
    \label{fig:RK_V0frac_lambda_xi16_LL-1_Nphi18_spinpol}
\end{figure}

In Fig.~\ref{fig:RK_V0frac_lambda_xi16_LL-1_Nphi18_spinpol}, we show $N_\Phi=18$ and $\lambda=1.0$ results for the \emph{Dirac} $n=1$ LL. In the $S_z=0$ sector, we observe that compared to the LLL case (Fig.~\ref{fig:RK_V0frac_lambda_xi16}), the FTI phase exists at a significantly lower range of $V_0$ percentage suppression. For large $l_\text{RK}$, the FTI survives very close to the limit where $V_0$ is not suppressed at all.  Our calculations for different valley sectors reveals that for some of the parameters where the $S_z=0$ sector is an FTI, the global ground state is actually magnetized at $\nu_++\nu_-=2/3$. However, a sizable portion of the parameters where the $S_z=0$ sector yields an FTI remains non-magnetized.

\clearpage

\section{Review of $t$MoTe$_2$ model}

In this section, we review the interacting model of $t$MoTe$_2$ used in this work. In App.~\ref{subsecapp:continuum}, we describe the single-particle continuum model. In App.~\ref{subsecapp:interactions}, we consider the general form of the interaction term.

\subsection{Single-particle continuum model}\label{subsecapp:continuum}

Our description of the single-particle model closely follows the presentation of Ref.~\onlinecite{jia2023moire}. The single-particle model, which captures the moir\'e band structure of the valence bands of twisted homobilayer MoTe$_2$, is defined in terms of continuum degrees of freedom associated with electron creation operators $c^\dagger_{\eta,l,\bm{r}}$. The index $\eta=\pm$ labels the valleys $\pm K$, which are locked to the spins $\uparrow,\downarrow$ due to the strong spin-orbit coupling, while $l=t,b$ labels the two layers, and $\bm{r}$ labels the in-plane position $\bm{r}=(x,y)$. The top (bottom) layer is rotated by $-\theta/2$ ($\theta/2$) starting from AA-stacking, such that the rotated $K$-points lie at the momenta
\begin{equation}
    K_b=\frac{4\pi}{3a_0}\begin{pmatrix}
        \cos\theta/2 \\ \sin\theta/2
    \end{pmatrix},\quad
    K_t=\frac{4\pi}{3a_0}\begin{pmatrix}
        \cos\theta/2 \\ -\sin\theta/2
    \end{pmatrix},
\end{equation}
where $a_0=0.352\,\text{nm}$ is the lattice constant of MoTe$_2$. $K_b$ ($K_t$) maps onto the $K_M$ ($K'_M$) point of the moir\'e Brillouin zone (mBZ).
We also define the following wavevectors
\begin{gather}
\bm{q}_1=K_b-K_t,\quad \bm{q}_2=C_3\bm{q}_1,\quad \bm{q}_3=C_3^2\bm{q}_1,
\end{gather}
where $C_3$ is a counter-clockwise rotation matrix by $2\pi/3$. In terms of these, we define the basis moir\'e reciprocal lattice vectors
\begin{equation}
    \bm{b}_1=\bm{q}_3-\bm{q}_2,\quad \bm{b}_2=\bm{q}_1-\bm{q}_2.
\end{equation}

The continuum model for valley $\eta$ takes the general form
\begin{equation}
H_{\eta,0}=\int d^2\bm{r}\begin{pmatrix}
c^\dagger_{\eta,b,\bm{r}}&c^\dagger_{\eta,t,\bm{r}}
\end{pmatrix}
\begin{pmatrix}
    h_{\eta,b}(\bm{r})&t_\eta(\bm{r})\\
    t_\eta^*(\bm{r})&h_{\eta,t}(\bm{r})
\end{pmatrix}
\begin{pmatrix}
    c_{\eta,b,\bm{r}}\\
    c_{\eta,t,\bm{r}}
\end{pmatrix}.
\end{equation}
The intralayer term is
\begin{equation}
    h_{\eta,l}(\bm{r})=\frac{\hbar^2\nabla^2}{2m^*}+V_{\eta,l}(\bm{r})+(-1)^l\frac{D}{2},
\end{equation}
where $m^*=0.6m_e$ is the effective mass of the valence band maximum of monolayer MoTe$_2$, $V_{\eta,l}(\bm{r})$ is the intralayer potential to be specified shortly, and $D$ corresponds to an externally applied displacement field. Unless otherwise stated, we set $D=0$. Note that the continuum model operators are defined with a layer-dependent momentum boost. In particular, under the action of translation $T_{\bm{R}_M}$ by a moir\'e lattice vector $\bm{R}_M$, we have
\begin{equation}
    T_{\bm{R}_M}c^\dagger_{\eta,l,\bm{r}}T^{-1}_{\bm{R}_M}=c^\dagger_{\eta,l,\bm{r}+\bm{R}_M}e^{-i\eta \bm{R}_M\cdot K_l}.
\end{equation}

Both the intralayer moir\'e potential $V_{\eta,l}(\bm{r})$ and interlayer hopping $t_\eta(\bm{r})$ are expanded to the first harmonics
\begin{gather}
    V_{\eta,l}(\bm{r})=Ve^{-(-1)^li\psi}\sum_{i=1,2,3}e^{i\bm{g}_i\cdot\bm{r}}+Ve^{(-1)^li\psi}\sum_{i=1,2,3}e^{-i\bm{g}_i\cdot\bm{r}}\\
    t_\eta(\bm{r})=w\sum_{i=1,2,3}e^{-i\eta\bm{q}_i\cdot\bm{r}}
\end{gather}
where $(-1)^t=1$ and $(-1)^b=-1$, and $\bm{g}_i=C_3^{i-1}\bm{b}_1$. Unless otherwise stated, we use the parameters $w=-18.8\,\text{meV},V=16.5\,\text{meV}$ and $\psi=-105.9^\circ$, which are determined by fitting the above parameterization to DFT calculations~\cite{jia2023moire}.

\subsection{Interactions}
\label{subsecapp:interactions}

In this work we consider generally layer-dependent density-density interactions. Since we are considering hole-doping of the moir\'e valence bands (fully filled valence bands corresponds to $\nu=0$, i.e.~charge neutrality), the interaction Hamiltonian is more naturally expressed in the hole basis as explained in Ref.~\cite{Yu2024MFCI0}. We denote by $\tilde{c}^\dagger_{\eta,l,\bm{r}}$ the hole creation operator,
which is related to the electron creation operator by
\eq{
\label{eq:c_ctilde_complexconjugate}
\cc   c_{\eta,l,\bsl{r}}^\dagger    \cc^{-1} =  \widetilde{c}_{\eta,l,\bsl{r}},
}
with $\cc$ the complex conjugate.
We then define the hole density operator $\tilde{\rho}_{\eta,l}(\bm{q})$ for a fixed valley and layer
\begin{gather}
\tilde{\rho}_{\eta,l}(\bsl{q}) = \sum_{\bsl{k}\in \text{mBZ},\bsl{Q}\in\Q_{\eta,l}} \tilde{c}^\dagger_{\eta,l,\bsl{k}-\bsl{Q}+\bsl{q}} \tilde{c}_{\eta,l,\bsl{k}-\bsl{Q}}   \\
\tilde{c}^\dagger_{\eta,l,\bm{r}}=\frac{1}{\sqrt{\mathcal{V}}}\sum_{\bm{k}}\sum_{\bm{\Q}\in\Q_l^\eta}e^{-i(\bm{k}-\bm{Q})\cdot\bm{r}}\tilde{c}^\dagger_{\eta,l,\bm{k}-\bm{Q}}\\
\Q_l^\eta = \{ \bsl{G}_M + \eta (-)^l \bsl{q}_1 \},
\end{gather}
where $\mathcal{V}$ is the total area of the system, and $\bm{G}_M$ is a moir\'e reciprocal lattice vector. We have introduced the hole creation operator in momentum space $\tilde{c}^\dagger_{\eta,l,\bm{k}-\bm{Q}}$.  The interaction term takes the form
\begin{equation}\label{eqapp:Hint_normal}
    H_\text{int}=\frac{1}{2\V}\sum_{\bsl{q}ll'\eta\eta'}V_{ll'}(\bm{q}):\tilde{\rho}_{\eta,l}(\bsl{q})\tilde{\rho}_{\eta',l'}(-\bsl{q}):
\end{equation}
where $V_{ll'}(\bm{q})$ is the Fourier transform of the layer-dependent interaction potential. The total Hamiltonian is $H=H_\text{int}+\sum_{\eta}H_{\eta,0}$.

The notation $:\hat{O}:$ in Eq.~\eqref{eqapp:Hint_normal} denotes a normal-ordering of the operators in $\hat{O}$ with respect to the \emph{hole vacuum}. This means that all hole creation operators $\tilde{c}^\dagger$ appear to the \emph{left} of all hole annihilation operators $\tilde{c}$. Therefore, $H_\text{int}$ annihilates the state $\ket{\nu=0}$ at charge neutrality at filling factor $\nu=0$, which corresponds to having no holes in the system. 
This form of the interaction is sensible because the continuum model parameters are extracted from DFT calculations performed at charge neutrality.   

In this work, we consider two contributions to the interaction potential $V_{ll'}(\bm{q})$. The main contribution is the long-range Coulomb interaction, which is screened by the metallic gates and the non-local dielectric environment of the $t$MoTe$_2$ device. This is discussed in detail in App.~\ref{secapp:dielectric}. To help stabilize the FTI, we also add an on-site intralayer attractive interaction with amplitude $g$ (see Eq.~\eqref{eq:g} in the main text and Eq.~\eqref{eq:H_g}). Due to fermion antisymmetry, this only has an effect between particles with different spins. One possible origin of an attractive $g$ is the electron-phonon coupling, which is studied in App.~\ref{secapp:EPC}.

We can characterize the interaction potential $V_{ll'}(\bm{q})$ by calculating the Haldane pseudopotentials in the case that the it is projected into the LLL. This is carried out by considering Eq.~\ref{appeq:Haldane_PP} with the effective magnetic length $\ell_B^*=2.02\,\text{nm}$. This value of $\ell_B^*$ is chosen to satisfy~\cite{Liu2015characterization} $2\pi {\ell_B^*}^2=A_M$, with $A_M$ the moir\'e unit cell area of $t$MoTe$_2$ at $\theta=3.7^\circ$. 

\clearpage

\section{Contribution to short-range interaction from electron-phonon coupling in $t$MoTe$_2$}
\label{secapp:EPC}

In this section, we discuss a possible physical origin for the onsite interaction $g$ in \eqnref{eq:g}. 
More explicitly, the onsite interaction has the following form
\eqa{
\label{eq:H_g}
H_g = \frac{g}{2} \sum_{l\eta\eta'} \int d^2 r:\tilde{\rho}_{\eta,l}(\bsl{r})\tilde{\rho}_{\eta',l}(\bsl{r}): =  \frac{g}{2\V}\sum_{\bsl{q}l\eta\eta'}:\tilde{\rho}_{\eta,l}(\bsl{q})\tilde{\rho}_{\eta',l}(-\bsl{q}): \ ,
}
where the hole density operator $\tilde{\rho}_{\eta,l}(\bm{q})$ and the normal-ordering notation are defined in App.~\ref{subsecapp:interactions}, and $\tilde{\rho}_{\eta,l}(\bsl{r})$ is the Fourier transformation of $\tilde{\rho}_{\eta,l}(\bsl{q})$:
\eq{
\tilde{\rho}_{\eta,l}(\bsl{r}) = \widetilde{c}_{\eta,l,\bsl{r}}^\dagger \widetilde{c}_{\eta,l,\bsl{r}}\ .
}
We note that the intravalley component of \eqnref{eq:H_g} is ineffective due to the onsite and intralayer nature of the interaction, as fermionic statistics leads to
\eq{
:\tilde{\rho}_{\eta,l}(\bsl{r})\tilde{\rho}_{\eta,l}(\bsl{r}): = \widetilde{c}^\dagger_{\bsl{r},\eta,l}\widetilde{c}^\dagger_{\bsl{r},\eta,l} \widetilde{c}_{\bsl{r},\eta,l} \widetilde{c}_{\bsl{r},\eta,l} = 0\ .
}
As such, $g$ mimics the effect of the $V_0$ Haldane pseudopotential in the LLL model discussion of App.~\ref{secapp:LLL}. As shown in the main text and App.~\ref{secapp:additional_tMoTe2}, such a short-range interaction with attractive $g<0$ is required to stabilize the FTI phase in the ED calculations of $t$MoTe$_2$.
In the following, we will derive the contribution to $g$ from the electron-phonon coupling (EPC).

The EPC that we consider is within each individual MoTe$_2$ layer.
The low-energy physics of the monolayer MoTe$_2$ can be well captured by $d$ orbitals on Mo atoms and the $p$ orbitals on the Te atoms~\cite{Wu2019,jia2023moire}.
The single-layer MoTe$_2$ has a mirror symmetry with the mirror plane lying in the 2D system, namely $m_z$ where $z$ is perpendicular to the sample.
The low-energy states at and near $\pm\K$ mainly come from the $d_{x^2-y^2}$ and $d_{xy}$ orbitals at Mo atoms, which are even under $m_z$.
Therefore, we will only consider the fermionic $m_z$-even sector of the Hamiltonian, leading to the following general EPC Hamiltonian
\eq{
\label{eq:EPC_665_real_space}
H_{\text{el-ph}} = \sum_{\bsl{R}_1\bsl{R}_2\bsl{R}}\sum_{\bsl{\tau}_1\bsl{\tau}_2\bsl{\tau}}^{\bsl{\tau}_{\text{Mo}},\bsl{\tau}_{\text{Te},1},\bsl{\tau}_{\text{Te},2}}  \sum_{\alpha_{\bsl{\tau}_1} \alpha_{\bsl{\tau}_2}' i} \sum_{s }\hat{c}^\dagger_{\bsl{R}_1+\bsl{\tau}_1,\alpha_{\bsl{\tau}_1},s}\hat{c}_{\bsl{R}_2+\bsl{\tau}_2,\alpha_{\bsl{\tau}_2}',s} u_{\bsl{R}+\bsl{\tau},i} \  F^{\alpha_{\bsl{\tau}_1}\alpha_{\bsl{\tau}_2}'i}_{ \bsl{\tau}_1 , \bsl{R}_2 -\bsl{R}_1 + \bsl{\tau}_2,  \bsl{R} - \bsl{R}_1 + \bsl{\tau}}\ ,
}
where $s=\uparrow,\downarrow$ labels the spin, $\bsl{\tau}_{\text{Mo}}$ labels the position of the Mo atom in one unit cell, $\bsl{\tau}_{\text{Te},1}$ and $\bsl{\tau}_{\text{Te},2}$ label the positions of the two $\Te$ atoms in one unit cell, $\alpha_{\bsl{\tau}}$ labels the orbitals on the atom $\bsl{\tau}$, $\hat{c}^\dagger_{\bsl{R}_1+\bsl{\tau}_1,\alpha_{\bsl{\tau}_1},s}$ creates an electron at $\bsl{R}_1+\bsl{\tau}_1$ with orbital $\alpha_{\bsl{\tau}_1}$ and spin $s$, and $u_{\bsl{R}+\bsl{\tau},i}$ labels the ion motion at $\bsl{R}+\bsl{\tau}$ along the $i$th direction.
Here we use the fact that the dominant contribution from spin-orbit coupling usually only enters in an on-site way, and thus the EPC is diagonal in spin.
The lattice translations lead to
\eq{
\label{eq:F_latt}
F^{\alpha_{\bsl{\tau}_1}\alpha_{\bsl{\tau}_2}'i}_{\bsl{R}_1+\bsl{R}_0 + \bsl{\tau}_1 , \bsl{R}_2+\bsl{R}_0 + \bsl{\tau}_2,  \bsl{R} +\bsl{R}_0 + \bsl{\tau}} = F^{\alpha_{\bsl{\tau}_1}\alpha_{\bsl{\tau}_2}'i}_{\bsl{R}_1 + \bsl{\tau}_1 , \bsl{R}_2 + \bsl{\tau}_2,  \bsl{R} + \bsl{\tau}}\ \forall\ \text{lattice vector }\bsl{R}_0\ .
}
Transforming \eqnref{eq:EPC_665_real_space} to the momentum space, we have
\eq{
H_{\text{el-ph}} = \frac{1}{{N}} \sum_{\bsl{\tau}_1,\bsl{\tau}_2,\bsl{\tau}}\sum_{\alpha_{\bsl{\tau}_1}\alpha_{\bsl{\tau}_2}'i} \sum_{s}\sum_{\bsl{k}_1,\bsl{k}_2}^{\text{\BZ}}  \hat{c}^\dagger_{\bsl{k}_1,\bsl{\tau}_1,\alpha_{\bsl{\tau}_1},s}\hat{c}_{\bsl{k}_2,\bsl{\tau}_2,\alpha_{\bsl{\tau}_2}',s} \left[ F_{\bsl{\tau}i}(\bsl{k}_1,\bsl{k}_2) \right]_{\bsl{\tau}_1\alpha_{\bsl{\tau}_1},\bsl{\tau}_2\alpha_{\bsl{\tau}_2}'}  u_{\bsl{k}_2-\bsl{k}_1,\bsl{\tau},i}^\dagger \ ,
}
where $\text{\BZ}$ is the first Brillouin zone of the monolayer MoTe$_2$, and
\eqa{
\label{eq:f_k}
& \left[ F_{\bsl{\tau}i}(\bsl{k}_1,\bsl{k}_2) \right]_{\bsl{\tau}_1\alpha_{\bsl{\tau}_1},\bsl{\tau}_2\alpha_{\bsl{\tau}_2}'}  =  \sum_{\bsl{R}_1\bsl{R}_2 }e^{-\ii \bsl{k}_1\cdot(\bsl{R}_1+\bsl{\tau}_1-\bsl{\tau})} e^{\ii \bsl{k}_2\cdot(\bsl{R}_2+\bsl{\tau}_2-\bsl{\tau})} F^{\alpha_{\bsl{\tau}_1}\alpha_{\bsl{\tau}_2}'i}_{\bsl{R}_1 + \bsl{\tau}_1 , \bsl{R}_2 + \bsl{\tau}_2,  \bsl{\tau}} \ .
}
For the moment, we focus on momenta $\pm \K$ for the fermions, since the relevant low-energy fermionic degrees of freedom lie around the $K$ valleys. We then have 
\eq{
H_{\text{el-ph}} \approx \frac{1}{{N}}\sum_{\eta_1 \eta_2 } \sum_{\bsl{\tau}_1,\bsl{\tau}_2,\bsl{\tau}}\sum_{\alpha_{\bsl{\tau}_1}\alpha_{\bsl{\tau}_2}'i} \sum_{s} \hat{c}^\dagger_{\eta_1\K,\bsl{\tau}_1,\alpha_{\bsl{\tau}_1},s}\hat{c}_{ \eta_2\K,\bsl{\tau}_2,\alpha_{\bsl{\tau}_2}',s} u_{(\eta_2-\eta_1)\K,\bsl{\tau},i}^\dagger \left[ F_{\bsl{\tau}i}(\eta_1\K,\eta_2\K) \right]_{\bsl{\tau}_1\alpha_{\bsl{\tau}_1},\bsl{\tau}_2\alpha_{\bsl{\tau}_2}'}.
}

After integrating out the fermions, the EPC would eventually lead to an effective attractive interaction.
To show this, we first convert the EPC Hamiltonian to the eigenbasis of electrons $\gamma^\dagger$ and phonons $b^\dagger$, which reads 
\eqa{
\label{eq:H_el_ph_gen}
& H_{\text{el-ph}} = \frac{1}{{N}} \sum_{nmls} \sum_{\eta_1\eta_2=\pm}  G_{nml}^{s}(\eta_1\K,\eta_2\K) \gamma^\dagger_{\eta_1\K,n,s}\gamma_{\eta_2\K,m,s} (b^\dagger_{(\eta_2-\eta_1)\K,a}+b_{-(\eta_2-\eta_1)\K,a})\ ,
}
where
\eq{
\label{eq:ft_l}
\widetilde{F}_{a}(\eta_1\K,\eta_2\K) = \sum_{\bsl{\tau}',i'}  F_{\bsl{\tau}'i'}(\eta_1\K,\eta_2\K)\frac{1}{{m_{\bsl{\tau}'}}} [v_a^*(\eta_2\K-\eta_1\K)]_{\bsl{\tau}'i'}
}
\eq{
\label{eq:Gt_nml}
\widetilde{G}_{nma}^{s}(\eta_1\K,\eta_2\K) = U_{n,s}^\dagger(\eta_1\K)  \widetilde{F}_{a}(\eta_1\K,\eta_2\K) U_{m,s}(\eta_2\K)\ ,
}
\eq{
\label{eq:G_nml}
G_{nma}^{s}(\eta_1\K,\eta_2\K) = { \frac{\hbar }{2\omega_a(\eta_2\K-\eta_1\K)} } \widetilde{G}_{nma}^{s}(\eta_1\K,\eta_2\K)\ ,
} 
and $\omega_a(\bsl{q})$ and $v_a(\bsl{q})$ are the frequency and eigenvector of the phonon in the $a$th phonon band and momentum $\bsl{q}$.
Here, we have used the spin-$\U(1)$ symmetry of the single-particle fermionic Hamiltonian, $U_{m,s}(\bsl{k})$ labels the eigenvector of the $m$th electronic band with spin $s$ at $\bsl{k}$, and 
\eq{
\gamma^\dagger_{\bsl{k},n,s} = \sum_{\bsl{\tau}_1,\alpha_{\bsl{\tau}_1} } \hat{c}^\dagger_{\bsl{k},\bsl{\tau}_1,\alpha_{\bsl{\tau}_1},s} \left[ U_{m,s}(\bsl{k}) \right]_{\bsl{\tau}_1,\alpha_{\bsl{\tau}_1}}\ .
}
Then, based on the second-order perturbation theory, the EPC can mediate an attractive interaction as the following:
\eqa{
\label{eq:eff_int_ph_induced}
H_{\text{eff-int}} & = \frac{1}{N} \sum_{\eta_1,a}\sum_{nn'mm' ss'}\sum_{\eta,\eta'}\frac{G_{n'm'a}^{s'}(\eta'\K,\eta'\K+\eta\K-\eta_1\K) G_{nma}^{s}(\eta\K,\eta_1\K) \hbar \omega_a(\eta_1\K-\eta\K) }{(E_{ns}(\eta\K) - E_{ms}(\eta_1\K))^2 -(\hbar \omega_a(\eta_1\K-\eta\K))^2 } \\
& \ \times \gamma^\dagger_{\eta\K,n,s} \gamma_{\eta_1\K,m,s} \gamma^\dagger_{\eta'\K,n',s'} \gamma_{\eta'\K+\eta\K-\eta_1\K,m',s'}\ ,
}
where $E_{ns}(\bsl{k})$ is the energy of the $n$th electronic band with spin $s$ at $\bsl{k}$.
Owing to the spin-orbit coupling, the low-energy electronic states in the top valence band are spin-valley locked---at low energies, we only need to consider the spin-up state at $\K$ valley and the spin-down state at $-\K$ valley.
As a result, \eqnref{eq:eff_int_ph_induced} becomes 
\eqa{
H_{\text{eff-int}} & = \frac{1}{N} \sum_{\bsl{p}\bsl{p}'\bsl{q}} \sum_{a}\sum_{\eta,\eta'}\frac{G_{00a}^{\eta'}(\eta'\K,\eta'\K) G_{00a}^{\eta}(\eta\K,\eta\K)  }{ -(\hbar \omega_a(0)) }  \gamma^\dagger_{\eta\K+\bsl{p}+\bsl{q},0,\eta} \gamma_{\eta\K+\bsl{p}',0,\eta} \gamma^\dagger_{\eta'\K+\bsl{p}'-\bsl{q},0,\eta'} \gamma_{\eta'\K+\bsl{p},0,\eta'} \ ,
}
where $n=0$ corresponds to the top valence band with spin $s$. Note that we have re-introduced the momentum dependence in the fermionic operators, and made the approximation that the EPC-induced interaction is independent of the momentum deviations from $\pm K$.
Owing to the TR symmetry, we have 
\eqa{
G_{00a}^{-}(-\K,-\K) = G_{00a}^{+}(\K,\K) \in \dsR\ . 
}
By defining \eq{
\label{eq:g_EPC}
g_{EPC} = -2 \Omega \sum_{a}\frac{ \left| G_{00a}^{+}(\K,\K) \right|^2 }{ \hbar \omega_a(0) }
}
with $\Omega$ the unit cell area of monolayer MoTe$_2$, we arrive at 
\eqa{
H_{\text{eff-int}} & = \frac{g_{EPC}}{2\V} \sum_{\bsl{p} \bsl{p}'\bsl{q} } \sum_{\eta,\eta'}\gamma^\dagger_{\eta\K+\bsl{p}+\bsl{q},0,\eta} \gamma_{\eta\K+\bsl{p}',0,\eta} \gamma^\dagger_{\eta'\K+\bsl{p}'-\bsl{q},0,\eta'} \gamma_{\eta'\K+\bsl{p},0,\eta'} \ .
}
The above derivation holds for each individual layer. 
By restoring the layer index $l$, we would have $\gamma_{\eta\K+\bsl{p},0,\eta, l }^\dagger$ for each layer $l$, which is equivalent to the continuum electron creation operator $c^\dagger_{\eta,l,\bsl{p}}$ in momentum space.
After including the layer index and by converting $c^\dagger_{\eta,l,\bsl{p}}$ to the real space, $H_{\text{eff-int}}$ becomes
\eqa{
H_{\text{eff-int}} & = \frac{g_{EPC}}{2} \int d^2 r \sum_{l,\eta,\eta'}c^\dagger_{\eta,l,\bsl{r}} c_{\eta,l,\bsl{r}} c^\dagger_{\eta',l,\bsl{r}} c_{\eta',l,\bsl{r}} \ ,
}
which, in the hole basis  (\eqnref{eq:c_ctilde_complexconjugate}) reads
\eqa{
H_{\text{eff-int}} & = \frac{g_{EPC}}{2} \int d^2 r \sum_{l,\eta,\eta'}\widetilde{c}_{\eta,l,\bsl{r}} \widetilde{c}_{\eta,l,\bsl{r}}^\dagger \widetilde{c}_{\eta',l,\bsl{r}} \widetilde{c}_{\eta',l,\bsl{r}}^\dagger \ .
}
In this work, we consider the normal-ordered part of the interaction (see App.~\ref{subsecapp:interactions}).
After the normal-ordering, $H_{\text{eff-int}}$ becomes
\eq{
: H_{\text{eff-int}} : = \frac{g_{EPC}}{2} \int d^2 r \sum_{l,\eta,\eta'}\widetilde{c}_{\eta,l,\bsl{r}}^\dagger  \widetilde{c}_{\eta',l,\bsl{r}}^\dagger  \widetilde{c}_{\eta',l,\bsl{r}} \widetilde{c}_{\eta,l,\bsl{r}}  \ ,
}
which has the same form as \eqnref{eq:H_g}.
Therefore, 
$g_{EPC}$ is the contribution to $g$ from the EPC.

Numerically, \eqnref{eq:g_EPC} is hard to evaluate owing to the fact that $\omega_a(0)$ is zero for acoustic modes.
Therefore, we use the dimensionless EPC constant $\Lambda$ to approximately calculate $g_{EPC}$.
Specifically, $\Lambda$ has the following expression
\eq{
\Lambda = \frac{2}{D(\mu) N} \sum_{\bsl{k},\bsl{k}'}^{\BZ} \sum_{nma s}\frac{|G_{nma}^s(\bsl{k},\bsl{k}')|^2}{\hbar \omega_a(\bsl{k}'-\bsl{k})} \delta\left(\mu - E_n^s(\bsl{k}) \right) \delta\left(\mu - E_m^s(\bsl{k}') \right)\ ,
}
where $E_n^s(\bsl{k})$ is the $n$th electron band in the spin-$s$ subspace, and $D(\mu)=\sum_{\bsl{k} n s} \delta\left(\mu - E_n^s(\bsl{k}) \right)$ is the density of states.
For slightly hole-doped case that is relevant for the experiments, the chemical potential is close to the valence band top such that we would have two Fermi surfaces centered around $\pm\K$, and $\Lambda$ becomes
\eqa{
\Lambda & = \frac{2}{D(\mu) N} \sum_{\bsl{p}\bsl{p}'} \sum_{\eta \eta'} \sum_{a s}\frac{|G_{00a}^s(\eta \K+\bsl{p}, \eta' \K +\bsl{p}')|^2}{\hbar \omega_a(\eta' \K +\bsl{p}' - \eta \K -\bsl{p})} \delta\left(\mu - E_0^s(\eta \K+\bsl{p}) \right) \delta\left(\mu - E_0^s(\eta' \K+\bsl{p}') \right) \\
& = \frac{2}{D(\mu) N} \sum_{\bsl{p}\bsl{p}'} \sum_{\eta } \sum_{a }\frac{|G_{00a}^\eta(\eta \K+\bsl{p}, \eta \K +\bsl{p}')|^2}{\hbar \omega_a(\eta \K +\bsl{p}' - \eta \K -\bsl{p})} \delta\left(\mu - E_0^\eta(\eta \K+\bsl{p}) \right) \delta\left(\mu - E_0^\eta(\eta \K+\bsl{p}') \right) \\
& \approx \frac{2}{D(\mu) N} \sum_{\bsl{p}\bsl{p}'} \sum_{\eta } \sum_{a }\frac{|G_{00a}^\eta(\eta \K, \eta \K )|^2}{\hbar \omega_a(0)} \delta\left(\mu - E_0^\eta(\eta \K+\bsl{p}) \right) \delta\left(\mu - E_0^\eta(\eta \K+\bsl{p}') \right) \\
& = \frac{1}{D(\mu) N} \sum_{\eta } \frac{-g_{EPC}}{\Omega} D_\eta(\mu)^2 \ ,
}
where we have used the fact that $\bsl{p}$ on the Fermi surface is small when the filling factor $\nu$ is close to the valence band top, and $D_s(\mu)=\sum_{\bsl{k}}\delta\left(\mu - E_n^s(\bsl{k}) \right)$.
Owing to the TR symmetry, we have $D_+(\mu) = D_-(\mu) = D(\mu)/2$, leading to 
\eq{
\Lambda \approx \frac{D(\mu)}{ N}  \frac{-g_{EPC}}{2\Omega} \Rightarrow g_{EPC} \approx - \frac{2 \Omega}{D(\mu)/N} \Lambda\ .
}
Using the software package EPW~\cite{lee2023EPW}, one can numerically calculate $\Lambda$.
For $\mu=-20$meV (we choose the valence band top to correspond to zero energy), we obtain~\cite{EPCNbSe2} 
\eq{
D(\mu)/N \approx 0.306 \eV^{-1}\ ,\text{ and }\Lambda \approx 0.0233\ .
}
Combined with $\Omega\approx 0.11 \text{nm}^2$, we arrive at
\eq{
g_{EPC} \approx -16.8 \,\text{meV}\cdot \text{nm}^2 \ .
}

\clearpage

\section{Dielectric screening of interaction potential}\label{secapp:dielectric}
In this appendix, we discuss the impact of the dielectric environment on the effective interaction potential between electrons in the $t$MoTe$_2$ sample. 
We consider a system that is homogeneous in-plane, but is constructed by stacking distinct `slabs' (for instance, various dielectric materials and the $t$MoTe$_2$ sample itself) in the out-of-plane $z$-direction.
We first discuss the general formulation of the electrostatics problem and explain how the layer-dependent potential $V_{ll'}(q)$ used in the ED calculations of the main text is generated. We also consider some analytical examples. Then, we perform calculations on various system geometries to understand the impact of the dielectric environment on $V_0/V_1$, including configurations with additional `spacer' dielectric slabs.

\begin{figure}[h!]
    \centering
    \includegraphics[width=10cm]{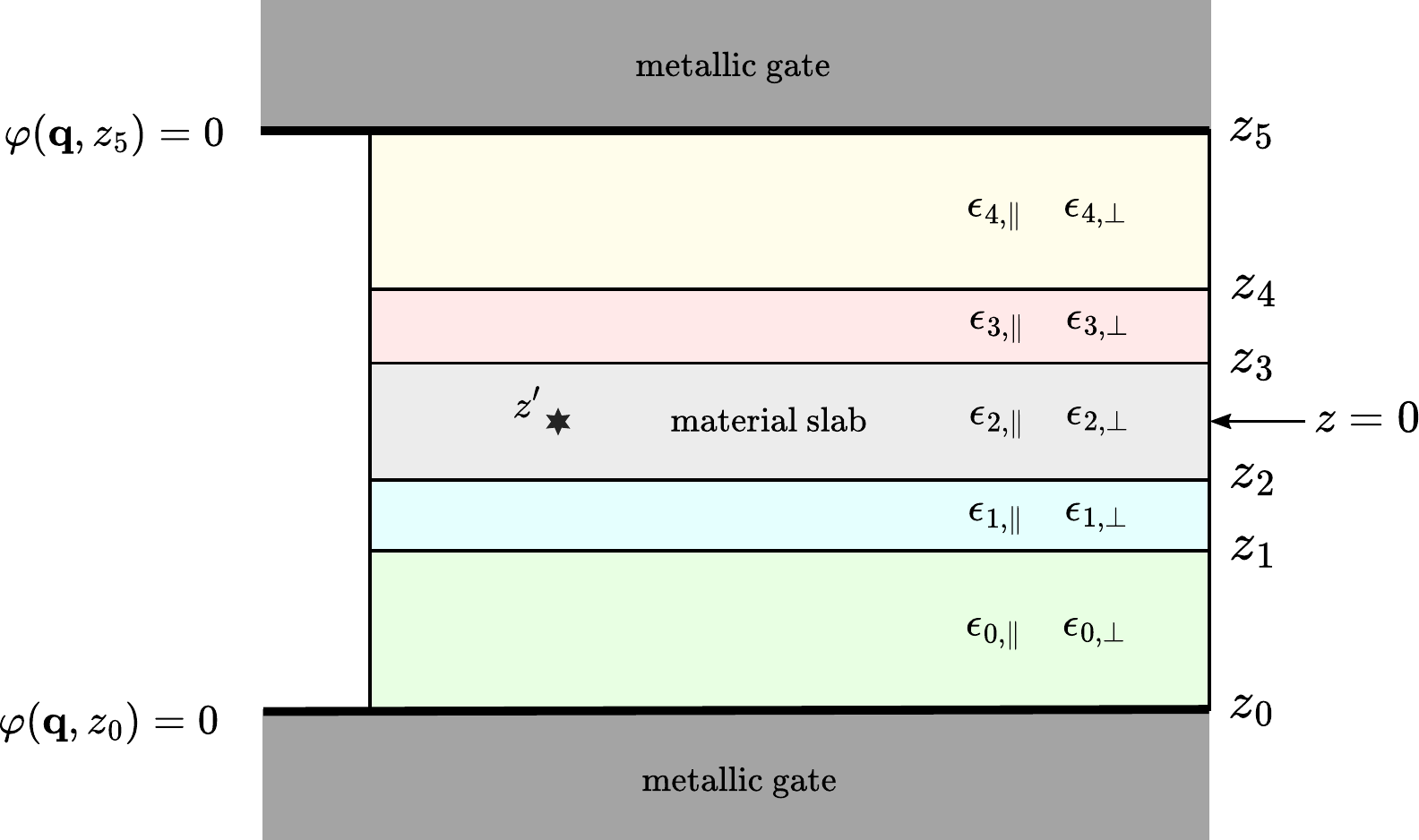}
    \caption{\textbf{Schematic of sample geometry.} This figure considers an example system with $N_\text{slab}=5$ slabs, where the material slab is chosen as the middle $i=2$ slab. The $z$-axis is centered so that the middle of the material slab is at $z=0$. Each slab $i$ is characterized by homogeneous dielectric constants $\epsilon_{i,\parallel},\epsilon_{i,\perp}$, and has its bottom face at $z=z_i$. Perfect metallic gates are positioned at $z=z_0$ and $z=z_5$, and enforce the vanishing of the 2D Fourier-transformed interaction potential $\varphi(\mbf{q},z_0)=\varphi(\mbf{q},z_5)=0$. To compute the screened interaction potential within the material slab, we solve the Poisson equation sourced by a point test charge at $z=z'$.}
    \label{figapp:dielectric_schematic}
\end{figure}

\subsection{General setup}\label{app:subsec:general setup}

The time-independent Maxwell equations for electrostatics in dielectric media are
\begin{equation}\label{eqapp:Maxwell}
    \nabla_{\tilde{\mbf{r}}}\cdot\mbf{D}=\rho_f,\quad \nabla_{\tilde{\mbf{r}}}\times \mbf{E}=0,
\end{equation}
where $\rho_f$ is the free charge, $\tilde{\mbf{r}}=(\mbf{r},z)$ is the 3D position, and $\mbf{r}=(x,y)$ is the 2D in-plane position. To solve the second equation above, we express
\begin{equation}
    \mbf{E}=\nabla_{\tilde{\mbf{r}}}\phi
\end{equation}
in terms of the electric potential $\phi$. 

We consider a system consisting of $N_\text{slab}$ slabs of varying thickness $d_i$ that are vertically stacked and indexed by $i=0,\ldots,N_\text{slab}-1$ in order of increasing $z$. Let $z_i$ be the $z$-coordinate of the bottom face of slab $i$, and $z_{N_\text{slab}}$ be the $z$-coordinate of the top face of slab $N_\text{slab}-1$ (the top slab), so that $d_i=z_{i+1}-z_i$. We place ideal metallic gates at $z_0$ and $z_{N_\text{slab}}$, i.e.~at the bottom and top of the stack. For a sample with no metallic gates, we send $z_0$ ($z_{N_\text{slab}}$) to $-\infty$ ($+\infty$). For a single-gated sample, we send $z_{N_\text{slab}}$ to $+\infty$. The slab $i=i_m$ is designated as the `material slab', i.e.~$t$MoTe$_2$, whose screened interaction potential $V_{ll'}(q)$ is of interest. By convention, we will center the $z$-axis so that the material slab is centered at $z=0$. The setup for $N_\text{slab}=5$ is illustrated in Fig.~\ref{figapp:dielectric_schematic}.

Each slab is associated with anisotropic and homogeneous dielectric constants $\epsilon_{i,\parallel}$ and $\epsilon_{i,\perp}$. In terms of these, we define the following constants for convenience
\begin{equation}\label{appeq:kappaeta}
    \kappa_i=\sqrt{\frac{\epsilon_{i,\parallel}}{\epsilon_{i,\perp}}},\quad \eta_i=\sqrt{\epsilon_{i,\parallel}\epsilon_{i,\perp}}.
\end{equation}
Within slab $i$, the electric displacement field $D$ is related to the electric field $E$ as
\begin{equation}
    D_x=\epsilon_0\epsilon_{i,\parallel}E_x,\quad D_y=\epsilon_0\epsilon_{i,\parallel}E_y,\quad D_z=\epsilon_0\epsilon_{i,\perp}E_z,
\end{equation}
where $\epsilon_0$ is the vacuum permittivity.
To account for the in-plane translation invariance, we define the 2D Fourier transform of the in-plane coordinates
\begin{equation}
    \phi(\mbf{r},z) = \int \frac{d^2\bm{q}}{(2\pi)^2} e^{i\bm{q}\cdot \bm{r}}\varphi(\bm{q},z),
\end{equation}
where $\mbf{q}$ is the 2D in-plane momentum. Due to the discontinuities in the dielectric environment as a function of $z$, we define the electric potential within each slab
\begin{equation}
    \varphi_i(\mbf{q},z)=\varphi(\mbf{q},z)\text{ for }z_i\leq z\leq z_{i+1}.
\end{equation}
Within slab $i$, we obtain the Poisson equation from the first equation of Eq.~\eqref{eqapp:Maxwell}
\begin{equation}
    \epsilon_{i,\perp}\partial_z^2\varphi_i(\mbf{q},z)-\epsilon_{i,\parallel}q^2\varphi_i(\mbf{q},z)=\rho_f(\mbf{r},z).
\end{equation}
To stitch together the solutions for different slabs, we utilize the boundary conditions
\begin{gather}
    \varphi_i(\mbf{q},z_{i+1})=\varphi_{i+1}(\mbf{q},z_{i+1})\\
    \epsilon_{i,\perp}\partial_z\varphi_i(\mbf{q},z_{i+1})= \epsilon_{i+1,\perp}\partial_z\varphi_{i+1}(\mbf{q},z_{i+1})
\end{gather}
for $i=0,\ldots N_\text{slab}-2$. For the outermost faces of the system, we assume perfect metallic gates which impose
\begin{equation}\label{eqapp:metalBC}
    \varphi_0(\mbf{q},z_0)=\varphi_{N_\text{slab}-1}(\mbf{q},z_{N_\text{slab}})=0.
\end{equation}
Note that we could have considered instead metallic gates characterized by e.g.~a finite screening length. However, the resulting Poisson equation would be non-local and considerably more complicated to solve. Furthermore, the screening length of graphite is $\sim1\,\text{nm}$~\cite{miyazaki2008inter}, which is smaller than the typical thicknesses of both the encapsulating hBN substrates and graphite gates in experiments.

In the following, we consider $\rho_f$ to consist of a test charge of magnitude $Q$ positioned at $\mbf{r}=0$ and height $z'$ within the material slab $i=i_m$
\begin{equation}
    \rho_f(\mbf{r},z)=Q\delta(\mbf{r})\delta(z-z').
\end{equation}
Hence for $i\neq i_m$, we have the general solution
\begin{equation}
    \varphi_i(\mbf{q},z)=A_i(\mbf{q})e^{\kappa_{i}qz}+B_i(\mbf{q})e^{-\kappa_{i}qz},
\end{equation}
while for $i=i_m$, we also need to include the particular integral
\begin{equation}
    \varphi_i(\mbf{q},z)=A_i(\mbf{q})e^{\kappa_{i}qz}+B_i(\mbf{q})e^{-\kappa_{i}qz}-\frac{Q}{2\epsilon_0\eta_iq}e^{-\kappa_{i}q|z-z'|}.
\end{equation}
For each $\mbf{q}$, we have $2N_\text{slab}$ unknowns $A_i(\mbf{q}),B_i(\mbf{q})$. There are $2N_\text{slab}$ boundary conditions, with $2N_\text{slab}-2$ from the internal slab interfaces, and $2$ from the outermost faces. The resulting linear system of equations can be solved to obtain $\varphi(\mbf{q},z)$. 

In the $t$MoTe$_2$ continuum model, we take the microscopic layers to lie at the Mo planes at $z=\pm z_\text{Mo}=\pm 0.365\,\text{nm}$. For the ED calculations of $t$MoTe$_2$ in the main text, the layer-dependent interaction potential $V_{ll'}(q)$ is computed by numerically solving the Poisson problem outlined above. The intralayer interaction [$l=l'$ in $V_{ll'}(q)$] is computed by taking $z=z'=z_\text{Mo}$, while the interlayer interaction [$l\neq l'$ in $V_{ll'}(q)$] is computed by taking $z=-z'=z_\text{Mo}$. These interactions in turn can also be used to compute the LLL Haldane pseudopotentials in Eq.~\eqref{appeq:Haldane_PP} with the effective magnetic length $\ell_B^*=2.02\,\text{nm}$ appropriate for $t$MoTe$_2$ at $\theta=3.7^\circ$. Note that for $\bm{q}=0$, the electric potential within the material slab is $\varphi_{i_m}(\bm{q}=0,z)=\frac{Q|z-z'
|}{2\epsilon_0\epsilon_{i_m,\perp}}$. 

\subsubsection{Isotropic material embedded in isotropic dielectric substrate}

In this subsection, we discuss the analytic form of the interaction potential for a simple configuration as presented in Ref.~\onlinecite{rytova2018screened}. We consider three slabs $(N_\text{slab}=3)$ and no metallic gates. The middle material slab $i_m=1$ has width $L$, occupies $-L/2\leq z\leq L/2$, and has isotropic dielectric constant $\epsilon_m$. The rest of space consists of a background isotropic dielectric environment with dielectric constant $\epsilon_b$. Define the ratio $\epsilon=\epsilon_m/\epsilon_b$. A free test charge $Q$ is positioned at $\mbf{r}=0$ and $z=z'$. Using the formalism presented above, the solution to the electric potential in the material slab can be obtained analytically:
\begin{equation}
    \varphi_1(q,z,z')=-\frac{Q}{2\epsilon_0 \epsilon_mq}\bigg[e^{-q|z-z'|}+\frac{2\delta}{e^{2qL}-\delta^2}\Big(\delta \cosh(q(z-z'))+e^{qL}\cosh(q(z+z')\Big)\bigg],
\end{equation}
where we defined $\delta=\frac{\epsilon-1}{\epsilon+1}$. 

In the long-wavelength limit where $qL\ll 1$, the interaction becomes independent of $z$ and $z'$:
\begin{equation}
    \varphi_1(q) = -\frac{Q}{2\epsilon_0 \epsilon_m q} \frac{e^{qL}+\delta}{e^{qL}-\delta}.
\end{equation}
Finally, taking the limit as $\epsilon \gg 1$, we get the Rytova-Keldysh (RK) interaction
\begin{equation}
    \varphi_1(q) =-\frac{Q}{2\epsilon_0 \epsilon_bq(1+l_{\text{RK}} q)},
\end{equation}
where we defined the length scale $l_{\text{RK}}=\frac{\epsilon L}{2}$. For long distances corresponding to small $q$, the interaction is inversely proportional to $\epsilon_b$ as the electric field lines are predominantly in the background dielectric and outside of the material slab. For short distances (but still longer than $L$) corresponding to large $q$, the interaction is inversely proportional to $\epsilon_m$ as the electric field lines are mostly contained within the material slab. In the context of $t$MoTe$_2$ encapsulated by hBN, we have $\eta_{\text{MoTe}_2}\simeq 14.5$ and $\eta_{\text{hBN}}\simeq 5$ (see Sec.~\ref{subsecapp:V0V1results} for a discussion of the anisotropic dielectric constants), from which we estimate $\epsilon=\eta_{\text{MoTe}_2}/\eta_{\text{hBN}}\simeq 2.9$.  Combining this with $L\simeq 1.4\,\text{nm}$, we estimate $l_\text{RK}\simeq 2.0\,\text{nm}$.

If probing on distances much smaller than $L$, i.e.~$qL\gg1$, we find instead $\varphi_1(q)\simeq -Q/(2\epsilon_0\epsilon_mq)$. In the context of $t$MoTe$_2$ at $\theta\simeq 3.7^\circ$, since the MoTe$_2$ slab thickness of $\simeq 1.4\,\text{nm}$ is appreciably smaller than the moir\'e lattice constant $\simeq 6\,\text{nm}$, this $qL\gg1$ regime is not expected to be important for the low-energy physics.

\subsubsection{Anisotropic material embedded in anisotropic dielectric substrate}
The calculations in the previous subsection can be generalised to the more complicated three-slab set-up where an anisotropic material with dielectric constants $\epsilon_m^{\perp,\parallel}$ is embedded in an anisotropic dielectric background with $\epsilon_b^{\perp,\parallel}$. We define $\kappa_m=\sqrt{\frac{\epsilon_m^\parallel}{\epsilon_m^\perp}}$, $\kappa_b=\sqrt{\frac{\epsilon_b^\parallel}{\epsilon_b^\perp}}$ and $\epsilon=\frac{\epsilon^\perp_m}{\epsilon^\perp_b}$. The electric potential in the middle material slab can be obtained analytically:
\begin{equation}
    \varphi(q,z,z')=-\frac{Q}{2\epsilon_0\eta_m q}\left(e^{-\kappa_m q|z-z'|}+\frac{N_-+e^{\kappa_m q L} N_+}{N_0}\right),
\end{equation}
where we defined
\begin{align}
    &N_0 = \epsilon^2 \kappa_m^2 \left(e^{2 \kappa_m q L}-1\right)+2 \epsilon \kappa_b \kappa_m \left(e^{2 \kappa_m q L}+1\right)+\kappa_b^2 \left(e^{2 \kappa_m q L}-1\right),\\
    &N_\pm = (\epsilon \kappa_m-\kappa_b)e^{\kappa_m q(z\pm z')}\left[\epsilon \kappa_m (e^{\kappa_m q(2z'\mp L)}+1)\mp \kappa_b(e^{\kappa_m q(2z'\mp L)}-1)\right].
\end{align}
In the long-wavelength limit where $qL\ll 1$, and in the limit where $\epsilon\gg 1$, one recovers the RK interaction
\begin{equation}
    \varphi(q) = -\frac{Q}{2\epsilon_0 \eta_b q(1+l_\text{RK}'q)} , \ \text{where } l'_\text{RK}=\frac{1}{2}\epsilon L \frac{\kappa_m^2}{\kappa_b}.
\end{equation}
Note that compared to the isotropic case, the RK length is now multiplied by a $\frac{\kappa_m^2}{\kappa_b}$ factor. In the context of $t$MoTe$_2$ at $\theta\simeq 3.7^\circ$, we have $\kappa_{\text{MoTe}_2}=1.45$ and $\kappa_{\text{hBN}}=1.40$ (see Sec.~\ref{subsecapp:V0V1results} for a discussion of the anisotropic dielectric constants), from which we estimate $l_\text{RK}\simeq 3.0\,\text{nm}$.

\subsection{Numerical results for $V_0/V_1$}\label{subsecapp:V0V1results}

In this section, we present numerical results for sample geometries relevant to $t$MoTe$_2$, focusing on the impact of dielectric engineering in suppressing the pseudopotential ratio $V_0/V_1$. The material $t$MoTe$_2$ slab is fixed to have thickness $w_{\text{MoTe}_2}= 1.4$nm~\cite{ma2022growth}. First principles calculations of untwisted bilayers~\cite{laturia2018dielectric} yield $\epsilon^\parallel_{\text{MoTe}_2} =21$ and $  \epsilon^\perp_{\text{MoTe}_2} =10$, leading to $\eta_{\text{MoTe}_2}=\sqrt{\epsilon^\parallel_{\text{MoTe}_2}\epsilon^\perp_{\text{MoTe}_2}}\simeq 14.5$.
In some of the calculations below, we will treat $\eta_{\text{MoTe}_2}$ as a tunable parameter to capture the uncertainty in the appropriate dielectric parameters for $t$MoTe$_2$. The primary encapsulating substrate for $t$MoTe$_2$ is hBN, which has $\epsilon^\parallel_\text{hBN}=6.9$ and $\epsilon^\perp_\text{hBN}=3.5$ \cite{laturia2018dielectric}. Note that $\eta_\text{hBN}=\sqrt{\epsilon^\parallel_\text{hBN}\epsilon^\perp_\text{hBN}}\simeq 5$. To soften the short-range repulsion and suppress the ratio $V_0/V_1$, we consider inserting one or more high-dielectric `spacer' slabs on either side of the $t$MoTe$_2$, with dielectric constants $\epsilon_\text{spacer}^{\perp,\parallel}$.

\begin{figure}[h!]
    \centering
    \includegraphics[width=7cm]{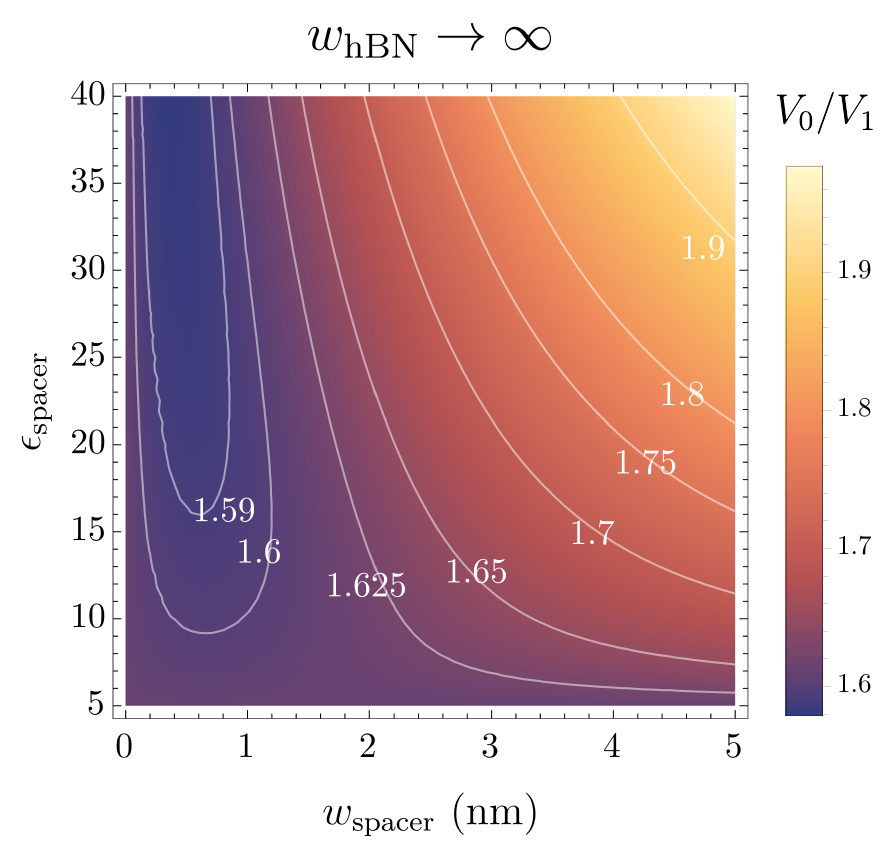}
    \caption{\textbf{LLL pseudopotential ratio $V_0/V_1$ as a function of the spacer thickness and dielectric constant.} $V_0/V_1$ is minimized when $\epsilon_\text{spacer}$ is large and $w_\text{spacer}<1\,$nm, suggesting that a thin, high-dielectric material would be ideal. Interaction potential computed assuming $z=z'=0$. We assume an isotropic spacer on both sides of the $t\text{MoTe}_2$, and set the metallic gates infinitely far away. The white curves are contours of varying $V_0/V_1$.}
    \label{fig:symmetric_eps_dep}
\end{figure}

In Fig.~\ref{fig:symmetric_eps_dep} we compute $V_0/V_1$ for the screened interaction when a spacer dielectric is added on either side of $t$MoTe$_2$. In particular, we study the dependence of $V_0/V_1$ on the spacer width $w_\text{spacer}$ and dielectric constant $\epsilon_\text{spacer}$ (assumed isotropic). We also set $w_\text{hBN} \rightarrow \infty$ and $z=z'=0$ for simplicity. We see that $V_0/V_1$ is minimized when $w_\text{spacer} \sim 0.5$nm roughly independent of $\epsilon_\text{spacer}$. Moreover, $V_0/V_1$ for small $w_\text{spacer}$ is roughly independent of $\epsilon_\text{spacer}$, although it seems to reach slightly smaller values for larger $\epsilon_\text{spacer}$. We note however that this behaviour does not persist for arbitrarily large $\epsilon_\text{spacer}$, and for a sufficiently large dielectric constant the $V_0/V_1$ ratio will begin to increase again. 
Thus, the ideal set-up would be to use a thin material with moderately-high dielectric constant as a spacer. 
For concreteness, one example material is HfO$_2$ which is isotropic with $\epsilon_{\text{HfO}_2}\simeq 26$~\cite{Wilk2001highk}, though other dielectrics can be considered.

\begin{figure}[h!]
    \centering
    \centerline{\includegraphics[width=18cm]{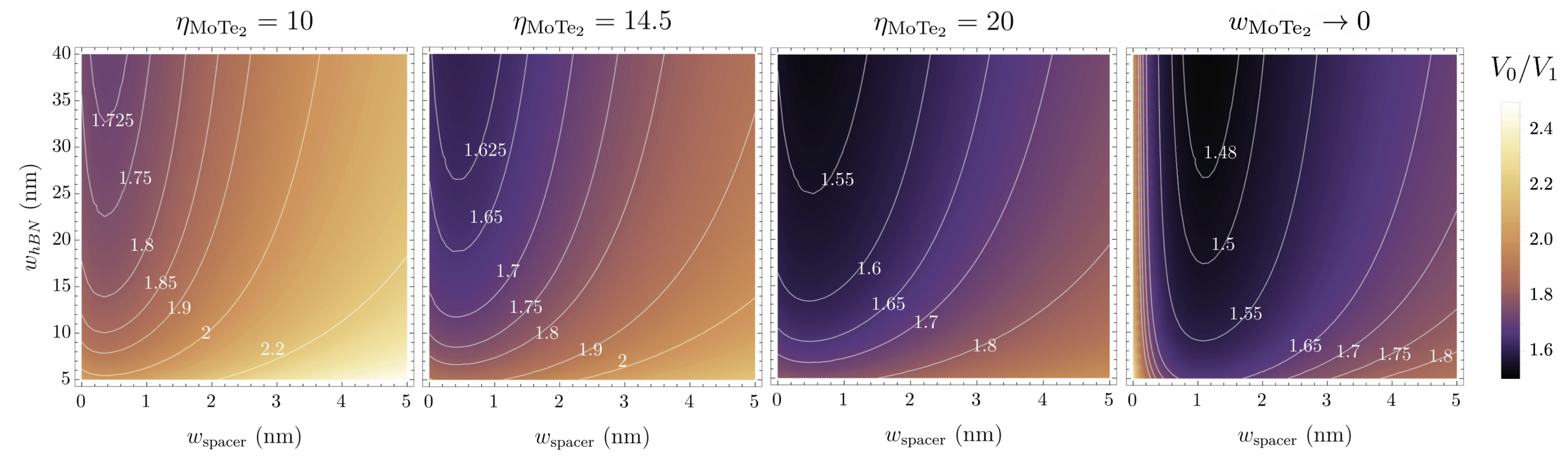}}
    \caption{\textbf{LLL pseudopotential ratio $V_0/V_1$ using a spacer with $\epsilon_\text{spacer}^{\perp,\parallel}=26$ on both sides of the $t$MoTe$_2$, as a function of $w_{\text{spacer}}$ and $w_{\text{hBN}}$.} The results in the three panels on the left are plotted for varying values of $\eta_{\text{MoTe}_2} = \sqrt{\epsilon^\parallel_{\text{MoTe}_2}\epsilon^\perp_{\text{MoTe}_2}}$, with a fixed $\kappa_{\text{MoTe}_2}=1.45$, showing that $V_0/V_1$ is smaller for larger values of $\eta_\mathrm{MoTe_2}$. From \cite{laturia2018dielectric}, we estimate $\eta_{\text{MoTe}_2}\approx 14.5$ for untwisted bilayer MoTe$_2$, but this value could be enhanced due to screening coming from the moiré bands. The results in the right-most panel neglect the finite thickness of the $t$MoTe$_2$ and show lower values of $V_0/V_1$ compared to the left three panels. Interaction potential computed assuming $z=z'=0$. The white curves are contours of varying $V_0/V_1$.}
    \label{fig:symmetric_w_vs_d}
\end{figure}

In Fig.~\ref{fig:symmetric_w_vs_d}, we plot $V_0/V_1$ as a function of $w_\text{hBN}$ and $w_{\text{spacer}}$ for different values of $\eta_{\text{MoTe}_2}$, fixing the dielectric anisotropy to $\kappa_{\text{MoTe}_2}=1.45$. For concreteness, we set $\epsilon_\text{spacer}^{\perp,\parallel}=26$ corresponding to HfO$_2$. We find that $V_0/V_1$ is overall lower for larger values of $\eta_{\text{MoTe}_2}$. For untwisted bilayer MoTe$_2$, we expect $\eta_{\text{MoTe}_2} \approx 14.5$. However, screening coming from the moiré bands could enhance this value in $t\text{MoTe}_2$. We also observe that larger $w_{\text{hBN}}$, and thus larger gate-to-gate distance, results in smaller values of $V_0/V_1$.

\begin{figure}[h!]
    \centering
    \includegraphics[width=11cm]{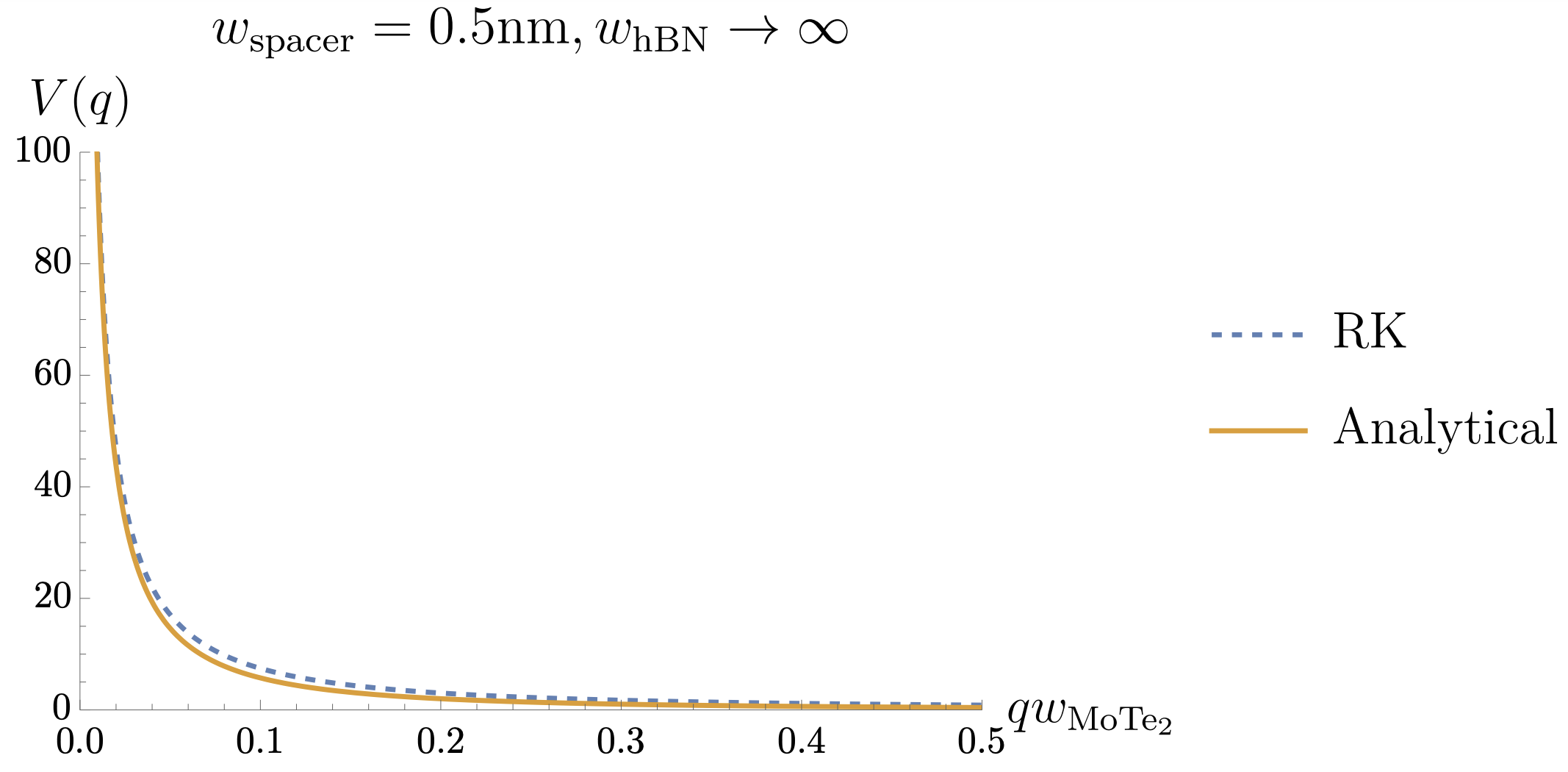}
    \caption{\textbf{Comparison of the analytical form of the interaction for $w_\text{hBN}\rightarrow \infty$ and $\epsilon_\text{spacer}^{\perp,\parallel}=26$ and its approximate asymptotic form as an RK interaction.} The RK interaction fits the analytical solution well for $qw_{\text{MoTe}_2}\ll1$.}
    \label{fig:RK_fit}
\end{figure}

In the $w_\text{hBN}\rightarrow \infty$ limit, the form of $V(q)$ is analytically tractable. Let $\kappa_1=\kappa_\text{hBN},\kappa_2=\kappa_{\text{MoTe}_2}$, $\eta_1=\eta_{\text{hBN}}$, $\eta_2=\eta_{\text{MoTe}_2}$, $\epsilon_1=\frac{\epsilon_{\text{spacer}}}{\epsilon_\text{hBN}^\perp}$, $\epsilon_2=\frac{\epsilon^\perp_{\text{MoTe}_2}}{\epsilon_{\text{spacer}}}$. We also let $w_{\text{MoTe}_2}=L$ and $w_\text{spacer}=w$ in the expressions that follow for sake of brevity. Then we find:
\begin{align}
    \varphi(q)&=-\frac{Q}{2\epsilon_0 \eta_2 q} \bigg[\frac{2 \left( ({\epsilon_2} \kappa_2-1)\chi(q)-4{\epsilon_2} \kappa_2 \left(\kappa_1- {\epsilon_1}\right) e^{\kappa_2 q L}\right)}{\left(\left( {\epsilon_2} \kappa_2+1\right) e^{\kappa_2 q L}- {\epsilon_2} \kappa_2+1\right)\chi(q)}+1\bigg],\\
    \chi(q)&={\epsilon_2} \kappa_1 \kappa_2 \left(e^{\kappa_2 q L}-1\right) \left(e^{2 q  w}-1\right)+\kappa_1 \left(e^{\kappa_2 q L}+1\right) \left(e^{2 q  w}+1\right)+ \nonumber \\ 
&\quad {\epsilon_1} \left[\left( {\epsilon_2} \kappa_2+1\right) e^{q  \left(\kappa_2L+2 w\right)}+\left(1- {\epsilon_2} \kappa_2\right) e^{2 q  w}+\left( {\epsilon_2} \kappa_2-1\right) e^{\kappa_2 q L}- {\epsilon_2} \kappa_2-1\right]. \nonumber
\end{align}
This interaction can be expanded as a power series in $q$ for $qL\ll1$ and $qw\ll 1$, and taking terms up to $O(q)$ we find:
\begin{equation}
    \varphi(q) \approx -\frac{Q}{2\epsilon_0 \eta_1 q}\Big(1-\underbrace{\frac{\epsilon_1^2\epsilon_2^2\kappa_2L - 2(\kappa_1 \epsilon_2- \epsilon_1^2 \epsilon_2)w-\kappa_1^2L}{2
    \kappa_1 \epsilon_1 \epsilon_2}}_{l_\text{RK}}q+o(q^2)\Big) \approx -\frac{Q}{2\epsilon_0 \eta_1 q(1+l_\text{RK}q)},
\end{equation}
where we defined an effective RK length scale:
\begin{equation}\label{appeq:eff_lRK_expansion}
    l_\text{RK} = \frac{\epsilon_1^2\epsilon_2^2\kappa_2 L - 2(\kappa_1 \epsilon_2 - \epsilon_1^2 \epsilon_2)w-\kappa_1^2L}{2
    \kappa_1 \epsilon_1 \epsilon_2} =\frac{1}{\kappa_1}\left(l_\text{RK}'\kappa_2^2 +\left(\epsilon_1-\frac{\kappa_1^2}{\epsilon_1}\right)w-\frac{\kappa_1^2}{2\epsilon_1 \epsilon_2}L\right),
\end{equation}
where $l_\text{RK}'=\epsilon_1\epsilon_2\tfrac{L}{2}=\tfrac{\epsilon_{\text{MoTe}_2}^\perp}{\epsilon_\text{hBN}}\tfrac{L}{2}$ is the RK length in the absence of a spacer. So for $w_\text{spacer}\lesssim w_{\text{MoTe}_2}$, the long wavelength interaction takes the form of an RK interaction with a renormalized length scale $l_\text{RK}'$. If $\kappa_1<\epsilon_1$, then $l_\text{RK}'$ will increase with $w_\text{spacer}$ until roughly $w_\text{spacer} \sim w_{\text{MoTe}_2}$. 
On the other hand, when $w_{\text{MoTe}_2} \rightarrow \infty$, a sensible approximation of the RK length should tend to $\epsilon_2 \frac{L}{2}$ (this cannot be seen from Eq.~\eqref{appeq:eff_lRK_expansion} which assumes $qw_{\text{MoTe}_2}\ll 1$), which is smaller than $l_\text{RK}$ with optimal $w_\text{spacer}$. Note that $V_0/V_1$ decreases as $l_\text{RK}$ increases. Consequently, $V_0/V_1$ should be minimized for some value $w_\text{spacer} \lesssim w_{\text{MoTe}_2}$, and then increase again for large $w_\text{spacer}$. This explains why the optimal spacer thickness is bounded above by the $t$MoTe$_2$ thickness of $w_{\text{MoTe}_2}=1.4$nm. In Fig.~\ref{fig:RK_fit} we compare the analytical solution to its RK asymptotic form, showing that the RK interaction is a good approximation at long wavelengths compared to the MoTe$_2$ thickness.

We note that the intrinsic finite thickness of MoTe$_2$ suppresses the impact of the dielectric spacer in decreasing $V_0/V_1$. This is clearly shown in the rightmost panel of Fig.~\ref{fig:symmetric_w_vs_d}, where lower values of $V_0/V_1$ can be obtained when MoTe${}_2$ is treated as a 2D material with $w_{\text{MoTe}_2}\rightarrow 0$ (this also ignores any contribution due to the in-plane polarizability~\cite{cudazzo2011dielectric}). 

\begin{figure}[h!]
    \centering
    \includegraphics[width=16cm]{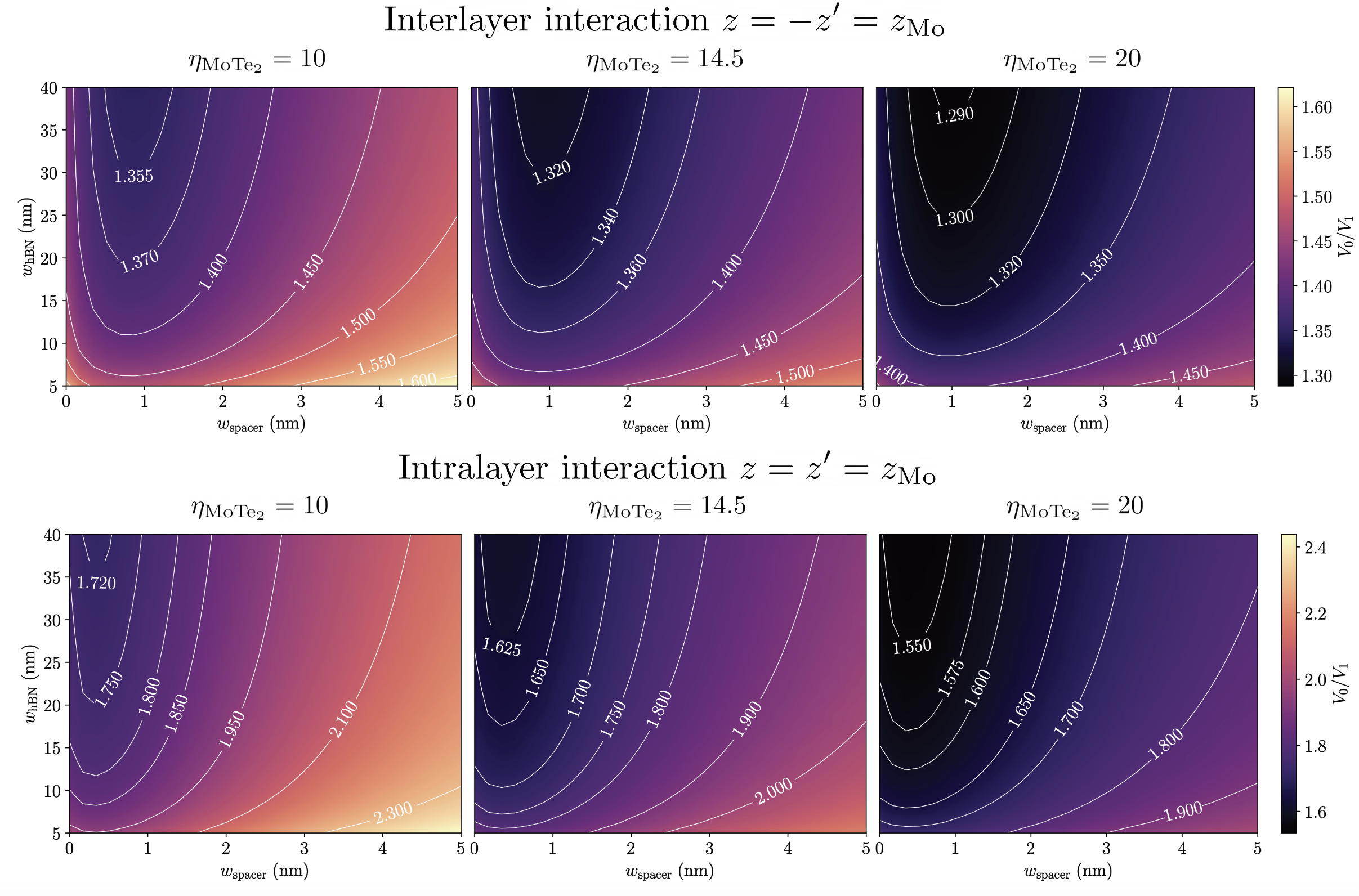}
    \caption{\textbf{LLL pseudopotential ratio $V_0/V_1$ for the interlayer (top row) and intralayer (bottom row) interaction as a function of $w_{\text{spacer}}$ and $w_{\text{hBN}}$.} The results are plotted for varying values of $\eta_{\text{MoTe}_2} = \sqrt{\epsilon^\parallel_{\text{MoTe}_2}\epsilon^\perp_{\text{MoTe}_2}}$, with a fixed $\kappa_{\text{MoTe}_2}=1.45$, and assuming an isotropic spacer with $\epsilon_\text{spacer}^{\perp,\parallel}=26$ placed on both sides of $t\text{MoTe}_2$. We set $z_\text{Mo}=\pm 0.365\,\text{nm}$ for the position of the Mo planes. Note that the colorbar ranges are different for the top and bottom row. The white curves are contours of varying $V_0/V_1$.}
    \label{fig:3D_interaction_V0V1}
\end{figure}

\begin{figure}[h!]
    \centering
    \includegraphics[width=11cm]{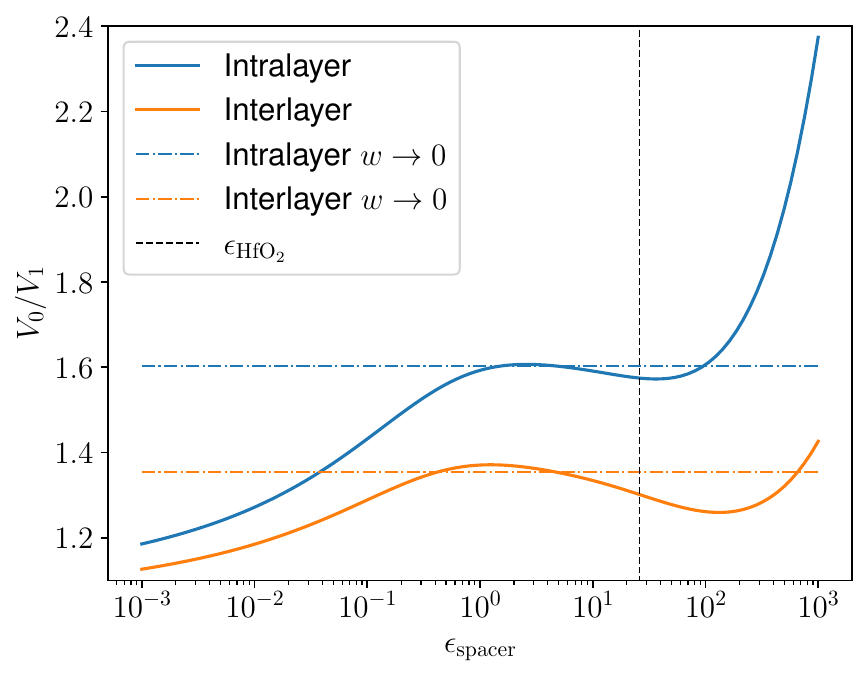}
    \caption{\textbf{$V_0/V_1$ ratio for the intralayer (blue) and interlayer (orange) interaction as a function of the spacer dielectric constant.} The hBN slabs are taken to be infinitely thick, and the spacer dielectric is taken to be isotropic. We set $z_\text{Mo}=\pm 0.365\,\text{nm}$ for the position of the Mo planes, fix $w_\text{spacer}=0.5\,\text{nm}$, and take $\epsilon^\parallel_{\text{MoTe}_2} =21,  \epsilon^\perp_{\text{MoTe}_2} =10, \epsilon^\parallel_\text{hBN}=6.9$ and $\epsilon^\perp_\text{hBN}=3.5$. The horizontal blue and orange dot-dashed lines indicate the corresponding $V_0/V_1$ values in the absence of spacers $w_\text{spacer}=0$. Vertical black dashed line corresponds to $\epsilon_\text{spacer}$ appropriate for HfO$_2$.}
    \label{fig:nonmonotonic}
\end{figure}

In Fig.~\ref{fig:3D_interaction_V0V1}, we account for the electrons residing on the Mo planes at $\pm z_\text{Mo}=\pm 0.365$nm. We find that $V_0/V_1$ is reduced for the interlayer interaction, while the intralayer interaction is largely unaffected (this follows from the fact that the intralayer interaction is only weakly dependent on $z$ in the long-wavelength limit, as discussed in App.~\ref{app:subsec:general setup}). In Fig.~\ref{fig:nonmonotonic}, we plot the $V_0/V_1$ ratios as a function of the spacer dielectric constant $\epsilon_\text{spacer}$, which exhibits a suppression for moderately high spacer dielectric constants, compared to the case without any spacers. $V_0/V_1$ increases again for sufficiently large $\epsilon_\text{spacer}$, since the spacer acts like metallic gates for $\epsilon_\text{spacer}\rightarrow\infty$. $V_0/V_1$ decreases significantly for small $\epsilon_\text{spacer}<1$ as the electric field lines are confined to the $t$MoTe$_2$, but such values of the spacer dielectric constant are not physically realistic.

We summarize the main messages of the calculations in this subsection. In the main text, our exact diagonalization calculations show that the FTI is stabilized in $\theta=3.7^\circ$ $t$MoTe$_2$ when the intralyer interaction has $V_0/V_1\simeq 1.1$ (see e.g.~Fig.~\ref{fig:main_tMoTe2_dielectric}b and c). Recall that the unscreened Coulomb interaction has $V_0/V_1=2$, which only increases in the presence of gate-screening (Fig.~\ref{fig:double_gated_pseudopotentials}). By accounting for the finite thickness of the MoTe$_2$ and using physically realistic values of the MoTe$_2$ and hBN dielectric constants, we find that $V_0/V_1$ can be reduced to $\sim 1.6$ for a typical hBN thickness of $\simeq 30\,\text{nm}$ (see bottom row of Fig.~\ref{fig:3D_interaction_V0V1} for $\eta_{\text{MoTe}_2}=14.5$). While this represents a substantial reduction compared to the unscreened Coulomb interaction, this does not yet reach the regime of $V_0/V_1$ that we expect is required for the FTI. Introduction of spacer dielectric slabs can further reduce $V_0/V_1$. Considering $\epsilon^{\perp,\parallel}_\text{spacer}=26$, which is relevant for HfO$_2$, we find that $V_0/V_1$ can be further reduced for a spacer width $w_\text{spacer}\lesssim 0.5\,\text{nm}$. However, at least for the parameters relevant to $\theta=3.7^\circ$ $t$MoTe$_2$, we find that this reduction is small (Fig.~\ref{fig:3D_interaction_V0V1}).

\clearpage

\section{Additional numerical data for $t$MoTe$_2$}
\label{secapp:additional_tMoTe2}

In this appendix, we present further numerical results on $t$MoTe$_2$ beyond those presented in the main text, in particular with regards to different system sizes, dependence on other physical parameters, and band mixing. For simplicity, unless otherwise stated, the calculations in this section use the 2D gate-screened Coulomb interaction with a RK lengthscale $l_\text{RK}$
\begin{equation}\label{eqapp:lRK}
    V_{ll'}(q)=\frac{e^2}{2\epsilon_0\epsilon q(1+l_\text{RK}q)}\tanh{\frac{q\xi}{2}},
\end{equation}
where $\xi$ is the distance between the gates, and $\epsilon$ is the relative permittivity which is assumed isotropic. Note that Eq.~\ref{eqapp:lRK} is independent of the layer indices $l,l'$. Based on non-moir\'e TMD studies~\cite{berkelbach2013theory,chernikov2014exciton,wang2018colloq,tuan2018coulomb,zhao2023probing}, we estimate $l_\text{RK}$ to be a few nm. The additional short-range interaction takes the form of Eq.~\eqref{eq:g} in the main text. 

\begin{figure}
    \centering  \includegraphics[width=\linewidth]{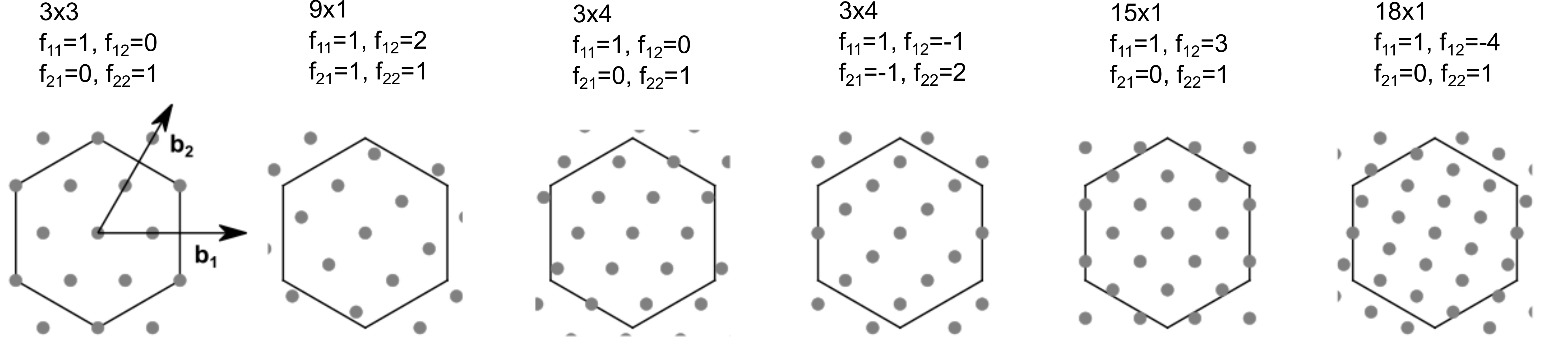}
    \caption{\textbf{Momentum grids in the mBZ for different tilted finite-size lattices.} The lattices shown correspond to those listed in Tab.~\ref{tab:systemsizes}. The basis moir\'e reciprocal lattice vectors $\mbf{b}_1$ and $\mbf{b}_2$ are shown in the first panel.}
    \label{fig:mesh_schematics}
\end{figure}

We list all the system sizes studied in this work in Tab.~\ref{tab:systemsizes}. In particular, we employ tilted lattices for different system sizes since, as known in finite-size studies of FCIs, lattice tilting can affect the results. We define the new reciprocal lattice vectors $\mathbf B_{1,2}$ as 
\begin{align}
    \mathbf B_{1}&=f_{11}\mathbf b_{1}+f_{12}\mathbf b_{2}\label{eq:tilting}\\
    \mathbf B_{2}&=f_{21}\mathbf b_{1}+f_{22}\mathbf b_{2},\nonumber
\end{align}
where $\mathbf b_{1,2}$ are the standard basis moir\'e reciprocal lattice vectors (see Fig.~\ref{fig:mesh_schematics}, left), and the $f_{ij}$ are integers. The set of inequivalent momenta that are not related by a reciprocal lattice vector can be generated as $n_1\mbf{B}_1/N_1+n_2\mbf{B}_2/N_2$, with $n_1=0,\ldots,N_1-1$ and $n_2=0,\ldots,N_2-1$.
\begin{table}
    \centering
    \begin{tabular}{cccccccc}
        $N_s$ & $N_1$ & $N_2$ & $f_{11}$ & $f_{12}$ & $f_{21}$ & $f_{22}$ \\ \hline\hline
        9 & 3 & 3 & 1 & 0 & 0 & 1 & \\
        9 & 9 & 1 & 1 & 2 & 1 & 1 & \\ \hline
        12 & 3 & 4 & 1 & 0 & 0 & 1 & \\
        12 & 3 & 4 & 1 & -1 & -1 & 2 & \\ \hline
        15 & 15 & 1 & 1 & 3 & 0 & 1 & \\ \hline
        18 & 18 & 1 & 1 & -4 & 0 & 1 & \\
         &  &  &  &  &  & \\
    \end{tabular}
    \caption{\textbf{System sizes studied in this work.} $N_s$ is the number of moiré unit cells, $N_1$ and $N_2$ are the numbers of points along the axes defined by the (tilted) reciprocal lattice vectors. $f_{ij}$ are parameters describing the tilting of the lattice. In particular, the $f_{ij}$ are the coordinates of new reciprocal lattice vectors $\mathbf{B}_{1,2}$ in terms of the standard basis reciprocal lattice vectors $\mathbf{b}_{1,2}$ (see Eq.~\eqref{eq:tilting}). $f_{ij}=\delta_{ij}$ for an untilted lattice.}
    \label{tab:systemsizes}
\end{table}
The momentum grids in the mBZ are shown in Fig.~\ref{fig:mesh_schematics}.

\subsection{Effects of system parameters}\label{subsecapp:effectsparameters}

In this subsection, we consider the properties of the FTI phase, and how it is impacted by changing various system parameters.

\begin{figure}
    \centering  \includegraphics[width=\linewidth]{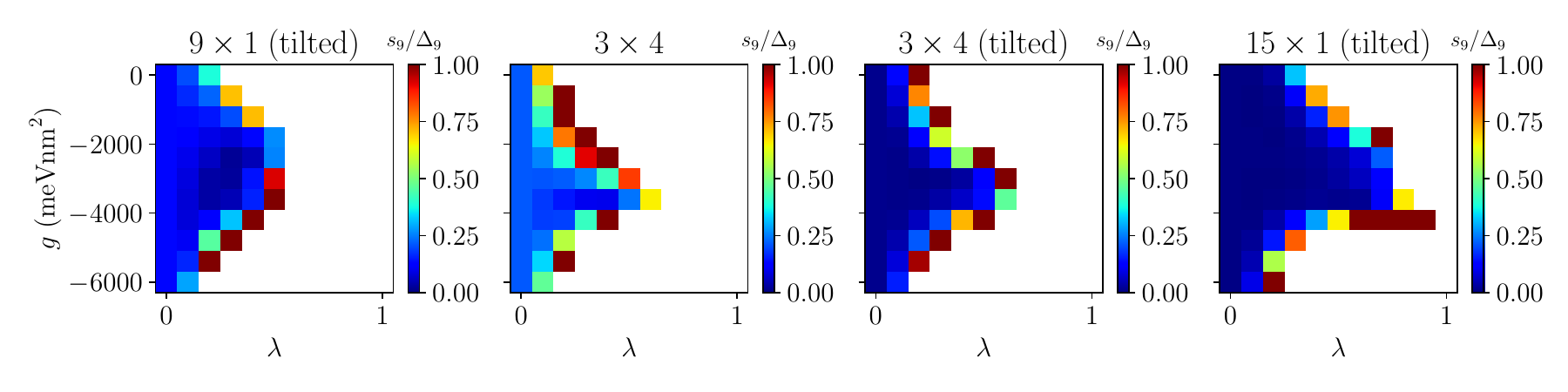}
    \caption{\textbf{FTI stability region for different system sizes.} For four different lattices, we plot the spacing/gap ratio $s_9/\Delta_9$ where $\Delta_9>0$ (white regions correspond to where $\Delta_9<0$). The FTI stability region is largest for the tilted $15\times1$ lattice (which is the lattice that is primarily used in the main text). We choose $10/\epsilon=2$ and $l_\textrm{RK}=0$nm.}
    \label{fig:ratio_compiled_sizes}
\end{figure}

\begin{figure}
    \centering  \includegraphics[width=\linewidth]{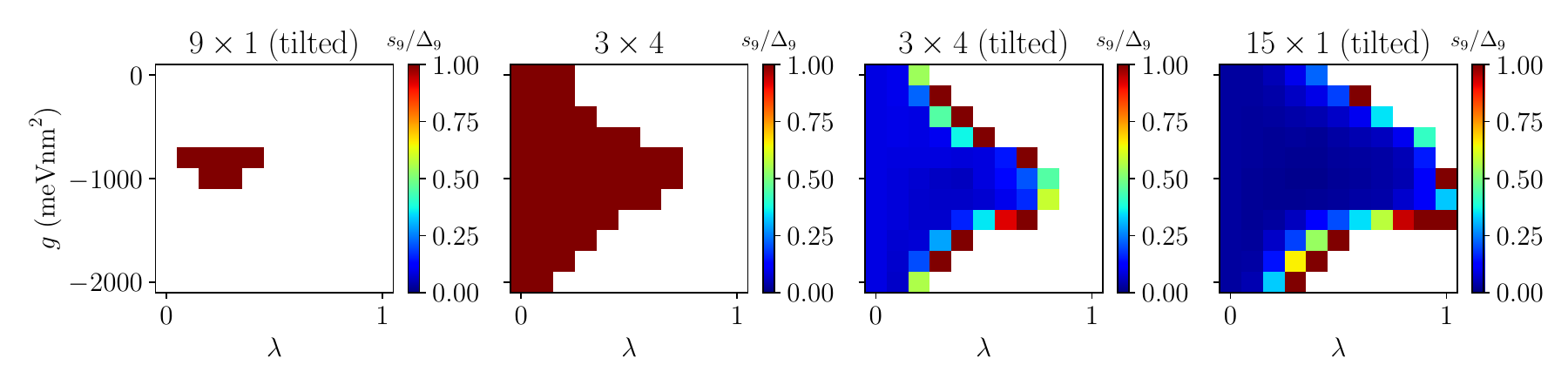}
    \caption{\textbf{FTI stability region for different system sizes with finite $l_\text{RK}$.} For four different lattices, we plot the spacing/gap ratio $s_9/\Delta_9$ where $\Delta_9>0$ (white regions correspond to where $\Delta_9<0$). The FTI stability region is largest for the tilted $15\times1$ lattice (which is the lattice that is primarily used in the main text). We choose $10/\epsilon=2$ and $l_\textrm{RK}=3$nm.}
    \label{fig:ratio_compiled_sizes_l_RK_3}
\end{figure}

We compare the $\lambda - g$ phase diagram for four different system sizes in Fig.~\ref{fig:ratio_compiled_sizes}, which shows that larger system sizes tend to favor the FTI state. There are also variations for fixed $N_s$ depending on the lattice tilting (compare the untilted and tilted $3\times 4$ lattices). The results in Fig.~\ref{fig:ratio_compiled_sizes} are for the bare gate-screened Coulomb interaction ($l_\textrm{RK}=0$nm), since the optimal RK length depends on the system size. We observe that for $g=0$, the FTI survives for only a small range of $\lambda\lesssim0.2$ (recall that $\lambda$ is a parameter that multiplies the intervalley part of the interaction, see Eq.~\eqref{app:eq:lambda}), but the stability window can be substantially enhanced for finite $g$. 
In Fig.~\ref{fig:ratio_compiled_sizes_l_RK_3} we show the same results but for $l_\textrm{RK}=3$nm. Note the difference in the $g$-axis scale compared to Fig.~\ref{fig:ratio_compiled_sizes}: in the presence of finite $l_\text{RK}$, the magnitude of $|g|$ required to obtain the FTI is reduced. The FTI phase for the $15\times1$ lattice persists to $\lambda=1$. Note also the adiabatic continuity of the gapped phase to $\lambda=g=0$, where the ground state in the spin-unpolarized sector is a decoupled product of two FCIs.

\begin{figure}
    \centering  \includegraphics[width=0.5\linewidth]{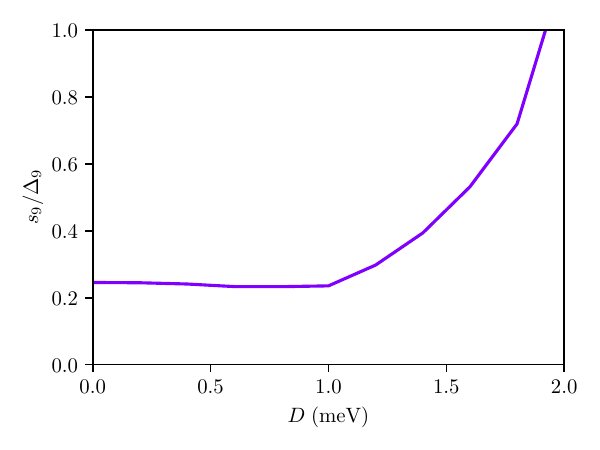}
    \caption{\textbf{Stability of FTI under displacement field.} For the tilted $15\times1$ lattice we plot the FTI spacing/gap ratio under an applied displacement field $D$. Applying a displacement field destroys the FTI.   We choose $10/\epsilon=2$ and $l_\textrm{RK}=3$nm and $g=-1250$meVnm$^2$.}
    \label{fig:displacement_field}
\end{figure}

\begin{figure}
    \centering  \includegraphics[width=0.35\linewidth]{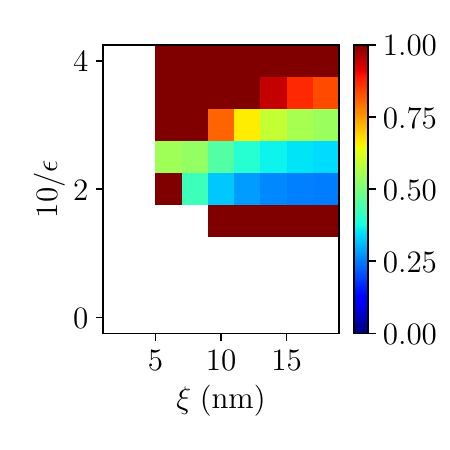}
    \caption{\textbf{Dependence on screening length.} We plot $s_9/\Delta_9$ for the tilted $15\times 1$ lattice at $\theta=3.7^\circ$. Smaller values of the screening length $\xi$ are worse for the FTI.  White regions correspond to where $\Delta_9<0$. Parameters: $\theta=3.7^\circ$, $(\epsilon/10)\times g=-625$meVnm$^2$ and $l_\textrm{RK}=3$nm.}
    \label{fig:xi}
\end{figure}

\begin{figure}
    \centering  \includegraphics[width=0.7\linewidth]{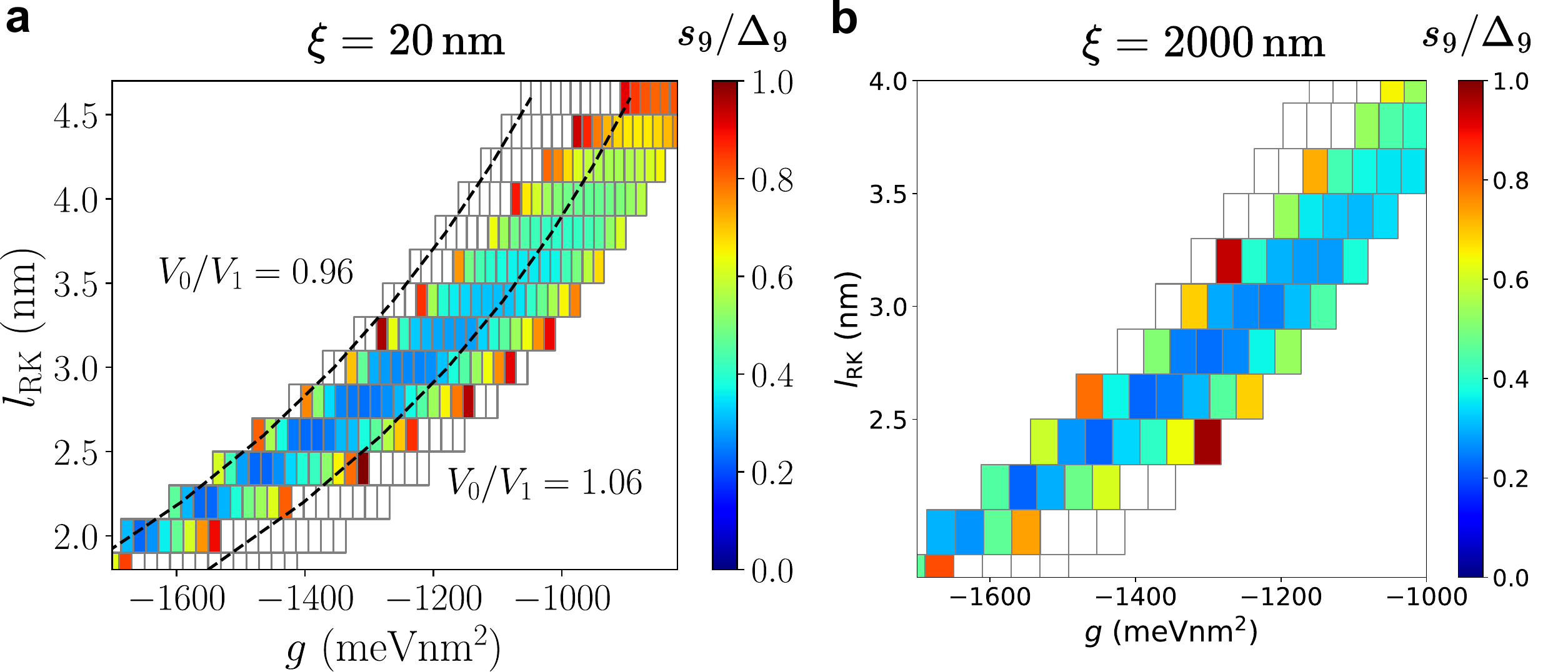}
    \caption{\textbf{Dependence on $l_\text{RK}$ and $g$ for different screening lengths.} 
    White regions correspond to where $\Delta_9<0$ or $s_9/\Delta_9>1$. a) $\xi=20\,\text{nm}$. Contours of constant $V_0/V_1$ are indicated. b) $\xi=2000\,\text{nm}$. In both cases, calculations are performed on the tilted $15\times 1$ lattice ($N_s=15$) with  $\theta=3.7^\circ$ and $\epsilon=5$.}
    \label{fig:tMoTe2_ED_lRK_vs_g_15x1_xi2000}
\end{figure}

In Figs.~\ref{fig:displacement_field} and \ref{fig:xi} we show the dependence of the FTI stability on the displacement field $D$ and the gate-screening length $\xi$ respectively. The displacement field is deleterious to the FTI, and we find that the spacing/gap ratio improves for larger $\xi$. In Fig.~\ref{fig:tMoTe2_ED_lRK_vs_g_15x1_xi2000}, we show $s_9/\Delta_9$ as a function of $l_\text{RK}$ and $g$ for $\xi=20$ and $2000\,\text{nm}$. The spacing/gap ratio is minimal approximately along a stripe of constant $V_0/V_1$. The FTI does not survive for much larger or smaller $l_\text{RK}$ than those shown in Fig.~\ref{fig:tMoTe2_ED_lRK_vs_g_15x1_xi2000}, implying that the higher pseudopotentials $V_{m>1}$ still play a non-trivial role in the stability of the FTI. The results for $\xi=20$ and $2000\,\text{nm}$ are very similar, demonstrating that $\xi=20\,\text{nm}$ already yields results close to the limit $\xi\rightarrow\infty$.

In Fig.~\ref{fig:occupation} we show that the momentum space occupation number fluctuations $\sigma_{n_k}$ of the FTI correlate with the spacing/gap ratio $s_9/\Delta_9$. In particular, a smaller $\sigma_{n_k}$ is correlated with a smaller $s_9/\Delta_9$.

\subsection{Spin depolarization}\label{subsecapp:spindepolarization}
A key criterion for obtaining a Time-Reversal FTI at $\nu=-4/3$ is that the global ground state of the system has zero spin polarization at this filling. Other FTI's with topological order but not necessarily edge states can be obtained if Time Reversal is broken.  In this subsection, we provide numerical results pertaining to the spin polarization as a function of system parameters and band mixing.

\begin{figure}
    \centering  \includegraphics[width=0.7\linewidth]{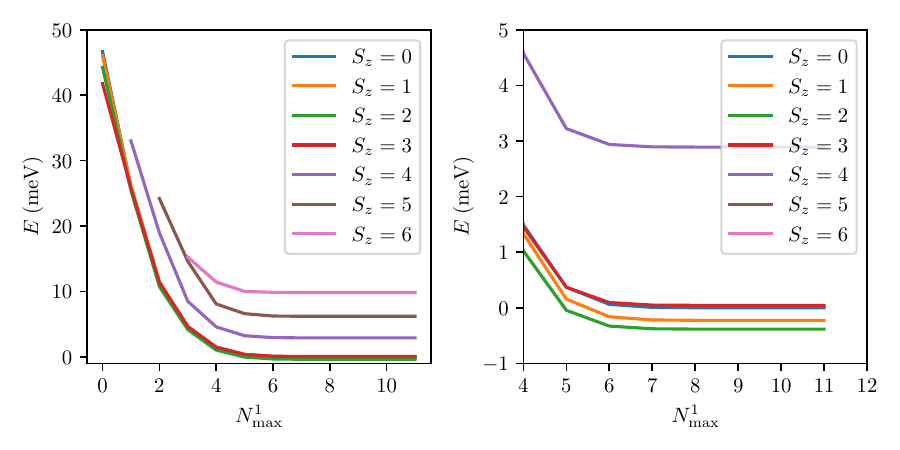}
    \caption{\textbf{Effect of band mixing on ground state energy across different spin sectors for the $3\times3$ lattice.} The ground state energies in the different spin sectors are plotted as a function of the maximum number of holes $N^1_\text{max}$ allowed in band 1 according to our truncation scheme (see main text). The energies converge as $N^1_\text{max}$ is increased. The right plot is a zoomed-in version of the left. We choose $10/\epsilon=0.6$, $g=0$meVnm$^2$ and $l_\textrm{RK}=0$nm. }
    \label{fig:remote_band_depol_3x3}
\end{figure}

\begin{figure}
    \centering  \includegraphics[width=0.7\linewidth]{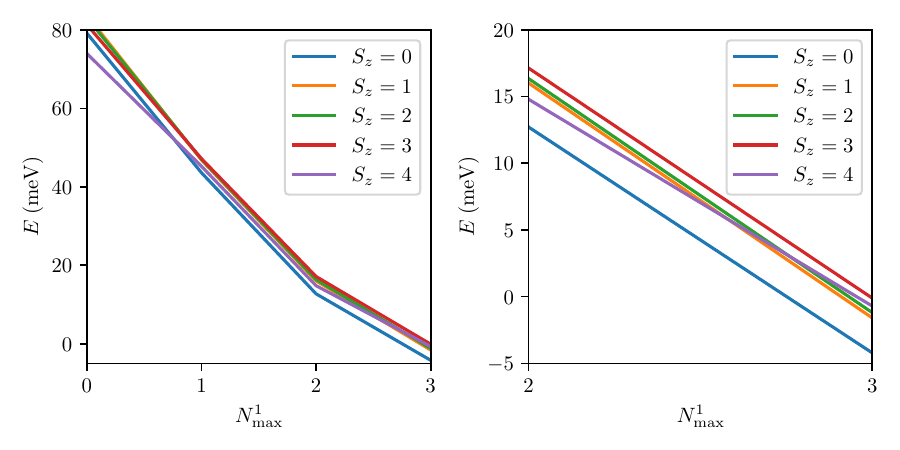}
    \caption{\textbf{Spin depolarization from band mixing for the $3\times4$ lattice}. Same as Fig.~\ref{fig:remote_band_depol_3x3} but for the $3\times4$ lattice. As the allowed number of holes $N^1_{\text{max}}$ in the remote band is increased, the ground state changes from a spin-polarized state to a spin-unpolarized state. Since the FTI state occurs in the spin-unpolarized sector, this shows that a necessary (but not sufficient) condition for the FTI to emerge for this particular choice of parameters ($10/\epsilon=0.6$, $g=0$meVnm$^2$ and $l_\textrm{RK}=0$nm) is band mixing.  The right plot is a zoomed-in version of the left.}
    \label{fig:remote_band_depol_3x4}
\end{figure}

\begin{figure}
    \centering
    \includegraphics[width=0.35\linewidth]{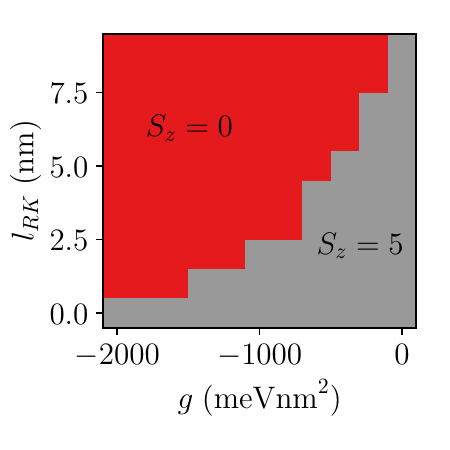}
    \caption{\textbf{Spin depolarization with $g$ and $l_\textrm{RK}$}. We choose $10/\epsilon=2$ and study the $15\times 1$ tilted lattice with $N^1_\text{max}=0$. Both $g$ and $l_\textrm{RK}$ have a depolarizing effect on the ground state.}
    \label{fig:depol}
\end{figure}

In Fig.~\ref{fig:remote_band_depol_3x3} we show how band mixing affects the ground state energies across different spin sectors for the untilted $3\times 3 $ lattice. The total energies in every $S_z$ sector are converged at $N^1_\text{max}=7$. In Fig.~\ref{fig:remote_band_depol_3x4} we show that the band mixing can lead to complete depolarization of the spin of the global ground state for the larger untilted $3\times 4$ lattice. In Fig.~\ref{fig:depol} we show that large, negative $g$ as well as finite $l_\textrm{RK}$ have a similar depolarizing effect on the ground state.  

\subsection{Insights from Hartree-Fock}\label{subsecapp:insights}

\begin{figure}
    \centering  \includegraphics[width=0.7\linewidth]{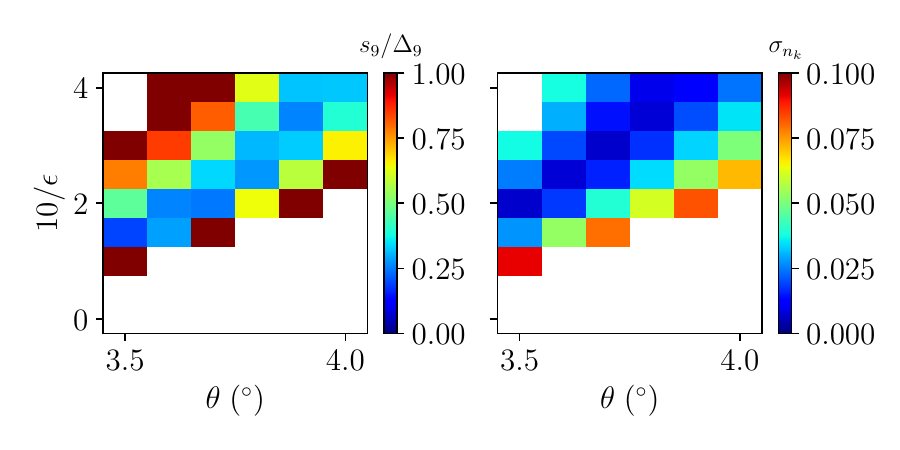}
    \caption{\textbf{Occupation number fluctuations.} For the tilted $15\times1$ lattice we plot the FTI spacing/gap ratio (left; white regions correspond to where $\Delta_9<0$) and compare to the occupation number fluctuations $\sigma_{n_k}$ (right) as a function of interaction strength $10/\epsilon$ and twist angle $\theta$. The FTI spacing/gap ratio is smallest where the fluctuations in the occupation number are smallest. We choose $l_\textrm{RK}=3$nm and $(\epsilon/10)\times g=-625$meVnm$^2$.}
    \label{fig:occupation}
\end{figure}

We expect that a narrow bandwidth is an important requirement for stabilizing an FTI phase. For instance, Fig.~\ref{fig:occupation} demonstrates that a small FTI spacing/gap ratio is correlated with the FTI having small number occupation fluctuations across the mBZ. A significant dispersion would penalize having such a homogeneous number occupation distribution. However, the non-interacting bandwidth is not necessarily a reliable proxy for judging the effective dispersion, since interaction effects may significantly renormalize the bandwidth. In this subsection, we discuss our estimations of the Hartree-renormalized bandwidth and its correlation with the continuum model parameters and the FTI phase. We also comment on the fate of the $|C|=1$ insulator at $\nu=-1$ within Hartree-Fock calculations.

\begin{figure}
    \centering  \includegraphics[width=0.8\linewidth]{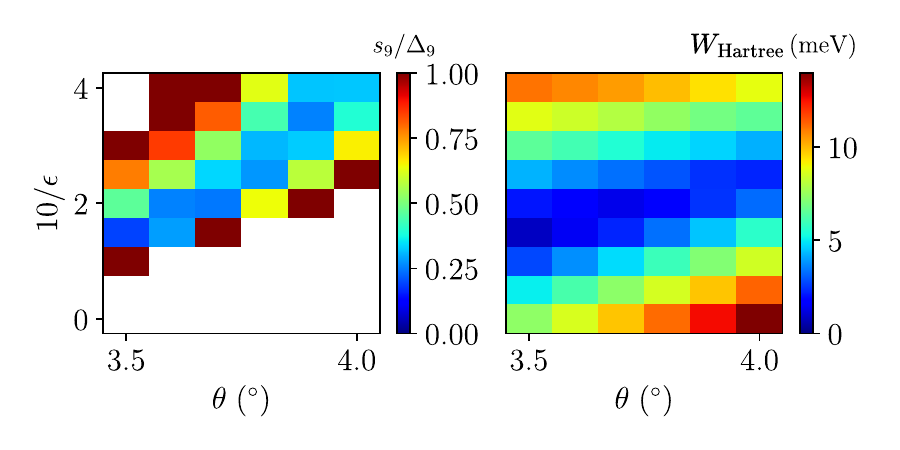}
    \caption{\textbf{Comparison of $\theta-(10/\epsilon)$ phase diagram with Hartree renormalized bandstructure}. Left: FTI spacing/gap ratio as a function of $10/\epsilon$ and $\theta$. White regions correspond to where $\Delta_9<0$. Right: Bandwidth $W_\text{Hartree}$ of the Hartree renormalized bandstructure. We perform a single-shot (i.e.~not self-consistent) Hartree calculation for the tilted $15\times1$ lattice with the hole density matrix $\nu_s\mathds{1}$ where the average hole occupation per spin is $\nu_s=2/3$. All the parameters in the Hartree calculation are the same as the ED [$(\epsilon/10)\times g=-625$meVnm$^2$, $l_\mathrm{RK}=3$nm]. }
    \label{fig:theta_U_with_Hartree}
\end{figure}

\begin{figure}
    \centering  \includegraphics[width=0.8\linewidth]{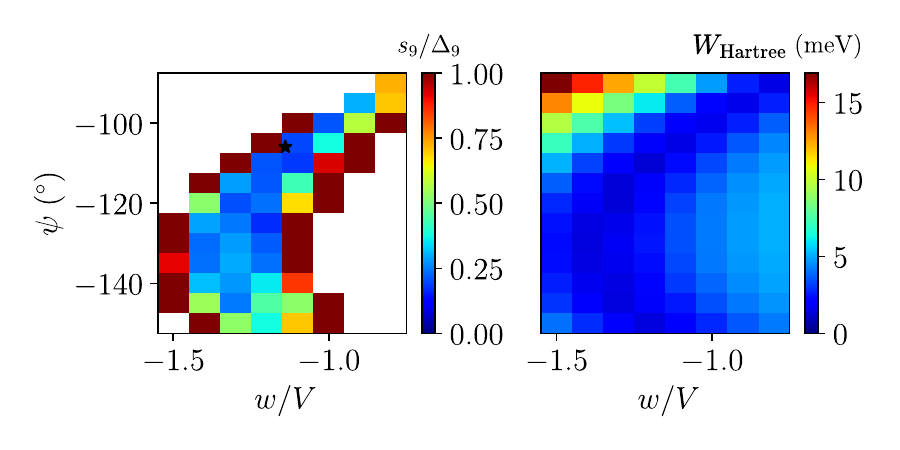}
    \caption{\textbf{Comparison of $w/V-\psi$ phase diagram with Hartree renormalized bandstructure}. Left: FTI spacing/gap ratio as a function of continuum model parameters. White regions correspond to where $\Delta_9<0$. The black star indicates the continuum model parameters of Ref.~\cite{jia2023moire}.  
    Right: Bandwidth $W_\text{Hartree}$ of the Hartree renormalized bandstructure. We perform a single-shot (i.e.~not self-consistent) Hartree calculation for the tilted $15\times1$ lattice with the hole density matrix $\nu_s\mathds{1}$, where the average hole occupation per spin is $\nu_s=2/3$. All the parameters in the Hartree calculation are the same as the ED (at fixed $w^2+V^2=25$meV, $g=-1250$meVnm$^2$, $l_\mathrm{RK}=3$nm, $\theta=3.7^\circ$, $10/\epsilon=2$). }
    \label{fig:wV_psi_only_Hartree}
\end{figure}

In Figs.~\ref{fig:theta_U_with_Hartree} and \ref{fig:wV_psi_only_Hartree}, we show that the minimum bandwidth $W_\text{Hartree}$ of the Hartree-renormalized band structure gives a good estimate of the region where the FTI has the lowest value of $s_9/\Delta_9$. To obtain the Hartree-renormalized band structure in a 1BPV model, we consider the Hartree potential generated by the momentum-independent one-body hole density matrix $P_0=\frac{2}{3}\mathds{1}$, where $\mathds{1}$ is the identity in valley space. Hence, $P_0$ corresponds to a state at physical filling $\nu=-4/3$. The Hartree-renormalized band structure is given by the sum of the non-interacting dispersion and the Hartree potential. The choice of $P_0$ is motivated from the fact that the occupation number in the FTI is mostly uniform across the mBZ (Fig.~\ref{fig:occupation}).

\begin{figure}
    \centering  \includegraphics[width=1\linewidth]{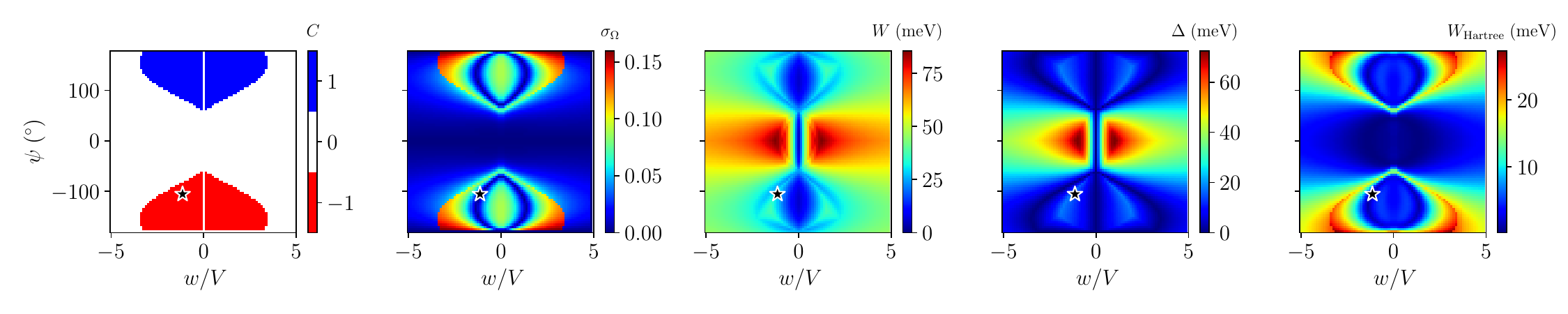}
    \caption{\textbf{Bandstructure properties of the highest valence band in the $w/V-\psi$ phase diagram}. From left to right, we show the Chern number $C$, Berry curvature fluctuations $\sigma_\Omega$, single-particle bandwidth $W$, bandgap $\Delta$ and Hartree-renormalized  bandwidth $W_\mathrm{Hartree}$ (at fixed $w^2+V^2=25$meV and $\theta=3.7^\circ$). $W_\mathrm{Hartree}$ is calculated from a single-shot (i.e.~not self-consistent) Hartree calculation for the tilted $15\times1$ lattice with density matrix $\nu_s\mathds{1}$, where the average occupation per spin is $\nu_s=2/3$. For the Hartree calculation we use a gate-screened Coulomb interaction with $10/\epsilon=2$, $g=-1250$meVnm$^2$ and $l_\mathrm{RK}=3$nm. The black star indicates the continuum model parameters of Ref.~\cite{jia2023moire}.}
    \label{fig:DFT_SP}
\end{figure}

In Fig.~\ref{fig:DFT_SP} we show properties of the band structure as a function of the continuum model parameters. We note that for parameters where the Chern number of the highest single-particle valence band is non-zero, the region where the Hartree-renormalized bandwidth $W_\text{Hartree}$ is minimal correlates with the region where the Berry curvature fluctuations are small. 

\begin{figure}
    \centering  \includegraphics[width=1\linewidth]{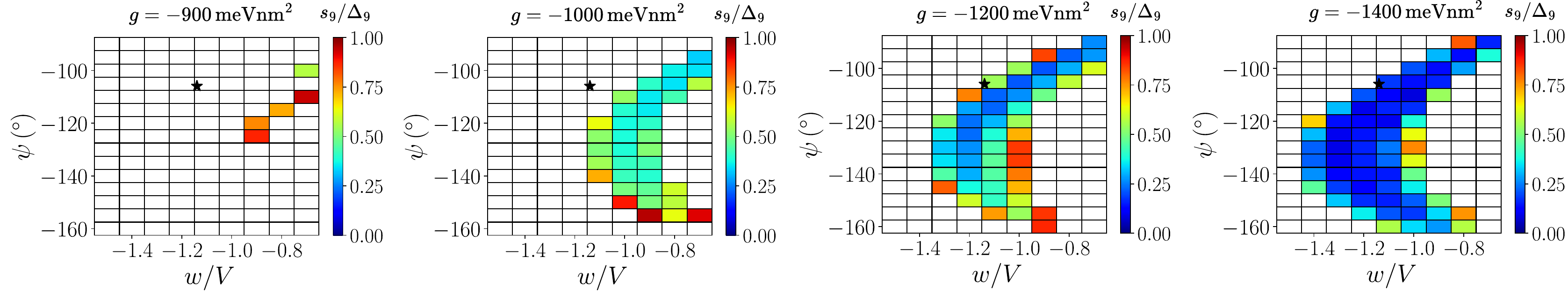}
    \caption{\textbf{Varying continuum model parameters for different $g$.} White regions correspond to where $\Delta_9<0$ or $s_9/\Delta_9>1$. From left to right, we show the analog of Fig.~\ref{fig:main_tMoTe2_ED}c in the main text for $g=-900,-1000,-1200,-1400\,$meVnm$^2$, on the tilted $15\times 1$ lattice. Note that we use the layer-dependent interaction obtained by solving the Poisson equation for the configuration in Fig.~\ref{fig:main_tMoTe2_dielectric}a with $\epsilon^{\perp}_{\text{MoTe}_2}=10$, $\epsilon^{\parallel}_{\text{MoTe}_2}=21$ and no spacer.}
    \label{fig:wVrat_g}
\end{figure}

In Fig.~\ref{fig:wVrat_g}, we repeat the calculation of Fig.~\ref{fig:main_tMoTe2_ED}c in the main text for smaller values of $|g|$. We find that for certain continuum model parameters with reduced $|w/V|$, the FTI can survive down to $g\sim -900\,\text{meVnm}^2$. 

\begin{figure}
    \centering  \includegraphics[width=1\linewidth]{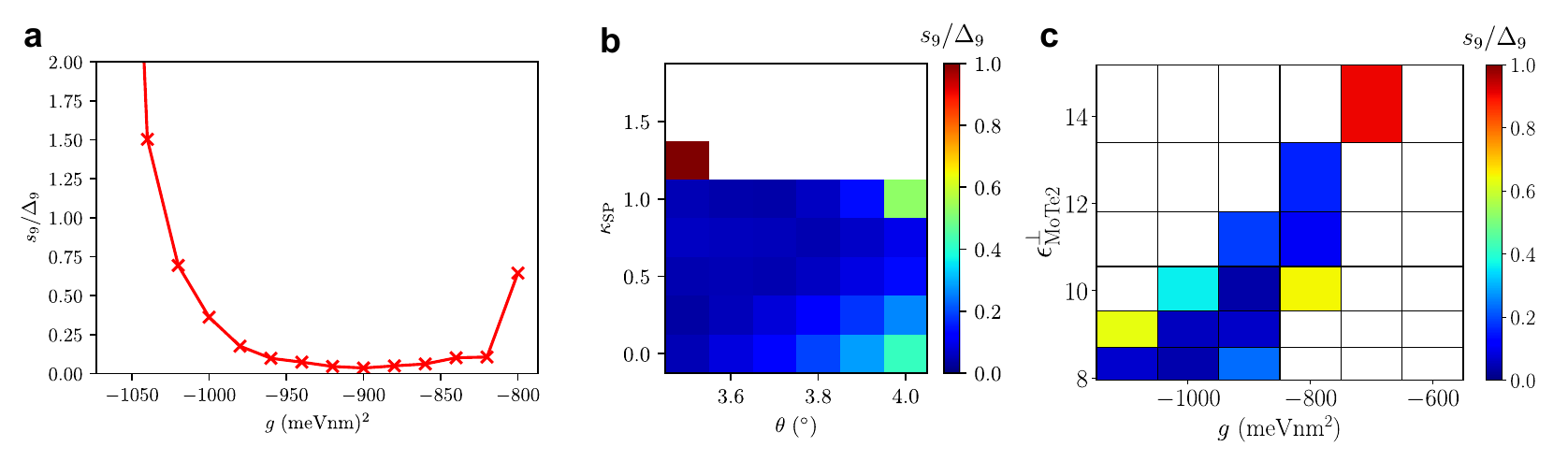}
    \caption{\textbf{Projection into band 1.} ED calculations are performed on the tilted $15\times 1$ lattice. Note that for all three plots, we use the layer-dependent interaction obtained by solving the Poisson equation for the configuration in Fig.~\ref{fig:main_tMoTe2_dielectric}a with no spacer. a) FTI spacing/gap ratio as a function of $g$. We set $\epsilon^{\perp}_{\text{MoTe}_2}=10$ and $\epsilon^{\parallel}_{\text{MoTe}_2}=21$. b) $s_9/\Delta_9$ as a function of twist angle $\theta$ and kinetic energy factor $\kappa_\text{SP}$ for $g=-900\,$meVnm$^2$. White regions correspond to where $\Delta_9<0$. We set $\epsilon^{\perp}_{\text{MoTe}_2}=10$ and $\epsilon^{\parallel}_{\text{MoTe}_2}=21$. c) $s_9/\Delta_9$ as a function of $\epsilon^\perp_{\text{MoTe}_2}$ and $g$ for fixed $\epsilon^\parallel_{\text{MoTe}_2}/\epsilon^\perp_{\text{MoTe}_2}=2.1$. }
    \label{fig:ED_band1}
\end{figure}

In Fig.~\ref{fig:ED_band1}, we repeat the $N_s=15$ 1BPV calculations of Figs.~\ref{fig:main_tMoTe2_ED}b,d and Fig.~\ref{fig:main_tMoTe2_dielectric}b, except that we project into band 1 instead of band 0. We find that the FTI can have a much lower $s_9/\Delta_9$ ratio, and survives to significantly lower values of $|g|$. Furthermore, the FTI phase extends over a larger range of kinetic factor $\kappa_\text{SP}$, including the flat-band limit $\kappa_\text{SP}=0$. The FTI similarly drifts to smaller $|g|$ when the dielectric constant of the MoTe$_2$ slab is increased.

\begin{figure}
    \centering  \includegraphics[width=0.76\linewidth]{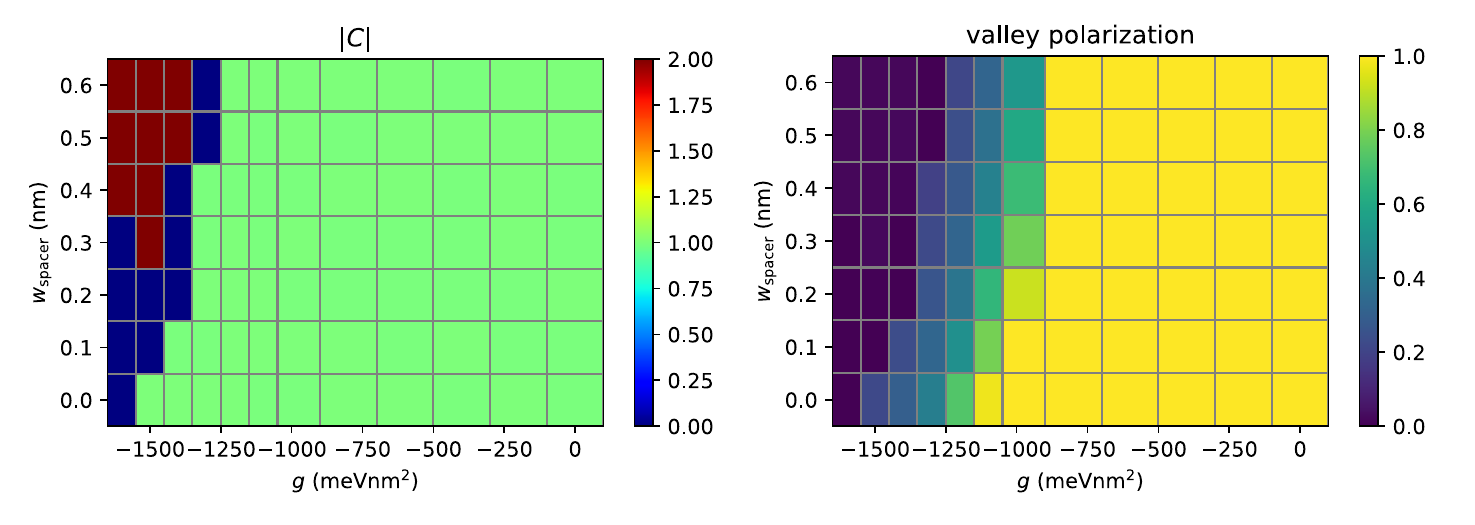}
    \caption{\textbf{Self-consistent Hartree-Fock calculations at $\nu=-1$.} Left: Chern number $C$. Right: Valley polarization. System size is $18\times18$, and 2 bands kept per valley. Note that we use the layer-dependent interaction obtained by solving the Poisson equation for the configuration in Fig.~\ref{fig:main_tMoTe2_dielectric}a with $\epsilon^{\perp}_{\text{MoTe}_2}=10$, $\epsilon^{\parallel}_{\text{MoTe}_2}=21$ and $\epsilon^{\perp,\parallel}_{\text{spacer}}=26$.}
    \label{fig:HF_wspacer}
\end{figure}

In Fig.~\ref{fig:HF_wspacer}, we perform self-consistent Hartree-Fock calculations at $\nu=-1$, which allow for intervalley coherence at wavevectors $\bm{q}=\Gamma_\text{M},K_\text{M},K'_\text{M}$. Note that the interaction we use here differs from previous theoretical HF studies of $t$MoTe$_2$. In particular, we use the layer-dependent interaction obtained by solving the Poisson equation (see App.~\ref{secapp:dielectric} for details) for the configuration in Fig.~\ref{fig:main_tMoTe2_dielectric}a with $\epsilon^{\perp}_{\text{MoTe}_2}=10$, $\epsilon^{\parallel}_{\text{MoTe}_2}=21$ and $\epsilon^{\perp,\parallel}_{\text{spacer}}=26$. We project onto both band 0 and band 1 (i.e.~a 2BPV calculation) --- a 1BPV calculation within band 0 would yield valley polarized Chern insulators for all the parameters shown. In the 2BPV case, we find that HF yields a $|C|=1$ Chern insulator at $\nu=-1$ for the parameters where 1BPV ED gives an FTI at $\nu=-4/3$ (see Fig.~\ref{fig:main_tMoTe2_dielectric}c in the main text). However, the HF state is only partially valley-polarized and has a finite amount of intervalley coherence.

\subsection{Fractional Chern insulators at $\nu=-2/3$}

The spin-polarized FCI phase at $\nu=-2/3$ has been extensively studied in previous works~\cite{Li2023,Crepel2023,moralesdurán2023pressureenhanced,wang2023fractional,reddy2023global,Reddy2023,Yu2024MFCI0,xu2024maximally,abouelkomsan2023band}. However, the interaction that we use (see App.~\ref{secapp:dielectric} for more details) for calculations of the FTI phase at $\nu=-4/3$ in the main text differs in several respects from these previous works. Firstly, we account for the layer-dependence of the interaction by considering the continuum model degrees of freedom to reside on the Mo planes at $z=\pm 0.365\,\text{nm}$. Secondly, our interaction potential is computed with the influence of the (anisotropic) dielectric environment imposed by the $t$MoTe$_2$ device geometry. This subsection addresses how these factors affect the FCI phase at $\nu=-2/3$.

\begin{figure}
    \centering  \includegraphics[width=0.8\linewidth]{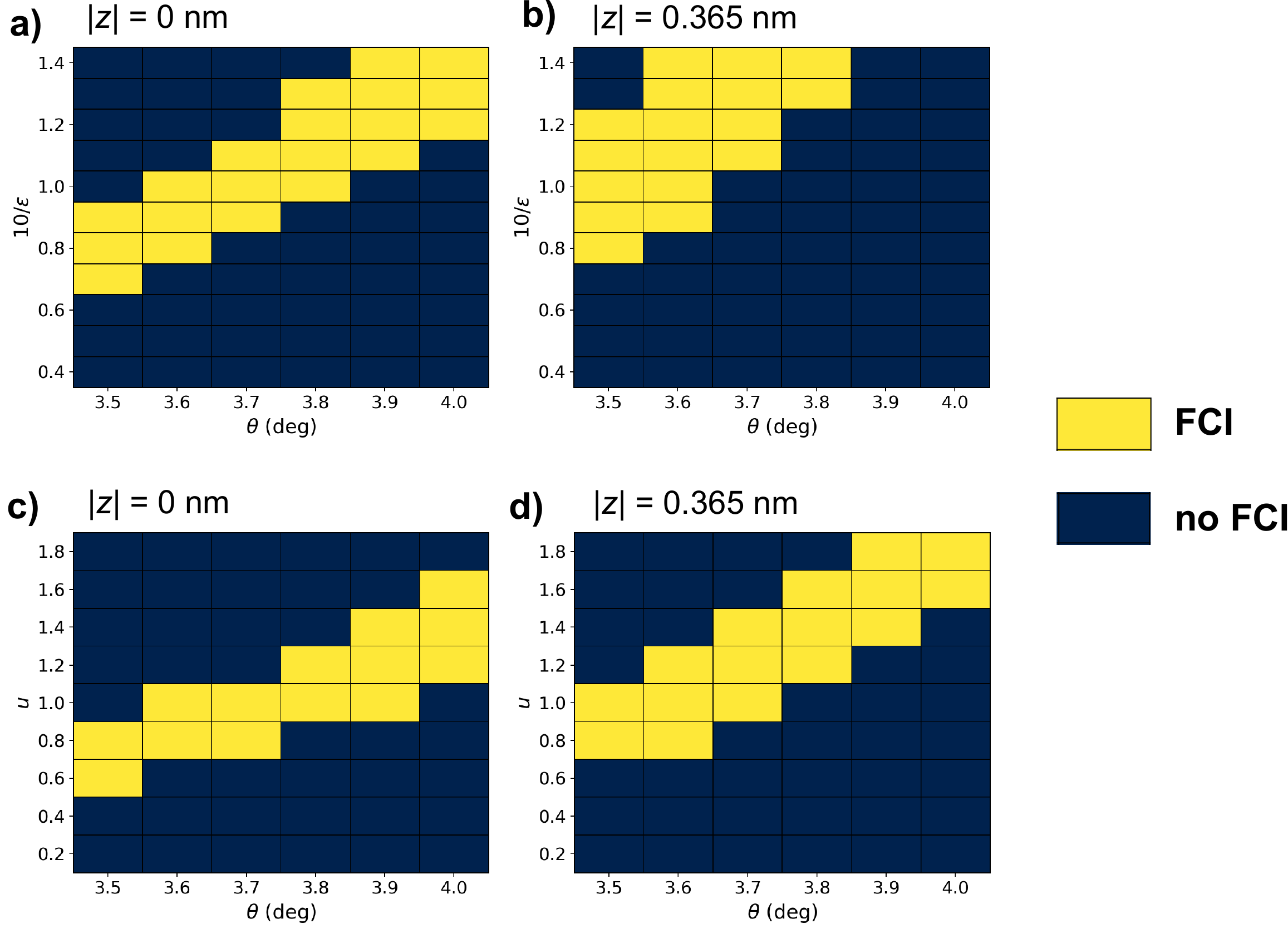}
    \caption{\textbf{ED calculations of FCIs at $\nu=-2/3$.} 
    All calculations are performed on the untilted $3\times 4$ lattice, and projected into band 0 and band 1 (see Fig~\ref{fig:main_bandstruct_flux}a), with no restriction on the band occupations (i.e.~2BPV calculations). We label a region an FCI if the spread of the FCI ground manifold is less than the gap to higher energy states.
    a,b) Interaction is computed by assuming an isotropic dielectric with dielectric constant $\epsilon$. The metallic gates are at $z=\pm 10\,\text{nm}$. In a), the continuum model layers are at $|z|=0\,\text{nm}$, while in b), the continuum model layers are at $|z|=\pm 0.365\,\text{nm}$.
    c,d) The interaction is obtained by solving the Poisson equation for the configuration in Fig.~\ref{fig:main_tMoTe2_dielectric}a with $\epsilon^{\perp}_{\text{MoTe}_2}=10$, $\epsilon^{\parallel}_{\text{MoTe}_2}=21$ and no spacer. $u$ is a multiplicative factor that scales the overall strength of the interaction. In c), the continuum model layers are at $|z|=0\,\text{nm}$, while in d), the continuum model layers are at $|z|=\pm 0.365\,\text{nm}$.}
    \label{fig:FCIcomparison}
\end{figure}

In Fig.~\ref{fig:FCIcomparison}, we perform spin-polarized  ED calculations at $\nu=-2/3$ on the $3\times 4$ untilted lattice, using different interaction potentials. We project into band 0 and band 1, with no restriction on the band occupations (i.e.~2BPV calculations). We present phase diagrams as a function of twist angle $\theta$ and a multiplicative factor $10/\epsilon$ or $u$ that scales the overall strength of the interaction.  Note that here, we consider the calculation to yield an FCI if the \emph{spread} of the FCI ground state manifold is less than the gap to higher energy states, since this measure has been widely used in previous studies of the FCI at $\nu=-2/3$. In Fig.~\ref{fig:FCIcomparison}a,b, we consider a dielectric environment that is used in most theoretical treatments of $t$MoTe$_2$, i.e.~we assume a uniform isotropic dielectric with dielectric constant $\epsilon$ which is terminated by perfect metallic gates at $z=\pm \xi/2$, where we choose $\xi=20\,\text{nm}$. In Fig.~\ref{fig:FCIcomparison}a, we place the continuum model layers at $z=0$, yielding a layer-independent interaction $V(q)\propto \tanh(q\xi/2)/\epsilon q$. In Fig.~\ref{fig:FCIcomparison}b, we place the continuum model layers at the Mo planes at $z=\pm0.365\,\text{nm}$, leading to a layer-dependent interaction. In Fig.~\ref{fig:FCIcomparison}c,d, the interaction is obtained by solving the Poisson equation for the configuration in Fig.~\ref{fig:main_tMoTe2_dielectric}a with $\epsilon^{\perp}_{\text{MoTe}_2}=10,\epsilon^{\parallel}_{\text{MoTe}_2}=21$ and no spacer. 
In Fig.~\ref{fig:FCIcomparison}c, the continuum model layers are at $|z|=0\,\text{nm}$, while in Fig.~\ref{fig:FCIcomparison}d, the continuum model layers are at $|z|=\pm 0.365\,\text{nm}$. In all cases, we find that the FCI phase lies in a stripe-like region in the phase space of interaction strength and twist angle. This implies that the properties of the FCI are qualitatively unchanged by incorporating the layer-dependence of the interaction and finer details of the dielectric screening.

\clearpage

\section{Edge mode stability and transport}\label{secapp:edgemodes}

A strong experimental indication for the realization of a topologically ordered state in FQHE and FCI phases is the fractionally quantized charge or thermal Hall response. For FTIs topological transport (edge) responses may or may not be available depending on the symmetries of the system -- irrespective of whether they support topological anyonic excitations in the 2D bulk. Here, we discuss the $\nu=-4/3$ FTI from this perspective. First in App.~\ref{subsecapp:stability}, we clarify that topological edge modes here are only stable when spin U(1) symmetry is present along with TR symmetry and charge U(1). This is in contrast to e.g.~the FTI at filling factor $\nu=1/3+1/3$ whose edge modes are stable against perturbations that violate spin U(1) symmetry~\cite{Levin_Stern}. Subsequently in App.~\ref{subsecapp:Buttiker}, we list the fractionally quantized charge transport responses of a four-contact geometry, assuming all these symmetries are present. 
We choose units where $e=h=1$ for simplicity.

\subsection{Stability analysis}\label{subsecapp:stability}

The stability of the edge modes in FTIs has been studied in Ref.~\onlinecite{Neupert2011FTI} and we use the notation introduced there. The edge is modelled as a collection of chiral Luttinger liquids. Its universal properties are contained in a symmetric and invertible integer matrix $K$ as well as  an integer charge vector $Q$ and an integer spin vector $S$. The number of components of the latter vectors corresponds to the number of real chiral scalar fields that describe the edge. The component $Q_i$ denotes the electric charge of the $i$th scalar field, while the component $S_i$ denotes the spin associated with up or down projection along the $\hat{S}_z$ spin-axis which is conserved in our system. 

Time-reversal symmetry implies the structure~\cite{Neupert2011FTI}
\begin{equation}\label{eqapp:TRStopodata}
K=\mat{\kappa& \Delta\\ \Delta^\mathsf{T} & -\kappa},
\quad Q=\mat{\rho\\ \rho},
\qquad S=\mat{\rho\\ -\rho}
\end{equation}
in terms of matrices $\kappa$ and $\Delta$ as well as a vector $\rho$ for each spin component. Above, $\kappa$ is an $N\times N$ symmetric matrix, $\Delta$ is an $N\times N$ anti-symmetric matrix, and $\rho$ is an $N$-component vector. $2N$ is the number of scalar fields. Since spin is odd under time-reversal, it can be seen that in our conventions, time-reversal relates the top $N$ components with the bottom $N$ components. 

The edge of a $\nu=-4/3$ FTI state is represented by $N=2$ and 
\begin{equation}\label{eqapp:2/3topodata}
\kappa = \mat{
   -3   &  0 \\
   0  & 1
}\ ,\quad \Delta = 0\ 
,\quad
 \rho=
\mat{1\\1},
\end{equation}
i.e.~in the quantum Hall analogy, in each of the Kramers paired blocks of $K$ there is a full Landau level and a conjugate Laughlin 1/3 state to model $\nu=-2/3$ in that spin sector\footnote{Eq.~\ref{eqapp:2/3topodata} corresponds to the so-called symmetric basis representation of each Kramers paired block. As shown in Ref.~\onlinecite{Santos2011TRShierarchy}, this is equivalent to the hierarchical basis representation with $\kappa'=\begin{pmatrix}-2 & 1 \\ 1 & 1\end{pmatrix}$ and $\rho'=(0,1)^\mathsf{T}$. This can be related to the symmetric basis via the linear transformation $\kappa=W^\mathsf{T}\kappa'W$ and $\rho=W^\mathsf{T}\rho'$, where $W^\mathsf{T}=\begin{pmatrix}
    -1 & 1 \\ 0 & 1
\end{pmatrix}$.}. The data of Eq.~\ref{eqapp:TRStopodata} (using the parameters of Eq.~\ref{eqapp:2/3topodata}) is therefore consistent with an FTI that is connected to the limit of a decoupled product of $\nu=-2/3$ FCIs in each spin with opposite chiralities. We can verify the vanishing charge Hall conductance, mandated by time-reversal symmetry, by computing $Q^\mathsf{T}K^{-1}Q=0$. The `spin filling fraction'~\cite{Neupert2011FTI} is $\frac{1}{2}Q^\mathsf{T}K^{-1}S=2/3$, which simply reflects the fact that each spin sector individually has a Hall conductance of $\pm 2/3$ which is opposite in the two spin sectors.

We now ask whether the edge modes are stable against perturbations at the edge. For instance, if we allow all possible tunnelings of electronic charge among the different edge branches, some or all of the gapless modes could be gapped out. The answer will generally depend on what symmetries the perturbations preserve.  Following the stability analysis of Ref.~\onlinecite{Neupert2011FTI}, we find that our state at $\nu=-4/3$ retains protected gapless modes as long as we impose charge $\U(1)$, spin $\U(1)$, and TR symmetry. We consider this a fractional quantum spin Hall (FQSH) state as it preserves spin $\U(1)$.
Breaking spin $\U(1)$ will allow to gap out all the edge modes, but this is beyond FQSH which always assumes spin $\U(1)$. The MoTe$2$ samples, due to their valley-spin locking, maintain $U(1)$ symmetry. 
To demonstrate the above statements, we specify a possible choice of two tunneling vectors $T$ that together would gap out all the edge modes~\cite{Neupert2011FTI}
\eq{
T'=\mat{1\\-1\\-1\\1},\quad
T''=\mat{-1\\3\\1\\-3}.
}
Both tunneling vectors are charge-conserving since they satisfy $Q^\mathsf{T}T=0$. They also satisfy time-reversal since they obey $\Sigma_1 T=-T$, where $\Sigma_1=\begin{pmatrix}
0 & I_N \\ I_N & 0
\end{pmatrix}$ with $I_N$ the $N\times N$ identity matrix.
Using the spin vector $S^{\mathsf{T}}=(1,1,-1,-1)$, one obtains that while $T'$ is spin conserving ($S^{\mathsf{T}}T'=0$), the tunneling vector $T''$ is not ($S^{\mathsf{T}}T''\neq 0$). The latter therefore represents the gapless modes that are protected by spin U(1), charge U(1) and time reversal symmetry in a strongly interacting edge.  
Note that the charge per spin that is gapped out by $T'$ is zero since $(1,-1)\rho=0$, so that the total charge of the remaining gapless mode is $\pm 2/3$ per spin. This result will be used below in the next subsection on non-local transport measurements.

\subsection{Landauer-B\"uttiker analysis}
\label{subsecapp:Buttiker}

\begin{figure}[h!]
    \centering
    \includegraphics[width=5cm]{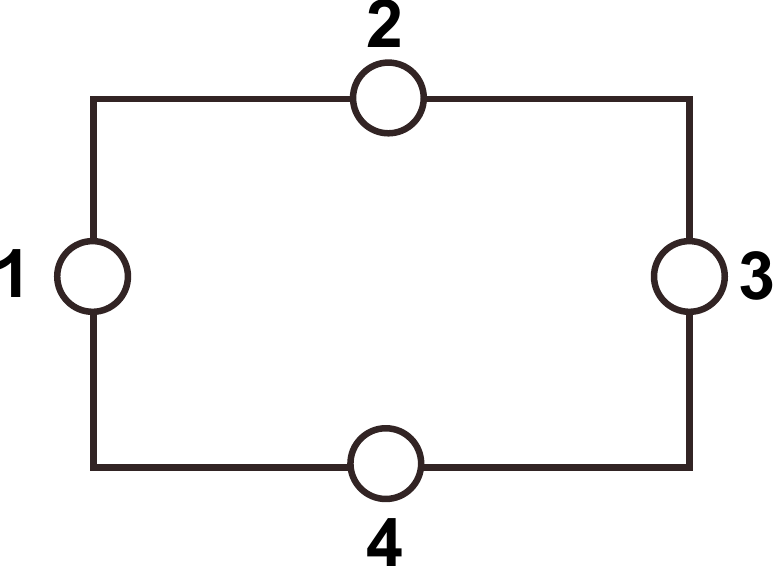}
    \caption{\textbf{Contact configuration for the Landauer-B\"uttiker analysis.} The four contacts are labelled.}
    \label{figapp:contact_4}
\end{figure}

While the FTI lacks a non-zero quantized charge Hall response, its presence can still be detected with non-local transport measurements. In this subsection, we consider the expected values of various transport experiments using the four-contact geometry in Fig.~\ref{figapp:contact_4}. Our focus is on low temperatures and the regime where transport is mediated by the gapless edge modes of the FTI at $\nu=-4/3$. We assume perfect spin $U(1)$ symmetry (such that we have a FQSH state) where the different spins cannot scatter into each other, perfect contacts, and full local equilibration between the possible multiple edge modes for a given spin projection. Within the Landauer-B\"uttiker framework, we have
\begin{equation}\label{eqapp:Gmatrix}
    I_\alpha=\sum_\beta \mathcal{G}_{\alpha\beta}V_\beta,\quad \mathcal{G}=\begin{pmatrix}
        -4/3 & 2/3 & 0 & 2/3\\
        2/3 & -4/3 & 2/3 & 0\\
        0 & 2/3 & -4/3 & 2/3\\
        2/3 & 0 & 2/3 & -4/3
    \end{pmatrix}
\end{equation}
where $I_\alpha$ and $V_\alpha$ are the current and voltage on lead $\alpha$ respectively. We use the convention that $I_\alpha$ denotes the current leaving the system via lead $\alpha$. We now explain how the conductance matrix $\mathcal{G}$ is determined. $\mathcal{G}$ has an entry $2/3$ between any pair of adjacent contacts, and $0$ for contacts that are not adjacent (every current goes through one adjacent contact to arrive to a non-adjacent one). This is because raising the voltage on lead $\alpha$ increases the chemical potential of its outgoing edge states which carry charge $2/3$. This is collected as current in the adjacent leads.
The diagonal entries are constrained by the fact that a uniform shift of all voltages should not change the current. This imposes that each row of $\mathcal{G}$ has vanishing sum. Consistent with current conservation for arbitrary voltages, each column of $\mathcal{G}$ has vanishing sum.

Eq.~\ref{eqapp:Gmatrix} can be solved with various constraints on $I_\alpha$ and $V_\alpha$ depending on the measurement configuration, as discussed below. We consider effective four-terminal and two-terminal measurements. The combination of these measurements would provide experimental evidence for the $\nu=-4/3$ FTI. We also note that since the FTI is incompressible, the below quantized responses should exhibit finite plateaus as a function of e.g.~filling factor and displacement field.

For the four-terminal measurements, we drive a current $I$ through two current contacts (i.e.~we impose $I_\alpha=-I_\beta=I$ for some $\alpha\neq\beta$), and measure the voltage drop $V$ across the voltage contacts (which are different than the current contacts). The resistance is determined as $R=V/I$. Because we impose the condition of zero current at the voltage leads, Eq.~\ref{eqapp:Gmatrix} can be solved for all the voltages up to an overall constant offset. We find, referring to Fig.~\ref{figapp:contact_4}:
\begin{itemize}
    \item \underline{Current contacts 1 \& 3; voltage contacts 2 \& 4 :} $R=0$. This corresponds to vanishing Hall conductance in an FTI.
    \item \underline{Current contacts 1 \& 4; voltage contacts 2 \& 3 :} $R=3/8$
\end{itemize}

For two-terminal measurements, we need to specify the two `terminals', where a terminal could involve shorting together several contacts, which equalizes their voltages. Other contacts not within a terminal are therefore floating, which means that they cannot source or sink any net current ($I_\alpha=0$). As an example, if we specify terminals $(1,3)$ \& 2, this means that contacts 1 and 3 are shorted together to make one terminal $(V_1=V_3)$, contact 2 forms the other terminal, and contact 4 is left floating $(I_4=0)$. By conservation of current, we have $I_1+I_3=-I_2=I$, where $I$ is the current driven through the terminals. The conductance $G=I/V$ is computed by dividing this current by the voltage drop $V$ across the terminals. $V$ can be uniquely determined by solving Eq.~\ref{eqapp:Gmatrix} with the above constraints. This analysis can be repeated for different choices of terminals. Referring again to Fig.~\ref{figapp:contact_4}, we list the conductances $G$ for $\nu=-4/3$ FTI using the four configurations considered in Ref.~\onlinecite{kang2024observation}:
\begin{itemize}
\item \underline{Terminals $(1,4)$ \& $(2,3)$ :} $G=4/3$
\item \underline{Terminals $(1)$ \& $(2,3)$ :} $G=1$
\item \underline{Terminals $(1)$ \& $(4)$ :} $G=8/9$
\item \underline{Terminals $(1)$ \& $(3)$ :} $G=2/3$
\end{itemize}

\end{appendix}


\begin{thebibliography}{102}%
\makeatletter
\providecommand \@ifxundefined [1]{%
 \@ifx{#1\undefined}
}%
\providecommand \@ifnum [1]{%
 \ifnum #1\expandafter \@firstoftwo
 \else \expandafter \@secondoftwo
 \fi
}%
\providecommand \@ifx [1]{%
 \ifx #1\expandafter \@firstoftwo
 \else \expandafter \@secondoftwo
 \fi
}%
\providecommand \natexlab [1]{#1}%
\providecommand \enquote  [1]{``#1''}%
\providecommand \bibnamefont  [1]{#1}%
\providecommand \bibfnamefont [1]{#1}%
\providecommand \citenamefont [1]{#1}%
\providecommand \href@noop [0]{\@secondoftwo}%
\providecommand \href [0]{\begingroup \@sanitize@url \@href}%
\providecommand \@href[1]{\@@startlink{#1}\@@href}%
\providecommand \@@href[1]{\endgroup#1\@@endlink}%
\providecommand \@sanitize@url [0]{\catcode `\\12\catcode `\$12\catcode `\&12\catcode `\#12\catcode `\^12\catcode `\_12\catcode `\%12\relax}%
\providecommand \@@startlink[1]{}%
\providecommand \@@endlink[0]{}%
\providecommand \url  [0]{\begingroup\@sanitize@url \@url }%
\providecommand \@url [1]{\endgroup\@href {#1}{\urlprefix }}%
\providecommand \urlprefix  [0]{URL }%
\providecommand \Eprint [0]{\href }%
\providecommand \doibase [0]{https://doi.org/}%
\providecommand \selectlanguage [0]{\@gobble}%
\providecommand \bibinfo  [0]{\@secondoftwo}%
\providecommand \bibfield  [0]{\@secondoftwo}%
\providecommand \translation [1]{[#1]}%
\providecommand \BibitemOpen [0]{}%
\providecommand \bibitemStop [0]{}%
\providecommand \bibitemNoStop [0]{.\EOS\space}%
\providecommand \EOS [0]{\spacefactor3000\relax}%
\providecommand \BibitemShut  [1]{\csname bibitem#1\endcsname}%
\let\auto@bib@innerbib\@empty
%</preamble>
\bibitem [{\citenamefont {Neupert}\ \emph {et~al.}(2011{\natexlab{a}})\citenamefont {Neupert}, \citenamefont {Santos}, \citenamefont {Chamon},\ and\ \citenamefont {Mudry}}]{neupert}%
  \BibitemOpen
  \bibfield  {author} {\bibinfo {author} {\bibfnamefont {T.}~\bibnamefont {Neupert}}, \bibinfo {author} {\bibfnamefont {L.}~\bibnamefont {Santos}}, \bibinfo {author} {\bibfnamefont {C.}~\bibnamefont {Chamon}},\ and\ \bibinfo {author} {\bibfnamefont {C.}~\bibnamefont {Mudry}},\ }\bibfield  {title} {\bibinfo {title} {Fractional quantum hall states at zero magnetic field},\ }\href {https://doi.org/10.1103/PhysRevLett.106.236804} {\bibfield  {journal} {\bibinfo  {journal} {Phys. Rev. Lett.}\ }\textbf {\bibinfo {volume} {106}},\ \bibinfo {pages} {236804} (\bibinfo {year} {2011}{\natexlab{a}})}\BibitemShut {NoStop}%
\bibitem [{\citenamefont {{Sheng}}\ \emph {et~al.}(2011)\citenamefont {{Sheng}}, \citenamefont {{Gu}}, \citenamefont {{Sun}},\ and\ \citenamefont {{Sheng}}}]{sheng}%
  \BibitemOpen
  \bibfield  {author} {\bibinfo {author} {\bibfnamefont {D.~N.}\ \bibnamefont {{Sheng}}}, \bibinfo {author} {\bibfnamefont {Z.-C.}\ \bibnamefont {{Gu}}}, \bibinfo {author} {\bibfnamefont {K.}~\bibnamefont {{Sun}}},\ and\ \bibinfo {author} {\bibfnamefont {L.}~\bibnamefont {{Sheng}}},\ }\bibfield  {title} {\bibinfo {title} {{Fractional quantum Hall effect in the absence of Landau levels}},\ }\href {https://doi.org/10.1038/ncomms1380} {\bibfield  {journal} {\bibinfo  {journal} {Nature Communications}\ }\textbf {\bibinfo {volume} {2}},\ \bibinfo {eid} {389} (\bibinfo {year} {2011})},\ \Eprint {https://arxiv.org/abs/1102.2658} {arXiv:1102.2658 [cond-mat.str-el]} \BibitemShut {NoStop}%
\bibitem [{\citenamefont {Regnault}\ and\ \citenamefont {Bernevig}(2011)}]{regnault}%
  \BibitemOpen
  \bibfield  {author} {\bibinfo {author} {\bibfnamefont {N.}~\bibnamefont {Regnault}}\ and\ \bibinfo {author} {\bibfnamefont {B.~A.}\ \bibnamefont {Bernevig}},\ }\bibfield  {title} {\bibinfo {title} {Fractional chern insulator},\ }\href {https://doi.org/10.1103/PhysRevX.1.021014} {\bibfield  {journal} {\bibinfo  {journal} {Phys. Rev. X}\ }\textbf {\bibinfo {volume} {1}},\ \bibinfo {pages} {021014} (\bibinfo {year} {2011})}\BibitemShut {NoStop}%
\bibitem [{\citenamefont {Sun}\ \emph {et~al.}(2011)\citenamefont {Sun}, \citenamefont {Gu}, \citenamefont {Katsura},\ and\ \citenamefont {Das~Sarma}}]{sun2011nearly}%
  \BibitemOpen
  \bibfield  {author} {\bibinfo {author} {\bibfnamefont {K.}~\bibnamefont {Sun}}, \bibinfo {author} {\bibfnamefont {Z.}~\bibnamefont {Gu}}, \bibinfo {author} {\bibfnamefont {H.}~\bibnamefont {Katsura}},\ and\ \bibinfo {author} {\bibfnamefont {S.}~\bibnamefont {Das~Sarma}},\ }\bibfield  {title} {\bibinfo {title} {Nearly flatbands with nontrivial topology},\ }\href {https://doi.org/10.1103/PhysRevLett.106.236803} {\bibfield  {journal} {\bibinfo  {journal} {Phys. Rev. Lett.}\ }\textbf {\bibinfo {volume} {106}},\ \bibinfo {pages} {236803} (\bibinfo {year} {2011})}\BibitemShut {NoStop}%
\bibitem [{\citenamefont {Yang}\ \emph {et~al.}(2012)\citenamefont {Yang}, \citenamefont {Gu}, \citenamefont {Sun},\ and\ \citenamefont {Das~Sarma}}]{yang2012arbitrary}%
  \BibitemOpen
  \bibfield  {author} {\bibinfo {author} {\bibfnamefont {S.}~\bibnamefont {Yang}}, \bibinfo {author} {\bibfnamefont {Z.-C.}\ \bibnamefont {Gu}}, \bibinfo {author} {\bibfnamefont {K.}~\bibnamefont {Sun}},\ and\ \bibinfo {author} {\bibfnamefont {S.}~\bibnamefont {Das~Sarma}},\ }\bibfield  {title} {\bibinfo {title} {Topological flat band models with arbitrary chern numbers},\ }\href {https://doi.org/10.1103/PhysRevB.86.241112} {\bibfield  {journal} {\bibinfo  {journal} {Phys. Rev. B}\ }\textbf {\bibinfo {volume} {86}},\ \bibinfo {pages} {241112} (\bibinfo {year} {2012})}\BibitemShut {NoStop}%
\bibitem [{\citenamefont {Tang}\ \emph {et~al.}(2011)\citenamefont {Tang}, \citenamefont {Mei},\ and\ \citenamefont {Wen}}]{tang2011high}%
  \BibitemOpen
  \bibfield  {author} {\bibinfo {author} {\bibfnamefont {E.}~\bibnamefont {Tang}}, \bibinfo {author} {\bibfnamefont {J.-W.}\ \bibnamefont {Mei}},\ and\ \bibinfo {author} {\bibfnamefont {X.-G.}\ \bibnamefont {Wen}},\ }\bibfield  {title} {\bibinfo {title} {High-temperature fractional quantum hall states},\ }\href {https://doi.org/10.1103/PhysRevLett.106.236802} {\bibfield  {journal} {\bibinfo  {journal} {Phys. Rev. Lett.}\ }\textbf {\bibinfo {volume} {106}},\ \bibinfo {pages} {236802} (\bibinfo {year} {2011})}\BibitemShut {NoStop}%
\bibitem [{\citenamefont {Cai}\ \emph {et~al.}(2023)\citenamefont {Cai}, \citenamefont {Anderson}, \citenamefont {Wang}, \citenamefont {Zhang}, \citenamefont {Liu}, \citenamefont {Holtzmann}, \citenamefont {Zhang}, \citenamefont {Fan}, \citenamefont {Taniguchi}, \citenamefont {Watanabe}, \citenamefont {Ran}, \citenamefont {Cao}, \citenamefont {Fu}, \citenamefont {Xiao}, \citenamefont {Yao},\ and\ \citenamefont {Xu}}]{cai2023signatures}%
  \BibitemOpen
  \bibfield  {author} {\bibinfo {author} {\bibfnamefont {J.}~\bibnamefont {Cai}}, \bibinfo {author} {\bibfnamefont {E.}~\bibnamefont {Anderson}}, \bibinfo {author} {\bibfnamefont {C.}~\bibnamefont {Wang}}, \bibinfo {author} {\bibfnamefont {X.}~\bibnamefont {Zhang}}, \bibinfo {author} {\bibfnamefont {X.}~\bibnamefont {Liu}}, \bibinfo {author} {\bibfnamefont {W.}~\bibnamefont {Holtzmann}}, \bibinfo {author} {\bibfnamefont {Y.}~\bibnamefont {Zhang}}, \bibinfo {author} {\bibfnamefont {F.}~\bibnamefont {Fan}}, \bibinfo {author} {\bibfnamefont {T.}~\bibnamefont {Taniguchi}}, \bibinfo {author} {\bibfnamefont {K.}~\bibnamefont {Watanabe}}, \bibinfo {author} {\bibfnamefont {Y.}~\bibnamefont {Ran}}, \bibinfo {author} {\bibfnamefont {T.}~\bibnamefont {Cao}}, \bibinfo {author} {\bibfnamefont {L.}~\bibnamefont {Fu}}, \bibinfo {author} {\bibfnamefont {D.}~\bibnamefont {Xiao}}, \bibinfo {author} {\bibfnamefont {W.}~\bibnamefont {Yao}},\ and\ \bibinfo {author} {\bibfnamefont {X.}~\bibnamefont {Xu}},\ }\bibfield  {title}
  {\bibinfo {title} {Signatures of fractional quantum anomalous hall states in twisted mote2},\ }\bibfield  {journal} {\bibinfo  {journal} {Nature}\ }\href {https://doi.org/10.1038/s41586-023-06289-w} {10.1038/s41586-023-06289-w} (\bibinfo {year} {2023})\BibitemShut {NoStop}%
\bibitem [{\citenamefont {Zeng}\ \emph {et~al.}(2023)\citenamefont {Zeng}, \citenamefont {Xia}, \citenamefont {Kang}, \citenamefont {Zhu}, \citenamefont {Kn{\"u}ppel}, \citenamefont {Vaswani}, \citenamefont {Watanabe}, \citenamefont {Taniguchi}, \citenamefont {Mak},\ and\ \citenamefont {Shan}}]{zeng2023integer}%
  \BibitemOpen
  \bibfield  {author} {\bibinfo {author} {\bibfnamefont {Y.}~\bibnamefont {Zeng}}, \bibinfo {author} {\bibfnamefont {Z.}~\bibnamefont {Xia}}, \bibinfo {author} {\bibfnamefont {K.}~\bibnamefont {Kang}}, \bibinfo {author} {\bibfnamefont {J.}~\bibnamefont {Zhu}}, \bibinfo {author} {\bibfnamefont {P.}~\bibnamefont {Kn{\"u}ppel}}, \bibinfo {author} {\bibfnamefont {C.}~\bibnamefont {Vaswani}}, \bibinfo {author} {\bibfnamefont {K.}~\bibnamefont {Watanabe}}, \bibinfo {author} {\bibfnamefont {T.}~\bibnamefont {Taniguchi}}, \bibinfo {author} {\bibfnamefont {K.~F.}\ \bibnamefont {Mak}},\ and\ \bibinfo {author} {\bibfnamefont {J.}~\bibnamefont {Shan}},\ }\bibfield  {title} {\bibinfo {title} {Thermodynamic evidence of fractional chern insulator in moir{\'e} mote2},\ }\bibfield  {journal} {\bibinfo  {journal} {Nature}\ }\href {https://doi.org/10.1038/s41586-023-06452-3} {10.1038/s41586-023-06452-3} (\bibinfo {year} {2023})\BibitemShut {NoStop}%
\bibitem [{\citenamefont {Park}\ \emph {et~al.}(2023)\citenamefont {Park}, \citenamefont {Cai}, \citenamefont {Anderson}, \citenamefont {Zhang}, \citenamefont {Zhu}, \citenamefont {Liu}, \citenamefont {Wang}, \citenamefont {Holtzmann}, \citenamefont {Hu}, \citenamefont {Liu}, \citenamefont {Taniguchi}, \citenamefont {Watanabe}, \citenamefont {Chu}, \citenamefont {Cao}, \citenamefont {Fu}, \citenamefont {Yao}, \citenamefont {Chang}, \citenamefont {Cobden}, \citenamefont {Xiao},\ and\ \citenamefont {Xu}}]{Park2023}%
  \BibitemOpen
  \bibfield  {author} {\bibinfo {author} {\bibfnamefont {H.}~\bibnamefont {Park}}, \bibinfo {author} {\bibfnamefont {J.}~\bibnamefont {Cai}}, \bibinfo {author} {\bibfnamefont {E.}~\bibnamefont {Anderson}}, \bibinfo {author} {\bibfnamefont {Y.}~\bibnamefont {Zhang}}, \bibinfo {author} {\bibfnamefont {J.}~\bibnamefont {Zhu}}, \bibinfo {author} {\bibfnamefont {X.}~\bibnamefont {Liu}}, \bibinfo {author} {\bibfnamefont {C.}~\bibnamefont {Wang}}, \bibinfo {author} {\bibfnamefont {W.}~\bibnamefont {Holtzmann}}, \bibinfo {author} {\bibfnamefont {C.}~\bibnamefont {Hu}}, \bibinfo {author} {\bibfnamefont {Z.}~\bibnamefont {Liu}}, \bibinfo {author} {\bibfnamefont {T.}~\bibnamefont {Taniguchi}}, \bibinfo {author} {\bibfnamefont {K.}~\bibnamefont {Watanabe}}, \bibinfo {author} {\bibfnamefont {J.-H.}\ \bibnamefont {Chu}}, \bibinfo {author} {\bibfnamefont {T.}~\bibnamefont {Cao}}, \bibinfo {author} {\bibfnamefont {L.}~\bibnamefont {Fu}}, \bibinfo {author} {\bibfnamefont {W.}~\bibnamefont {Yao}}, \bibinfo {author}
  {\bibfnamefont {C.-Z.}\ \bibnamefont {Chang}}, \bibinfo {author} {\bibfnamefont {D.}~\bibnamefont {Cobden}}, \bibinfo {author} {\bibfnamefont {D.}~\bibnamefont {Xiao}},\ and\ \bibinfo {author} {\bibfnamefont {X.}~\bibnamefont {Xu}},\ }\bibfield  {title} {\bibinfo {title} {Observation of fractionally quantized anomalous hall effect},\ }\href {https://doi.org/10.1038/s41586-023-06536-0} {\bibfield  {journal} {\bibinfo  {journal} {Nature}\ }\textbf {\bibinfo {volume} {622}},\ \bibinfo {pages} {74} (\bibinfo {year} {2023})}\BibitemShut {NoStop}%
\bibitem [{\citenamefont {Xu}\ \emph {et~al.}(2023)\citenamefont {Xu}, \citenamefont {Sun}, \citenamefont {Jia}, \citenamefont {Liu}, \citenamefont {Xu}, \citenamefont {Li}, \citenamefont {Gu}, \citenamefont {Watanabe}, \citenamefont {Taniguchi}, \citenamefont {Tong}, \citenamefont {Jia}, \citenamefont {Shi}, \citenamefont {Jiang}, \citenamefont {Zhang}, \citenamefont {Liu},\ and\ \citenamefont {Li}}]{Xu2023}%
  \BibitemOpen
  \bibfield  {author} {\bibinfo {author} {\bibfnamefont {F.}~\bibnamefont {Xu}}, \bibinfo {author} {\bibfnamefont {Z.}~\bibnamefont {Sun}}, \bibinfo {author} {\bibfnamefont {T.}~\bibnamefont {Jia}}, \bibinfo {author} {\bibfnamefont {C.}~\bibnamefont {Liu}}, \bibinfo {author} {\bibfnamefont {C.}~\bibnamefont {Xu}}, \bibinfo {author} {\bibfnamefont {C.}~\bibnamefont {Li}}, \bibinfo {author} {\bibfnamefont {Y.}~\bibnamefont {Gu}}, \bibinfo {author} {\bibfnamefont {K.}~\bibnamefont {Watanabe}}, \bibinfo {author} {\bibfnamefont {T.}~\bibnamefont {Taniguchi}}, \bibinfo {author} {\bibfnamefont {B.}~\bibnamefont {Tong}}, \bibinfo {author} {\bibfnamefont {J.}~\bibnamefont {Jia}}, \bibinfo {author} {\bibfnamefont {Z.}~\bibnamefont {Shi}}, \bibinfo {author} {\bibfnamefont {S.}~\bibnamefont {Jiang}}, \bibinfo {author} {\bibfnamefont {Y.}~\bibnamefont {Zhang}}, \bibinfo {author} {\bibfnamefont {X.}~\bibnamefont {Liu}},\ and\ \bibinfo {author} {\bibfnamefont {T.}~\bibnamefont {Li}},\ }\bibfield  {title} {\bibinfo {title}
  {Observation of integer and fractional quantum anomalous hall effects in twisted bilayer ${\mathrm{mote}}_{2}$},\ }\href {https://doi.org/10.1103/PhysRevX.13.031037} {\bibfield  {journal} {\bibinfo  {journal} {Phys. Rev. X}\ }\textbf {\bibinfo {volume} {13}},\ \bibinfo {pages} {031037} (\bibinfo {year} {2023})}\BibitemShut {NoStop}%
\bibitem [{\citenamefont {Park}\ \emph {et~al.}(2024)\citenamefont {Park}, \citenamefont {Cai}, \citenamefont {Anderson}, \citenamefont {Zhang}, \citenamefont {Liu}, \citenamefont {Holtzmann}, \citenamefont {Li}, \citenamefont {Wang}, \citenamefont {Hu}, \citenamefont {Zhao}, \citenamefont {Taniguchi}, \citenamefont {Watanabe}, \citenamefont {Yang}, \citenamefont {Cobden}, \citenamefont {Chu}, \citenamefont {Regnault}, \citenamefont {Bernevig}, \citenamefont {Fu}, \citenamefont {Cao}, \citenamefont {Xiao},\ and\ \citenamefont {Xu}}]{park2024ferromagnetism}%
  \BibitemOpen
  \bibfield  {author} {\bibinfo {author} {\bibfnamefont {H.}~\bibnamefont {Park}}, \bibinfo {author} {\bibfnamefont {J.}~\bibnamefont {Cai}}, \bibinfo {author} {\bibfnamefont {E.}~\bibnamefont {Anderson}}, \bibinfo {author} {\bibfnamefont {X.-W.}\ \bibnamefont {Zhang}}, \bibinfo {author} {\bibfnamefont {X.}~\bibnamefont {Liu}}, \bibinfo {author} {\bibfnamefont {W.}~\bibnamefont {Holtzmann}}, \bibinfo {author} {\bibfnamefont {W.}~\bibnamefont {Li}}, \bibinfo {author} {\bibfnamefont {C.}~\bibnamefont {Wang}}, \bibinfo {author} {\bibfnamefont {C.}~\bibnamefont {Hu}}, \bibinfo {author} {\bibfnamefont {Y.}~\bibnamefont {Zhao}}, \bibinfo {author} {\bibfnamefont {T.}~\bibnamefont {Taniguchi}}, \bibinfo {author} {\bibfnamefont {K.}~\bibnamefont {Watanabe}}, \bibinfo {author} {\bibfnamefont {J.}~\bibnamefont {Yang}}, \bibinfo {author} {\bibfnamefont {D.}~\bibnamefont {Cobden}}, \bibinfo {author} {\bibfnamefont {J.-H.}\ \bibnamefont {Chu}}, \bibinfo {author} {\bibfnamefont {N.}~\bibnamefont {Regnault}}, \bibinfo
  {author} {\bibfnamefont {B.~A.}\ \bibnamefont {Bernevig}}, \bibinfo {author} {\bibfnamefont {L.}~\bibnamefont {Fu}}, \bibinfo {author} {\bibfnamefont {T.}~\bibnamefont {Cao}}, \bibinfo {author} {\bibfnamefont {D.}~\bibnamefont {Xiao}},\ and\ \bibinfo {author} {\bibfnamefont {X.}~\bibnamefont {Xu}},\ }\href {https://arxiv.org/abs/2406.09591} {\bibinfo {title} {Ferromagnetism and topology of the higher flat band in a fractional chern insulator}} (\bibinfo {year} {2024}),\ \Eprint {https://arxiv.org/abs/2406.09591} {arXiv:2406.09591 [cond-mat.mes-hall]} \BibitemShut {NoStop}%
\bibitem [{\citenamefont {Xu}\ \emph {et~al.}(2024{\natexlab{a}})\citenamefont {Xu}, \citenamefont {Chang}, \citenamefont {Xiao}, \citenamefont {Zhang}, \citenamefont {Liu}, \citenamefont {Sun}, \citenamefont {Mao}, \citenamefont {Peshcherenko}, \citenamefont {Li}, \citenamefont {Watanabe}, \citenamefont {Taniguchi}, \citenamefont {Tong}, \citenamefont {Lu}, \citenamefont {Jia}, \citenamefont {Qian}, \citenamefont {Shi}, \citenamefont {Zhang}, \citenamefont {Liu}, \citenamefont {Jiang},\ and\ \citenamefont {Li}}]{xu2024interplay}%
  \BibitemOpen
  \bibfield  {author} {\bibinfo {author} {\bibfnamefont {F.}~\bibnamefont {Xu}}, \bibinfo {author} {\bibfnamefont {X.}~\bibnamefont {Chang}}, \bibinfo {author} {\bibfnamefont {J.}~\bibnamefont {Xiao}}, \bibinfo {author} {\bibfnamefont {Y.}~\bibnamefont {Zhang}}, \bibinfo {author} {\bibfnamefont {F.}~\bibnamefont {Liu}}, \bibinfo {author} {\bibfnamefont {Z.}~\bibnamefont {Sun}}, \bibinfo {author} {\bibfnamefont {N.}~\bibnamefont {Mao}}, \bibinfo {author} {\bibfnamefont {N.}~\bibnamefont {Peshcherenko}}, \bibinfo {author} {\bibfnamefont {J.}~\bibnamefont {Li}}, \bibinfo {author} {\bibfnamefont {K.}~\bibnamefont {Watanabe}}, \bibinfo {author} {\bibfnamefont {T.}~\bibnamefont {Taniguchi}}, \bibinfo {author} {\bibfnamefont {B.}~\bibnamefont {Tong}}, \bibinfo {author} {\bibfnamefont {L.}~\bibnamefont {Lu}}, \bibinfo {author} {\bibfnamefont {J.}~\bibnamefont {Jia}}, \bibinfo {author} {\bibfnamefont {D.}~\bibnamefont {Qian}}, \bibinfo {author} {\bibfnamefont {Z.}~\bibnamefont {Shi}}, \bibinfo {author} {\bibfnamefont
  {Y.}~\bibnamefont {Zhang}}, \bibinfo {author} {\bibfnamefont {X.}~\bibnamefont {Liu}}, \bibinfo {author} {\bibfnamefont {S.}~\bibnamefont {Jiang}},\ and\ \bibinfo {author} {\bibfnamefont {T.}~\bibnamefont {Li}},\ }\href {https://arxiv.org/abs/2406.09687} {\bibinfo {title} {Interplay between topology and correlations in the second moir\'e band of twisted bilayer mote2}} (\bibinfo {year} {2024}{\natexlab{a}}),\ \Eprint {https://arxiv.org/abs/2406.09687} {arXiv:2406.09687 [cond-mat.mes-hall]} \BibitemShut {NoStop}%
\bibitem [{\citenamefont {Li}\ \emph {et~al.}(2021)\citenamefont {Li}, \citenamefont {Kumar}, \citenamefont {Sun},\ and\ \citenamefont {Lin}}]{Li2023}%
  \BibitemOpen
  \bibfield  {author} {\bibinfo {author} {\bibfnamefont {H.}~\bibnamefont {Li}}, \bibinfo {author} {\bibfnamefont {U.}~\bibnamefont {Kumar}}, \bibinfo {author} {\bibfnamefont {K.}~\bibnamefont {Sun}},\ and\ \bibinfo {author} {\bibfnamefont {S.-Z.}\ \bibnamefont {Lin}},\ }\bibfield  {title} {\bibinfo {title} {Spontaneous fractional chern insulators in transition metal dichalcogenide moir\'e superlattices},\ }\href {https://doi.org/10.1103/PhysRevResearch.3.L032070} {\bibfield  {journal} {\bibinfo  {journal} {Phys. Rev. Res.}\ }\textbf {\bibinfo {volume} {3}},\ \bibinfo {pages} {L032070} (\bibinfo {year} {2021})}\BibitemShut {NoStop}%
\bibitem [{\citenamefont {Cr\'epel}\ and\ \citenamefont {Fu}(2023)}]{Crepel2023}%
  \BibitemOpen
  \bibfield  {author} {\bibinfo {author} {\bibfnamefont {V.}~\bibnamefont {Cr\'epel}}\ and\ \bibinfo {author} {\bibfnamefont {L.}~\bibnamefont {Fu}},\ }\bibfield  {title} {\bibinfo {title} {Anomalous hall metal and fractional chern insulator in twisted transition metal dichalcogenides},\ }\href {https://doi.org/10.1103/PhysRevB.107.L201109} {\bibfield  {journal} {\bibinfo  {journal} {Phys. Rev. B}\ }\textbf {\bibinfo {volume} {107}},\ \bibinfo {pages} {L201109} (\bibinfo {year} {2023})}\BibitemShut {NoStop}%
\bibitem [{\citenamefont {Morales-Durán}\ \emph {et~al.}(2023)\citenamefont {Morales-Durán}, \citenamefont {Wang}, \citenamefont {Schleder}, \citenamefont {Angeli}, \citenamefont {Zhu}, \citenamefont {Kaxiras}, \citenamefont {Repellin},\ and\ \citenamefont {Cano}}]{moralesdurán2023pressureenhanced}%
  \BibitemOpen
  \bibfield  {author} {\bibinfo {author} {\bibfnamefont {N.}~\bibnamefont {Morales-Durán}}, \bibinfo {author} {\bibfnamefont {J.}~\bibnamefont {Wang}}, \bibinfo {author} {\bibfnamefont {G.~R.}\ \bibnamefont {Schleder}}, \bibinfo {author} {\bibfnamefont {M.}~\bibnamefont {Angeli}}, \bibinfo {author} {\bibfnamefont {Z.}~\bibnamefont {Zhu}}, \bibinfo {author} {\bibfnamefont {E.}~\bibnamefont {Kaxiras}}, \bibinfo {author} {\bibfnamefont {C.}~\bibnamefont {Repellin}},\ and\ \bibinfo {author} {\bibfnamefont {J.}~\bibnamefont {Cano}},\ }\href@noop {} {\bibinfo {title} {Pressure--enhanced fractional chern insulators in moir\'e transition metal dichalcogenides along a magic line}} (\bibinfo {year} {2023}),\ \Eprint {https://arxiv.org/abs/2304.06669} {arXiv:2304.06669 [cond-mat.str-el]} \BibitemShut {NoStop}%
\bibitem [{\citenamefont {Wang}\ \emph {et~al.}(2023{\natexlab{a}})\citenamefont {Wang}, \citenamefont {Zhang}, \citenamefont {Liu}, \citenamefont {He}, \citenamefont {Xu}, \citenamefont {Ran}, \citenamefont {Cao},\ and\ \citenamefont {Xiao}}]{wang2023fractional}%
  \BibitemOpen
  \bibfield  {author} {\bibinfo {author} {\bibfnamefont {C.}~\bibnamefont {Wang}}, \bibinfo {author} {\bibfnamefont {X.-W.}\ \bibnamefont {Zhang}}, \bibinfo {author} {\bibfnamefont {X.}~\bibnamefont {Liu}}, \bibinfo {author} {\bibfnamefont {Y.}~\bibnamefont {He}}, \bibinfo {author} {\bibfnamefont {X.}~\bibnamefont {Xu}}, \bibinfo {author} {\bibfnamefont {Y.}~\bibnamefont {Ran}}, \bibinfo {author} {\bibfnamefont {T.}~\bibnamefont {Cao}},\ and\ \bibinfo {author} {\bibfnamefont {D.}~\bibnamefont {Xiao}},\ }\href@noop {} {\bibinfo {title} {Fractional chern insulator in twisted bilayer mote$_2$}} (\bibinfo {year} {2023}{\natexlab{a}}),\ \Eprint {https://arxiv.org/abs/2304.11864} {arXiv:2304.11864 [cond-mat.str-el]} \BibitemShut {NoStop}%
\bibitem [{\citenamefont {Reddy}\ and\ \citenamefont {Fu}(2023)}]{reddy2023global}%
  \BibitemOpen
  \bibfield  {author} {\bibinfo {author} {\bibfnamefont {A.~P.}\ \bibnamefont {Reddy}}\ and\ \bibinfo {author} {\bibfnamefont {L.}~\bibnamefont {Fu}},\ }\href@noop {} {\bibinfo {title} {Toward a global phase diagram of the fractional quantum anomalous hall effect}} (\bibinfo {year} {2023}),\ \Eprint {https://arxiv.org/abs/2308.10406} {arXiv:2308.10406 [cond-mat.mes-hall]} \BibitemShut {NoStop}%
\bibitem [{\citenamefont {Reddy}\ \emph {et~al.}(2023)\citenamefont {Reddy}, \citenamefont {Alsallom}, \citenamefont {Zhang}, \citenamefont {Devakul},\ and\ \citenamefont {Fu}}]{Reddy2023}%
  \BibitemOpen
  \bibfield  {author} {\bibinfo {author} {\bibfnamefont {A.~P.}\ \bibnamefont {Reddy}}, \bibinfo {author} {\bibfnamefont {F.}~\bibnamefont {Alsallom}}, \bibinfo {author} {\bibfnamefont {Y.}~\bibnamefont {Zhang}}, \bibinfo {author} {\bibfnamefont {T.}~\bibnamefont {Devakul}},\ and\ \bibinfo {author} {\bibfnamefont {L.}~\bibnamefont {Fu}},\ }\bibfield  {title} {\bibinfo {title} {Fractional quantum anomalous hall states in twisted bilayer ${\mathrm{mote}}_{2}$ and ${\mathrm{wse}}_{2}$},\ }\href {https://doi.org/10.1103/PhysRevB.108.085117} {\bibfield  {journal} {\bibinfo  {journal} {Phys. Rev. B}\ }\textbf {\bibinfo {volume} {108}},\ \bibinfo {pages} {085117} (\bibinfo {year} {2023})}\BibitemShut {NoStop}%
\bibitem [{\citenamefont {Yu}\ \emph {et~al.}(2024)\citenamefont {Yu}, \citenamefont {Herzog-Arbeitman}, \citenamefont {Wang}, \citenamefont {Vafek}, \citenamefont {Bernevig},\ and\ \citenamefont {Regnault}}]{Yu2024MFCI0}%
  \BibitemOpen
  \bibfield  {author} {\bibinfo {author} {\bibfnamefont {J.}~\bibnamefont {Yu}}, \bibinfo {author} {\bibfnamefont {J.}~\bibnamefont {Herzog-Arbeitman}}, \bibinfo {author} {\bibfnamefont {M.}~\bibnamefont {Wang}}, \bibinfo {author} {\bibfnamefont {O.}~\bibnamefont {Vafek}}, \bibinfo {author} {\bibfnamefont {B.~A.}\ \bibnamefont {Bernevig}},\ and\ \bibinfo {author} {\bibfnamefont {N.}~\bibnamefont {Regnault}},\ }\bibfield  {title} {\bibinfo {title} {Fractional chern insulators versus nonmagnetic states in twisted bilayer ${\mathrm{mote}}_{2}$},\ }\href {https://doi.org/10.1103/PhysRevB.109.045147} {\bibfield  {journal} {\bibinfo  {journal} {Phys. Rev. B}\ }\textbf {\bibinfo {volume} {109}},\ \bibinfo {pages} {045147} (\bibinfo {year} {2024})}\BibitemShut {NoStop}%
\bibitem [{\citenamefont {Xu}\ \emph {et~al.}(2024{\natexlab{b}})\citenamefont {Xu}, \citenamefont {Li}, \citenamefont {Xu}, \citenamefont {Bi},\ and\ \citenamefont {Zhang}}]{xu2024maximally}%
  \BibitemOpen
  \bibfield  {author} {\bibinfo {author} {\bibfnamefont {C.}~\bibnamefont {Xu}}, \bibinfo {author} {\bibfnamefont {J.}~\bibnamefont {Li}}, \bibinfo {author} {\bibfnamefont {Y.}~\bibnamefont {Xu}}, \bibinfo {author} {\bibfnamefont {Z.}~\bibnamefont {Bi}},\ and\ \bibinfo {author} {\bibfnamefont {Y.}~\bibnamefont {Zhang}},\ }\bibfield  {title} {\bibinfo {title} {Maximally localized wannier functions, interaction models, and fractional quantum anomalous hall effect in twisted bilayer mote<sub>2</sub>},\ }\href {https://doi.org/10.1073/pnas.2316749121} {\bibfield  {journal} {\bibinfo  {journal} {Proceedings of the National Academy of Sciences}\ }\textbf {\bibinfo {volume} {121}},\ \bibinfo {pages} {e2316749121} (\bibinfo {year} {2024}{\natexlab{b}})},\ \Eprint {https://arxiv.org/abs/https://www.pnas.org/doi/pdf/10.1073/pnas.2316749121} {https://www.pnas.org/doi/pdf/10.1073/pnas.2316749121} \BibitemShut {NoStop}%
\bibitem [{\citenamefont {Abouelkomsan}\ \emph {et~al.}(2023)\citenamefont {Abouelkomsan}, \citenamefont {Reddy}, \citenamefont {Fu},\ and\ \citenamefont {Bergholtz}}]{abouelkomsan2023band}%
  \BibitemOpen
  \bibfield  {author} {\bibinfo {author} {\bibfnamefont {A.}~\bibnamefont {Abouelkomsan}}, \bibinfo {author} {\bibfnamefont {A.~P.}\ \bibnamefont {Reddy}}, \bibinfo {author} {\bibfnamefont {L.}~\bibnamefont {Fu}},\ and\ \bibinfo {author} {\bibfnamefont {E.~J.}\ \bibnamefont {Bergholtz}},\ }\href@noop {} {\bibinfo {title} {Band mixing in the quantum anomalous hall regime of twisted semiconductor bilayers}} (\bibinfo {year} {2023}),\ \Eprint {https://arxiv.org/abs/2309.16548} {arXiv:2309.16548 [cond-mat.mes-hall]} \BibitemShut {NoStop}%
\bibitem [{\citenamefont {Jia}\ \emph {et~al.}(2023)\citenamefont {Jia}, \citenamefont {Yu}, \citenamefont {Liu}, \citenamefont {Herzog-Arbeitman}, \citenamefont {Qi}, \citenamefont {Regnault}, \citenamefont {Weng}, \citenamefont {Bernevig},\ and\ \citenamefont {Wu}}]{jia2023moire}%
  \BibitemOpen
  \bibfield  {author} {\bibinfo {author} {\bibfnamefont {Y.}~\bibnamefont {Jia}}, \bibinfo {author} {\bibfnamefont {J.}~\bibnamefont {Yu}}, \bibinfo {author} {\bibfnamefont {J.}~\bibnamefont {Liu}}, \bibinfo {author} {\bibfnamefont {J.}~\bibnamefont {Herzog-Arbeitman}}, \bibinfo {author} {\bibfnamefont {Z.}~\bibnamefont {Qi}}, \bibinfo {author} {\bibfnamefont {N.}~\bibnamefont {Regnault}}, \bibinfo {author} {\bibfnamefont {H.}~\bibnamefont {Weng}}, \bibinfo {author} {\bibfnamefont {B.~A.}\ \bibnamefont {Bernevig}},\ and\ \bibinfo {author} {\bibfnamefont {Q.}~\bibnamefont {Wu}},\ }\href@noop {} {\bibinfo {title} {Moir\'e fractional chern insulators i: First-principles calculations and continuum models of twisted bilayer mote$_2$}} (\bibinfo {year} {2023}),\ \Eprint {https://arxiv.org/abs/2311.04958} {arXiv:2311.04958 [cond-mat.mes-hall]} \BibitemShut {NoStop}%
\bibitem [{\citenamefont {Mao}\ \emph {et~al.}(2023)\citenamefont {Mao}, \citenamefont {Xu}, \citenamefont {Li}, \citenamefont {Bao}, \citenamefont {Liu}, \citenamefont {Xu}, \citenamefont {Felser}, \citenamefont {Fu},\ and\ \citenamefont {Zhang}}]{mao2023lattice}%
  \BibitemOpen
  \bibfield  {author} {\bibinfo {author} {\bibfnamefont {N.}~\bibnamefont {Mao}}, \bibinfo {author} {\bibfnamefont {C.}~\bibnamefont {Xu}}, \bibinfo {author} {\bibfnamefont {J.}~\bibnamefont {Li}}, \bibinfo {author} {\bibfnamefont {T.}~\bibnamefont {Bao}}, \bibinfo {author} {\bibfnamefont {P.}~\bibnamefont {Liu}}, \bibinfo {author} {\bibfnamefont {Y.}~\bibnamefont {Xu}}, \bibinfo {author} {\bibfnamefont {C.}~\bibnamefont {Felser}}, \bibinfo {author} {\bibfnamefont {L.}~\bibnamefont {Fu}},\ and\ \bibinfo {author} {\bibfnamefont {Y.}~\bibnamefont {Zhang}},\ }\href@noop {} {\bibinfo {title} {Lattice relaxation, electronic structure and continuum model for twisted bilayer mote$_2$}} (\bibinfo {year} {2023}),\ \Eprint {https://arxiv.org/abs/2311.07533} {arXiv:2311.07533 [cond-mat.str-el]} \BibitemShut {NoStop}%
\bibitem [{\citenamefont {Zhang}\ \emph {et~al.}(2024)\citenamefont {Zhang}, \citenamefont {Wang}, \citenamefont {Liu}, \citenamefont {Fan}, \citenamefont {Cao},\ and\ \citenamefont {Xiao}}]{zhang2024polarizationdriven}%
  \BibitemOpen
  \bibfield  {author} {\bibinfo {author} {\bibfnamefont {X.-W.}\ \bibnamefont {Zhang}}, \bibinfo {author} {\bibfnamefont {C.}~\bibnamefont {Wang}}, \bibinfo {author} {\bibfnamefont {X.}~\bibnamefont {Liu}}, \bibinfo {author} {\bibfnamefont {Y.}~\bibnamefont {Fan}}, \bibinfo {author} {\bibfnamefont {T.}~\bibnamefont {Cao}},\ and\ \bibinfo {author} {\bibfnamefont {D.}~\bibnamefont {Xiao}},\ }\href@noop {} {\bibinfo {title} {Polarization-driven band topology evolution in twisted mote$_2$ and wse$_2$}} (\bibinfo {year} {2024}),\ \Eprint {https://arxiv.org/abs/2311.12776} {arXiv:2311.12776 [cond-mat.mtrl-sci]} \BibitemShut {NoStop}%
\bibitem [{\citenamefont {Wang}\ \emph {et~al.}(2023{\natexlab{b}})\citenamefont {Wang}, \citenamefont {Wang}, \citenamefont {Kim}, \citenamefont {Louie}, \citenamefont {Fu},\ and\ \citenamefont {Zaletel}}]{wang2023topologydft}%
  \BibitemOpen
  \bibfield  {author} {\bibinfo {author} {\bibfnamefont {T.}~\bibnamefont {Wang}}, \bibinfo {author} {\bibfnamefont {M.}~\bibnamefont {Wang}}, \bibinfo {author} {\bibfnamefont {W.}~\bibnamefont {Kim}}, \bibinfo {author} {\bibfnamefont {S.~G.}\ \bibnamefont {Louie}}, \bibinfo {author} {\bibfnamefont {L.}~\bibnamefont {Fu}},\ and\ \bibinfo {author} {\bibfnamefont {M.~P.}\ \bibnamefont {Zaletel}},\ }\href@noop {} {\bibinfo {title} {Topology, magnetism and charge order in twisted mote2 at higher integer hole fillings}} (\bibinfo {year} {2023}{\natexlab{b}}),\ \Eprint {https://arxiv.org/abs/2312.12531} {arXiv:2312.12531 [cond-mat.str-el]} \BibitemShut {NoStop}%
\bibitem [{\citenamefont {Han}\ \emph {et~al.}(2023)\citenamefont {Han}, \citenamefont {Lu}, \citenamefont {Scuri}, \citenamefont {Sung}, \citenamefont {Wang}, \citenamefont {Han}, \citenamefont {Watanabe}, \citenamefont {Taniguchi}, \citenamefont {Park},\ and\ \citenamefont {Ju}}]{Han2023}%
  \BibitemOpen
  \bibfield  {author} {\bibinfo {author} {\bibfnamefont {T.}~\bibnamefont {Han}}, \bibinfo {author} {\bibfnamefont {Z.}~\bibnamefont {Lu}}, \bibinfo {author} {\bibfnamefont {G.}~\bibnamefont {Scuri}}, \bibinfo {author} {\bibfnamefont {J.}~\bibnamefont {Sung}}, \bibinfo {author} {\bibfnamefont {J.}~\bibnamefont {Wang}}, \bibinfo {author} {\bibfnamefont {T.}~\bibnamefont {Han}}, \bibinfo {author} {\bibfnamefont {K.}~\bibnamefont {Watanabe}}, \bibinfo {author} {\bibfnamefont {T.}~\bibnamefont {Taniguchi}}, \bibinfo {author} {\bibfnamefont {H.}~\bibnamefont {Park}},\ and\ \bibinfo {author} {\bibfnamefont {L.}~\bibnamefont {Ju}},\ }\bibfield  {title} {\bibinfo {title} {Correlated insulator and chern insulators in pentalayer rhombohedral-stacked graphene},\ }\bibfield  {journal} {\bibinfo  {journal} {Nature Nanotechnology}\ }\href {https://doi.org/10.1038/s41565-023-01520-1} {10.1038/s41565-023-01520-1} (\bibinfo {year} {2023})\BibitemShut {NoStop}%
\bibitem [{\citenamefont {Lu}\ \emph {et~al.}(2024)\citenamefont {Lu}, \citenamefont {Han}, \citenamefont {Yao}, \citenamefont {Reddy}, \citenamefont {Yang}, \citenamefont {Seo}, \citenamefont {Watanabe}, \citenamefont {Taniguchi}, \citenamefont {Fu},\ and\ \citenamefont {Ju}}]{lu2023fractional}%
  \BibitemOpen
  \bibfield  {author} {\bibinfo {author} {\bibfnamefont {Z.}~\bibnamefont {Lu}}, \bibinfo {author} {\bibfnamefont {T.}~\bibnamefont {Han}}, \bibinfo {author} {\bibfnamefont {Y.}~\bibnamefont {Yao}}, \bibinfo {author} {\bibfnamefont {A.~P.}\ \bibnamefont {Reddy}}, \bibinfo {author} {\bibfnamefont {J.}~\bibnamefont {Yang}}, \bibinfo {author} {\bibfnamefont {J.}~\bibnamefont {Seo}}, \bibinfo {author} {\bibfnamefont {K.}~\bibnamefont {Watanabe}}, \bibinfo {author} {\bibfnamefont {T.}~\bibnamefont {Taniguchi}}, \bibinfo {author} {\bibfnamefont {L.}~\bibnamefont {Fu}},\ and\ \bibinfo {author} {\bibfnamefont {L.}~\bibnamefont {Ju}},\ }\bibfield  {title} {\bibinfo {title} {Fractional quantum anomalous hall effect in multilayer graphene},\ }\href@noop {} {\bibfield  {journal} {\bibinfo  {journal} {Nature}\ }\textbf {\bibinfo {volume} {626}},\ \bibinfo {pages} {759} (\bibinfo {year} {2024})}\BibitemShut {NoStop}%
\bibitem [{\citenamefont {Dong}\ \emph {et~al.}(2023{\natexlab{a}})\citenamefont {Dong}, \citenamefont {Wang}, \citenamefont {Wang}, \citenamefont {Soejima}, \citenamefont {Zaletel}, \citenamefont {Vishwanath},\ and\ \citenamefont {Parker}}]{dong2023anomalous}%
  \BibitemOpen
  \bibfield  {author} {\bibinfo {author} {\bibfnamefont {J.}~\bibnamefont {Dong}}, \bibinfo {author} {\bibfnamefont {T.}~\bibnamefont {Wang}}, \bibinfo {author} {\bibfnamefont {T.}~\bibnamefont {Wang}}, \bibinfo {author} {\bibfnamefont {T.}~\bibnamefont {Soejima}}, \bibinfo {author} {\bibfnamefont {M.~P.}\ \bibnamefont {Zaletel}}, \bibinfo {author} {\bibfnamefont {A.}~\bibnamefont {Vishwanath}},\ and\ \bibinfo {author} {\bibfnamefont {D.~E.}\ \bibnamefont {Parker}},\ }\href@noop {} {\bibinfo {title} {Anomalous hall crystals in rhombohedral multilayer graphene i: Interaction-driven chern bands and fractional quantum hall states at zero magnetic field}} (\bibinfo {year} {2023}{\natexlab{a}}),\ \Eprint {https://arxiv.org/abs/2311.05568} {arXiv:2311.05568 [cond-mat.str-el]} \BibitemShut {NoStop}%
\bibitem [{\citenamefont {Zhou}\ \emph {et~al.}(2023)\citenamefont {Zhou}, \citenamefont {Yang},\ and\ \citenamefont {Zhang}}]{zhou2023fractional}%
  \BibitemOpen
  \bibfield  {author} {\bibinfo {author} {\bibfnamefont {B.}~\bibnamefont {Zhou}}, \bibinfo {author} {\bibfnamefont {H.}~\bibnamefont {Yang}},\ and\ \bibinfo {author} {\bibfnamefont {Y.-H.}\ \bibnamefont {Zhang}},\ }\href@noop {} {\bibinfo {title} {Fractional quantum anomalous hall effects in rhombohedral multilayer graphene in the moir\'eless limit and in coulomb imprinted superlattice}} (\bibinfo {year} {2023}),\ \Eprint {https://arxiv.org/abs/2311.04217} {arXiv:2311.04217 [cond-mat.str-el]} \BibitemShut {NoStop}%
\bibitem [{\citenamefont {Dong}\ \emph {et~al.}(2023{\natexlab{b}})\citenamefont {Dong}, \citenamefont {Patri},\ and\ \citenamefont {Senthil}}]{dong2023theory}%
  \BibitemOpen
  \bibfield  {author} {\bibinfo {author} {\bibfnamefont {Z.}~\bibnamefont {Dong}}, \bibinfo {author} {\bibfnamefont {A.~S.}\ \bibnamefont {Patri}},\ and\ \bibinfo {author} {\bibfnamefont {T.}~\bibnamefont {Senthil}},\ }\href@noop {} {\bibinfo {title} {Theory of fractional quantum anomalous hall phases in pentalayer rhombohedral graphene moir\'e structures}} (\bibinfo {year} {2023}{\natexlab{b}}),\ \Eprint {https://arxiv.org/abs/2311.03445} {arXiv:2311.03445 [cond-mat.str-el]} \BibitemShut {NoStop}%
\bibitem [{\citenamefont {Herzog-Arbeitman}\ \emph {et~al.}(2023)\citenamefont {Herzog-Arbeitman}, \citenamefont {Wang}, \citenamefont {Liu}, \citenamefont {Tam}, \citenamefont {Qi}, \citenamefont {Jia}, \citenamefont {Efetov}, \citenamefont {Vafek}, \citenamefont {Regnault}, \citenamefont {Weng} \emph {et~al.}}]{herzog2023moir}%
  \BibitemOpen
  \bibfield  {author} {\bibinfo {author} {\bibfnamefont {J.}~\bibnamefont {Herzog-Arbeitman}}, \bibinfo {author} {\bibfnamefont {Y.}~\bibnamefont {Wang}}, \bibinfo {author} {\bibfnamefont {J.}~\bibnamefont {Liu}}, \bibinfo {author} {\bibfnamefont {P.~M.}\ \bibnamefont {Tam}}, \bibinfo {author} {\bibfnamefont {Z.}~\bibnamefont {Qi}}, \bibinfo {author} {\bibfnamefont {Y.}~\bibnamefont {Jia}}, \bibinfo {author} {\bibfnamefont {D.~K.}\ \bibnamefont {Efetov}}, \bibinfo {author} {\bibfnamefont {O.}~\bibnamefont {Vafek}}, \bibinfo {author} {\bibfnamefont {N.}~\bibnamefont {Regnault}}, \bibinfo {author} {\bibfnamefont {H.}~\bibnamefont {Weng}}, \emph {et~al.},\ }\bibfield  {title} {\bibinfo {title} {Moir$\backslash$'e fractional chern insulators ii: First-principles calculations and continuum models of rhombohedral graphene superlattices},\ }\href@noop {} {\bibfield  {journal} {\bibinfo  {journal} {arXiv preprint arXiv:2311.12920}\ } (\bibinfo {year} {2023})}\BibitemShut {NoStop}%
\bibitem [{\citenamefont {Kwan}\ \emph {et~al.}(2023{\natexlab{a}})\citenamefont {Kwan}, \citenamefont {Yu}, \citenamefont {Herzog-Arbeitman}, \citenamefont {Efetov}, \citenamefont {Regnault},\ and\ \citenamefont {Bernevig}}]{kwan2023moire}%
  \BibitemOpen
  \bibfield  {author} {\bibinfo {author} {\bibfnamefont {Y.~H.}\ \bibnamefont {Kwan}}, \bibinfo {author} {\bibfnamefont {J.}~\bibnamefont {Yu}}, \bibinfo {author} {\bibfnamefont {J.}~\bibnamefont {Herzog-Arbeitman}}, \bibinfo {author} {\bibfnamefont {D.~K.}\ \bibnamefont {Efetov}}, \bibinfo {author} {\bibfnamefont {N.}~\bibnamefont {Regnault}},\ and\ \bibinfo {author} {\bibfnamefont {B.~A.}\ \bibnamefont {Bernevig}},\ }\href@noop {} {\bibinfo {title} {Moir\'e fractional chern insulators iii: Hartree-fock phase diagram, magic angle regime for chern insulator states, the role of the moir\'e potential and goldstone gaps in rhombohedral graphene superlattices}} (\bibinfo {year} {2023}{\natexlab{a}}),\ \Eprint {https://arxiv.org/abs/2312.11617} {arXiv:2312.11617 [cond-mat.str-el]} \BibitemShut {NoStop}%
\bibitem [{\citenamefont {Guo}\ \emph {et~al.}(2023)\citenamefont {Guo}, \citenamefont {Lu}, \citenamefont {Xie},\ and\ \citenamefont {Liu}}]{guo2023theory}%
  \BibitemOpen
  \bibfield  {author} {\bibinfo {author} {\bibfnamefont {Z.}~\bibnamefont {Guo}}, \bibinfo {author} {\bibfnamefont {X.}~\bibnamefont {Lu}}, \bibinfo {author} {\bibfnamefont {B.}~\bibnamefont {Xie}},\ and\ \bibinfo {author} {\bibfnamefont {J.}~\bibnamefont {Liu}},\ }\href@noop {} {\bibinfo {title} {Theory of fractional chern insulator states in pentalayer graphene moir\'e superlattice}} (\bibinfo {year} {2023}),\ \Eprint {https://arxiv.org/abs/2311.14368} {arXiv:2311.14368 [cond-mat.str-el]} \BibitemShut {NoStop}%
\bibitem [{\citenamefont {Bernevig}\ and\ \citenamefont {Zhang}(2006)}]{bernevigQSHE2006}%
  \BibitemOpen
  \bibfield  {author} {\bibinfo {author} {\bibfnamefont {B.~A.}\ \bibnamefont {Bernevig}}\ and\ \bibinfo {author} {\bibfnamefont {S.-C.}\ \bibnamefont {Zhang}},\ }\bibfield  {title} {\bibinfo {title} {Quantum spin hall effect},\ }\href {https://doi.org/10.1103/PhysRevLett.96.106802} {\bibfield  {journal} {\bibinfo  {journal} {Phys. Rev. Lett.}\ }\textbf {\bibinfo {volume} {96}},\ \bibinfo {pages} {106802} (\bibinfo {year} {2006})}\BibitemShut {NoStop}%
\bibitem [{\citenamefont {Levin}\ and\ \citenamefont {Stern}(2009)}]{Levin_Stern}%
  \BibitemOpen
  \bibfield  {author} {\bibinfo {author} {\bibfnamefont {M.}~\bibnamefont {Levin}}\ and\ \bibinfo {author} {\bibfnamefont {A.}~\bibnamefont {Stern}},\ }\bibfield  {title} {\bibinfo {title} {Fractional topological insulators},\ }\href {https://doi.org/10.1103/PhysRevLett.103.196803} {\bibfield  {journal} {\bibinfo  {journal} {Phys. Rev. Lett.}\ }\textbf {\bibinfo {volume} {103}},\ \bibinfo {pages} {196803} (\bibinfo {year} {2009})}\BibitemShut {NoStop}%
\bibitem [{\citenamefont {Neupert}\ \emph {et~al.}(2011{\natexlab{b}})\citenamefont {Neupert}, \citenamefont {Santos}, \citenamefont {Ryu}, \citenamefont {Chamon},\ and\ \citenamefont {Mudry}}]{Neupert2011FTI}%
  \BibitemOpen
  \bibfield  {author} {\bibinfo {author} {\bibfnamefont {T.}~\bibnamefont {Neupert}}, \bibinfo {author} {\bibfnamefont {L.}~\bibnamefont {Santos}}, \bibinfo {author} {\bibfnamefont {S.}~\bibnamefont {Ryu}}, \bibinfo {author} {\bibfnamefont {C.}~\bibnamefont {Chamon}},\ and\ \bibinfo {author} {\bibfnamefont {C.}~\bibnamefont {Mudry}},\ }\bibfield  {title} {\bibinfo {title} {Fractional topological liquids with time-reversal symmetry and their lattice realization},\ }\href {https://doi.org/10.1103/PhysRevB.84.165107} {\bibfield  {journal} {\bibinfo  {journal} {Phys. Rev. B}\ }\textbf {\bibinfo {volume} {84}},\ \bibinfo {pages} {165107} (\bibinfo {year} {2011}{\natexlab{b}})}\BibitemShut {NoStop}%
\bibitem [{\citenamefont {Stern}(2016)}]{Stern2015review}%
  \BibitemOpen
  \bibfield  {author} {\bibinfo {author} {\bibfnamefont {A.}~\bibnamefont {Stern}},\ }\bibfield  {title} {\bibinfo {title} {Fractional topological insulators: A pedagogical review},\ }\href {https://doi.org/10.1146/annurev-conmatphys-031115-011559} {\bibfield  {journal} {\bibinfo  {journal} {Annual Review of Condensed Matter Physics}\ }\textbf {\bibinfo {volume} {7}},\ \bibinfo {pages} {349} (\bibinfo {year} {2016})},\ \Eprint {https://arxiv.org/abs/https://doi.org/10.1146/annurev-conmatphys-031115-011559} {https://doi.org/10.1146/annurev-conmatphys-031115-011559} \BibitemShut {NoStop}%
\bibitem [{\citenamefont {Neupert}\ \emph {et~al.}(2015)\citenamefont {Neupert}, \citenamefont {Chamon}, \citenamefont {Iadecola}, \citenamefont {Santos},\ and\ \citenamefont {Mudry}}]{Neupert_2015}%
  \BibitemOpen
  \bibfield  {author} {\bibinfo {author} {\bibfnamefont {T.}~\bibnamefont {Neupert}}, \bibinfo {author} {\bibfnamefont {C.}~\bibnamefont {Chamon}}, \bibinfo {author} {\bibfnamefont {T.}~\bibnamefont {Iadecola}}, \bibinfo {author} {\bibfnamefont {L.~H.}\ \bibnamefont {Santos}},\ and\ \bibinfo {author} {\bibfnamefont {C.}~\bibnamefont {Mudry}},\ }\bibfield  {title} {\bibinfo {title} {Fractional (chern and topological) insulators},\ }\href {https://doi.org/10.1088/0031-8949/2015/T164/014005} {\bibfield  {journal} {\bibinfo  {journal} {Physica Scripta}\ }\textbf {\bibinfo {volume} {2015}},\ \bibinfo {pages} {014005} (\bibinfo {year} {2015})}\BibitemShut {NoStop}%
\bibitem [{\citenamefont {Levin}\ and\ \citenamefont {Stern}(2012)}]{Levin2012classification}%
  \BibitemOpen
  \bibfield  {author} {\bibinfo {author} {\bibfnamefont {M.}~\bibnamefont {Levin}}\ and\ \bibinfo {author} {\bibfnamefont {A.}~\bibnamefont {Stern}},\ }\bibfield  {title} {\bibinfo {title} {Classification and analysis of two-dimensional abelian fractional topological insulators},\ }\href {https://doi.org/10.1103/PhysRevB.86.115131} {\bibfield  {journal} {\bibinfo  {journal} {Phys. Rev. B}\ }\textbf {\bibinfo {volume} {86}},\ \bibinfo {pages} {115131} (\bibinfo {year} {2012})}\BibitemShut {NoStop}%
\bibitem [{\citenamefont {Kang}\ \emph {et~al.}(2024{\natexlab{a}})\citenamefont {Kang}, \citenamefont {Shen}, \citenamefont {Qiu}, \citenamefont {Zeng}, \citenamefont {Xia}, \citenamefont {Watanabe}, \citenamefont {Taniguchi}, \citenamefont {Shan},\ and\ \citenamefont {Mak}}]{kang2024evidence_nature}%
  \BibitemOpen
  \bibfield  {author} {\bibinfo {author} {\bibfnamefont {K.}~\bibnamefont {Kang}}, \bibinfo {author} {\bibfnamefont {B.}~\bibnamefont {Shen}}, \bibinfo {author} {\bibfnamefont {Y.}~\bibnamefont {Qiu}}, \bibinfo {author} {\bibfnamefont {Y.}~\bibnamefont {Zeng}}, \bibinfo {author} {\bibfnamefont {Z.}~\bibnamefont {Xia}}, \bibinfo {author} {\bibfnamefont {K.}~\bibnamefont {Watanabe}}, \bibinfo {author} {\bibfnamefont {T.}~\bibnamefont {Taniguchi}}, \bibinfo {author} {\bibfnamefont {J.}~\bibnamefont {Shan}},\ and\ \bibinfo {author} {\bibfnamefont {K.~F.}\ \bibnamefont {Mak}},\ }\bibfield  {title} {\bibinfo {title} {Evidence of the fractional quantum spin hall effect in moir{\'e} mote2},\ }\href@noop {} {\bibfield  {journal} {\bibinfo  {journal} {Nature}\ }\textbf {\bibinfo {volume} {628}},\ \bibinfo {pages} {522} (\bibinfo {year} {2024}{\natexlab{a}})}\BibitemShut {NoStop}%
\bibitem [{\citenamefont {Repellin}\ \emph {et~al.}(2014)\citenamefont {Repellin}, \citenamefont {Bernevig},\ and\ \citenamefont {Regnault}}]{Repellin2014FTI}%
  \BibitemOpen
  \bibfield  {author} {\bibinfo {author} {\bibfnamefont {C.}~\bibnamefont {Repellin}}, \bibinfo {author} {\bibfnamefont {B.~A.}\ \bibnamefont {Bernevig}},\ and\ \bibinfo {author} {\bibfnamefont {N.}~\bibnamefont {Regnault}},\ }\bibfield  {title} {\bibinfo {title} {Z$_{2}$ fractional topological insulators in two dimensions},\ }\href {https://doi.org/10.1103/PhysRevB.90.245401} {\bibfield  {journal} {\bibinfo  {journal} {Phys. Rev. B}\ }\textbf {\bibinfo {volume} {90}},\ \bibinfo {pages} {245401} (\bibinfo {year} {2014})}\BibitemShut {NoStop}%
\bibitem [{\citenamefont {Furukawa}\ and\ \citenamefont {Ueda}(2014{\natexlab{a}})}]{Furukawa2014bose}%
  \BibitemOpen
  \bibfield  {author} {\bibinfo {author} {\bibfnamefont {S.}~\bibnamefont {Furukawa}}\ and\ \bibinfo {author} {\bibfnamefont {M.}~\bibnamefont {Ueda}},\ }\bibfield  {title} {\bibinfo {title} {Global phase diagram of two-component bose gases in antiparallel magnetic fields},\ }\href {https://doi.org/10.1103/PhysRevA.90.033602} {\bibfield  {journal} {\bibinfo  {journal} {Phys. Rev. A}\ }\textbf {\bibinfo {volume} {90}},\ \bibinfo {pages} {033602} (\bibinfo {year} {2014}{\natexlab{a}})}\BibitemShut {NoStop}%
\bibitem [{\citenamefont {Crépel}\ and\ \citenamefont {Regnault}(2024)}]{crepel2024attractive}%
  \BibitemOpen
  \bibfield  {author} {\bibinfo {author} {\bibfnamefont {V.}~\bibnamefont {Crépel}}\ and\ \bibinfo {author} {\bibfnamefont {N.}~\bibnamefont {Regnault}},\ }\href {https://arxiv.org/abs/2403.05622} {\bibinfo {title} {Attractive haldane bilayers for trapping non-abelian anyons}} (\bibinfo {year} {2024}),\ \Eprint {https://arxiv.org/abs/2403.05622} {arXiv:2403.05622 [cond-mat.str-el]} \BibitemShut {NoStop}%
\bibitem [{\citenamefont {Chen}\ and\ \citenamefont {Yang}(2012)}]{Chen2012FQHtorusFTI}%
  \BibitemOpen
  \bibfield  {author} {\bibinfo {author} {\bibfnamefont {H.}~\bibnamefont {Chen}}\ and\ \bibinfo {author} {\bibfnamefont {K.}~\bibnamefont {Yang}},\ }\bibfield  {title} {\bibinfo {title} {Interaction-driven quantum phase transitions in fractional topological insulators},\ }\href {https://doi.org/10.1103/PhysRevB.85.195113} {\bibfield  {journal} {\bibinfo  {journal} {Phys. Rev. B}\ }\textbf {\bibinfo {volume} {85}},\ \bibinfo {pages} {195113} (\bibinfo {year} {2012})}\BibitemShut {NoStop}%
\bibitem [{\citenamefont {Mukherjee}\ and\ \citenamefont {Park}(2019)}]{Mukherjee2019FQHsphereFTI}%
  \BibitemOpen
  \bibfield  {author} {\bibinfo {author} {\bibfnamefont {S.}~\bibnamefont {Mukherjee}}\ and\ \bibinfo {author} {\bibfnamefont {K.}~\bibnamefont {Park}},\ }\bibfield  {title} {\bibinfo {title} {Spin separation in the half-filled fractional topological insulator},\ }\href {https://doi.org/10.1103/PhysRevB.99.115131} {\bibfield  {journal} {\bibinfo  {journal} {Phys. Rev. B}\ }\textbf {\bibinfo {volume} {99}},\ \bibinfo {pages} {115131} (\bibinfo {year} {2019})}\BibitemShut {NoStop}%
\bibitem [{\citenamefont {Bultinck}\ \emph {et~al.}(2020)\citenamefont {Bultinck}, \citenamefont {Chatterjee},\ and\ \citenamefont {Zaletel}}]{Bultinck2020mechanism}%
  \BibitemOpen
  \bibfield  {author} {\bibinfo {author} {\bibfnamefont {N.}~\bibnamefont {Bultinck}}, \bibinfo {author} {\bibfnamefont {S.}~\bibnamefont {Chatterjee}},\ and\ \bibinfo {author} {\bibfnamefont {M.~P.}\ \bibnamefont {Zaletel}},\ }\bibfield  {title} {\bibinfo {title} {Mechanism for anomalous hall ferromagnetism in twisted bilayer graphene},\ }\href {https://doi.org/10.1103/PhysRevLett.124.166601} {\bibfield  {journal} {\bibinfo  {journal} {Phys. Rev. Lett.}\ }\textbf {\bibinfo {volume} {124}},\ \bibinfo {pages} {166601} (\bibinfo {year} {2020})}\BibitemShut {NoStop}%
\bibitem [{\citenamefont {Furukawa}\ and\ \citenamefont {Ueda}(2014{\natexlab{b}})}]{furukawa2014global}%
  \BibitemOpen
  \bibfield  {author} {\bibinfo {author} {\bibfnamefont {S.}~\bibnamefont {Furukawa}}\ and\ \bibinfo {author} {\bibfnamefont {M.}~\bibnamefont {Ueda}},\ }\bibfield  {title} {\bibinfo {title} {Global phase diagram of two-component bose gases in antiparallel magnetic fields},\ }\href {https://doi.org/10.1103/PhysRevA.90.033602} {\bibfield  {journal} {\bibinfo  {journal} {Phys. Rev. A}\ }\textbf {\bibinfo {volume} {90}},\ \bibinfo {pages} {033602} (\bibinfo {year} {2014}{\natexlab{b}})}\BibitemShut {NoStop}%
\bibitem [{\citenamefont {Zhang}(2018)}]{zhang2018composite}%
  \BibitemOpen
  \bibfield  {author} {\bibinfo {author} {\bibfnamefont {Y.-H.}\ \bibnamefont {Zhang}},\ }\href@noop {} {\bibinfo {title} {Composite fermion insulator in opposite-fields quantum hall bilayers}} (\bibinfo {year} {2018}),\ \Eprint {https://arxiv.org/abs/1810.03600} {arXiv:1810.03600 [cond-mat.str-el]} \BibitemShut {NoStop}%
\bibitem [{\citenamefont {Kwan}\ \emph {et~al.}(2021)\citenamefont {Kwan}, \citenamefont {Hu}, \citenamefont {Simon},\ and\ \citenamefont {Parameswaran}}]{kwan2021exciton}%
  \BibitemOpen
  \bibfield  {author} {\bibinfo {author} {\bibfnamefont {Y.~H.}\ \bibnamefont {Kwan}}, \bibinfo {author} {\bibfnamefont {Y.}~\bibnamefont {Hu}}, \bibinfo {author} {\bibfnamefont {S.~H.}\ \bibnamefont {Simon}},\ and\ \bibinfo {author} {\bibfnamefont {S.~A.}\ \bibnamefont {Parameswaran}},\ }\bibfield  {title} {\bibinfo {title} {Exciton band topology in spontaneous quantum anomalous hall insulators: Applications to twisted bilayer graphene},\ }\href {https://doi.org/10.1103/PhysRevLett.126.137601} {\bibfield  {journal} {\bibinfo  {journal} {Phys. Rev. Lett.}\ }\textbf {\bibinfo {volume} {126}},\ \bibinfo {pages} {137601} (\bibinfo {year} {2021})}\BibitemShut {NoStop}%
\bibitem [{\citenamefont {Kwan}\ \emph {et~al.}(2022)\citenamefont {Kwan}, \citenamefont {Hu}, \citenamefont {Simon},\ and\ \citenamefont {Parameswaran}}]{kwan2022hierarchy}%
  \BibitemOpen
  \bibfield  {author} {\bibinfo {author} {\bibfnamefont {Y.~H.}\ \bibnamefont {Kwan}}, \bibinfo {author} {\bibfnamefont {Y.}~\bibnamefont {Hu}}, \bibinfo {author} {\bibfnamefont {S.~H.}\ \bibnamefont {Simon}},\ and\ \bibinfo {author} {\bibfnamefont {S.~A.}\ \bibnamefont {Parameswaran}},\ }\bibfield  {title} {\bibinfo {title} {Excitonic fractional quantum hall hierarchy in moir\'e heterostructures},\ }\href {https://doi.org/10.1103/PhysRevB.105.235121} {\bibfield  {journal} {\bibinfo  {journal} {Phys. Rev. B}\ }\textbf {\bibinfo {volume} {105}},\ \bibinfo {pages} {235121} (\bibinfo {year} {2022})}\BibitemShut {NoStop}%
\bibitem [{\citenamefont {Eugenio}\ and\ \citenamefont {Dağ}(2020)}]{eugenio2020DMRG}%
  \BibitemOpen
  \bibfield  {author} {\bibinfo {author} {\bibfnamefont {P.~M.}\ \bibnamefont {Eugenio}}\ and\ \bibinfo {author} {\bibfnamefont {C.~B.}\ \bibnamefont {Dağ}},\ }\bibfield  {title} {\bibinfo {title} {{DMRG study of strongly interacting $\mathbb{Z}_2$ flatbands: a toy model inspired by twisted bilayer graphene}},\ }\href {https://doi.org/10.21468/SciPostPhysCore.3.2.015} {\bibfield  {journal} {\bibinfo  {journal} {SciPost Phys. Core}\ }\textbf {\bibinfo {volume} {3}},\ \bibinfo {pages} {015} (\bibinfo {year} {2020})}\BibitemShut {NoStop}%
\bibitem [{\citenamefont {Stefanidis}\ and\ \citenamefont {Sodemann}(2020)}]{stefanidis2020excitonic}%
  \BibitemOpen
  \bibfield  {author} {\bibinfo {author} {\bibfnamefont {N.}~\bibnamefont {Stefanidis}}\ and\ \bibinfo {author} {\bibfnamefont {I.}~\bibnamefont {Sodemann}},\ }\bibfield  {title} {\bibinfo {title} {Excitonic laughlin states in ideal topological insulator flat bands and their possible presence in moir\'e superlattice materials},\ }\href {https://doi.org/10.1103/PhysRevB.102.035158} {\bibfield  {journal} {\bibinfo  {journal} {Phys. Rev. B}\ }\textbf {\bibinfo {volume} {102}},\ \bibinfo {pages} {035158} (\bibinfo {year} {2020})}\BibitemShut {NoStop}%
\bibitem [{\citenamefont {Chatterjee}\ \emph {et~al.}(2022)\citenamefont {Chatterjee}, \citenamefont {Ippoliti},\ and\ \citenamefont {Zaletel}}]{chatterjee2022dmrg}%
  \BibitemOpen
  \bibfield  {author} {\bibinfo {author} {\bibfnamefont {S.}~\bibnamefont {Chatterjee}}, \bibinfo {author} {\bibfnamefont {M.}~\bibnamefont {Ippoliti}},\ and\ \bibinfo {author} {\bibfnamefont {M.~P.}\ \bibnamefont {Zaletel}},\ }\bibfield  {title} {\bibinfo {title} {Skyrmion superconductivity: Dmrg evidence for a topological route to superconductivity},\ }\href {https://doi.org/10.1103/PhysRevB.106.035421} {\bibfield  {journal} {\bibinfo  {journal} {Phys. Rev. B}\ }\textbf {\bibinfo {volume} {106}},\ \bibinfo {pages} {035421} (\bibinfo {year} {2022})}\BibitemShut {NoStop}%
\bibitem [{\citenamefont {Myerson-Jain}\ \emph {et~al.}(2023)\citenamefont {Myerson-Jain}, \citenamefont {Jian},\ and\ \citenamefont {Xu}}]{myersonjain2023conjugate}%
  \BibitemOpen
  \bibfield  {author} {\bibinfo {author} {\bibfnamefont {N.}~\bibnamefont {Myerson-Jain}}, \bibinfo {author} {\bibfnamefont {C.-M.}\ \bibnamefont {Jian}},\ and\ \bibinfo {author} {\bibfnamefont {C.}~\bibnamefont {Xu}},\ }\href@noop {} {\bibinfo {title} {The conjugate composite fermi liquid}} (\bibinfo {year} {2023}),\ \Eprint {https://arxiv.org/abs/2311.16250} {arXiv:2311.16250 [cond-mat.str-el]} \BibitemShut {NoStop}%
\bibitem [{\citenamefont {Yang}(2023)}]{yang2023phase}%
  \BibitemOpen
  \bibfield  {author} {\bibinfo {author} {\bibfnamefont {K.}~\bibnamefont {Yang}},\ }\bibfield  {title} {\bibinfo {title} {Phase transition in bilayer quantum hall system with opposite magnetic field},\ }\href@noop {} {\bibfield  {journal} {\bibinfo  {journal} {Chinese Physics B}\ }\textbf {\bibinfo {volume} {32}},\ \bibinfo {pages} {097303} (\bibinfo {year} {2023})}\BibitemShut {NoStop}%
\bibitem [{\citenamefont {Wu}\ \emph {et~al.}(2024)\citenamefont {Wu}, \citenamefont {Shaffer}, \citenamefont {Wu},\ and\ \citenamefont {Santos}}]{Wu2024}%
  \BibitemOpen
  \bibfield  {author} {\bibinfo {author} {\bibfnamefont {Y.-M.}\ \bibnamefont {Wu}}, \bibinfo {author} {\bibfnamefont {D.}~\bibnamefont {Shaffer}}, \bibinfo {author} {\bibfnamefont {Z.}~\bibnamefont {Wu}},\ and\ \bibinfo {author} {\bibfnamefont {L.~H.}\ \bibnamefont {Santos}},\ }\bibfield  {title} {\bibinfo {title} {Time-reversal invariant topological moir\'e flat band: A platform for the fractional quantum spin hall effect},\ }\href {https://doi.org/10.1103/PhysRevB.109.115111} {\bibfield  {journal} {\bibinfo  {journal} {Phys. Rev. B}\ }\textbf {\bibinfo {volume} {109}},\ \bibinfo {pages} {115111} (\bibinfo {year} {2024})}\BibitemShut {NoStop}%
\bibitem [{\citenamefont {Shi}\ \emph {et~al.}(2024{\natexlab{a}})\citenamefont {Shi}, \citenamefont {Goldman}, \citenamefont {Dong},\ and\ \citenamefont {Senthil}}]{shi2024excitonic}%
  \BibitemOpen
  \bibfield  {author} {\bibinfo {author} {\bibfnamefont {Z.~D.}\ \bibnamefont {Shi}}, \bibinfo {author} {\bibfnamefont {H.}~\bibnamefont {Goldman}}, \bibinfo {author} {\bibfnamefont {Z.}~\bibnamefont {Dong}},\ and\ \bibinfo {author} {\bibfnamefont {T.}~\bibnamefont {Senthil}},\ }\href@noop {} {\bibinfo {title} {Excitonic quantum criticality: from bilayer graphene to narrow chern bands}} (\bibinfo {year} {2024}{\natexlab{a}}),\ \Eprint {https://arxiv.org/abs/2402.12436} {arXiv:2402.12436 [cond-mat.str-el]} \BibitemShut {NoStop}%
\bibitem [{\citenamefont {Kwan}\ \emph {et~al.}(2024)\citenamefont {Kwan}, \citenamefont {Wang}, \citenamefont {Wagner}, \citenamefont {Simon}, \citenamefont {Parameswaran},\ and\ \citenamefont {Bultinck}}]{kwan2024textured}%
  \BibitemOpen
  \bibfield  {author} {\bibinfo {author} {\bibfnamefont {Y.~H.}\ \bibnamefont {Kwan}}, \bibinfo {author} {\bibfnamefont {Z.}~\bibnamefont {Wang}}, \bibinfo {author} {\bibfnamefont {G.}~\bibnamefont {Wagner}}, \bibinfo {author} {\bibfnamefont {S.~H.}\ \bibnamefont {Simon}}, \bibinfo {author} {\bibfnamefont {S.~A.}\ \bibnamefont {Parameswaran}},\ and\ \bibinfo {author} {\bibfnamefont {N.}~\bibnamefont {Bultinck}},\ }\href {https://arxiv.org/abs/2406.15343} {\bibinfo {title} {Textured exciton insulators}} (\bibinfo {year} {2024}),\ \Eprint {https://arxiv.org/abs/2406.15343} {arXiv:2406.15343 [cond-mat.str-el]} \BibitemShut {NoStop}%
\bibitem [{\citenamefont {Bernevig}\ and\ \citenamefont {Regnault}(2012)}]{Bernevig2012emergent}%
  \BibitemOpen
  \bibfield  {author} {\bibinfo {author} {\bibfnamefont {B.~A.}\ \bibnamefont {Bernevig}}\ and\ \bibinfo {author} {\bibfnamefont {N.}~\bibnamefont {Regnault}},\ }\bibfield  {title} {\bibinfo {title} {Emergent many-body translational symmetries of abelian and non-abelian fractionally filled topological insulators},\ }\href {https://doi.org/10.1103/PhysRevB.85.075128} {\bibfield  {journal} {\bibinfo  {journal} {Phys. Rev. B}\ }\textbf {\bibinfo {volume} {85}},\ \bibinfo {pages} {075128} (\bibinfo {year} {2012})}\BibitemShut {NoStop}%
\bibitem [{\citenamefont {Rytova}(2018)}]{rytova2018screened}%
  \BibitemOpen
  \bibfield  {author} {\bibinfo {author} {\bibfnamefont {N.~S.}\ \bibnamefont {Rytova}},\ }\bibfield  {title} {\bibinfo {title} {Screened potential of a point charge in a thin film},\ }\href@noop {} {\bibfield  {journal} {\bibinfo  {journal} {arXiv preprint arXiv:1806.00976}\ } (\bibinfo {year} {2018})}\BibitemShut {NoStop}%
\bibitem [{\citenamefont {{Keldysh}}(1979)}]{Keldysh1979}%
  \BibitemOpen
  \bibfield  {author} {\bibinfo {author} {\bibfnamefont {L.~V.}\ \bibnamefont {{Keldysh}}},\ }\bibfield  {title} {\bibinfo {title} {{Coulomb interaction in thin semiconductor and semimetal films}},\ }\href@noop {} {\bibfield  {journal} {\bibinfo  {journal} {Soviet Journal of Experimental and Theoretical Physics Letters}\ }\textbf {\bibinfo {volume} {29}},\ \bibinfo {pages} {658} (\bibinfo {year} {1979})}\BibitemShut {NoStop}%
\bibitem [{\citenamefont {Cudazzo}\ \emph {et~al.}(2011)\citenamefont {Cudazzo}, \citenamefont {Tokatly},\ and\ \citenamefont {Rubio}}]{cudazzo2011dielectric}%
  \BibitemOpen
  \bibfield  {author} {\bibinfo {author} {\bibfnamefont {P.}~\bibnamefont {Cudazzo}}, \bibinfo {author} {\bibfnamefont {I.~V.}\ \bibnamefont {Tokatly}},\ and\ \bibinfo {author} {\bibfnamefont {A.}~\bibnamefont {Rubio}},\ }\bibfield  {title} {\bibinfo {title} {Dielectric screening in two-dimensional insulators: Implications for excitonic and impurity states in graphane},\ }\href {https://doi.org/10.1103/PhysRevB.84.085406} {\bibfield  {journal} {\bibinfo  {journal} {Phys. Rev. B}\ }\textbf {\bibinfo {volume} {84}},\ \bibinfo {pages} {085406} (\bibinfo {year} {2011})}\BibitemShut {NoStop}%
\bibitem [{\citenamefont {Wang}\ \emph {et~al.}(2018)\citenamefont {Wang}, \citenamefont {Chernikov}, \citenamefont {Glazov}, \citenamefont {Heinz}, \citenamefont {Marie}, \citenamefont {Amand},\ and\ \citenamefont {Urbaszek}}]{wang2018colloq}%
  \BibitemOpen
  \bibfield  {author} {\bibinfo {author} {\bibfnamefont {G.}~\bibnamefont {Wang}}, \bibinfo {author} {\bibfnamefont {A.}~\bibnamefont {Chernikov}}, \bibinfo {author} {\bibfnamefont {M.~M.}\ \bibnamefont {Glazov}}, \bibinfo {author} {\bibfnamefont {T.~F.}\ \bibnamefont {Heinz}}, \bibinfo {author} {\bibfnamefont {X.}~\bibnamefont {Marie}}, \bibinfo {author} {\bibfnamefont {T.}~\bibnamefont {Amand}},\ and\ \bibinfo {author} {\bibfnamefont {B.}~\bibnamefont {Urbaszek}},\ }\bibfield  {title} {\bibinfo {title} {Colloquium: Excitons in atomically thin transition metal dichalcogenides},\ }\href {https://doi.org/10.1103/RevModPhys.90.021001} {\bibfield  {journal} {\bibinfo  {journal} {Rev. Mod. Phys.}\ }\textbf {\bibinfo {volume} {90}},\ \bibinfo {pages} {021001} (\bibinfo {year} {2018})}\BibitemShut {NoStop}%
\bibitem [{\citenamefont {Zhao}\ \emph {et~al.}(2023)\citenamefont {Zhao}, \citenamefont {Huang}, \citenamefont {Crépel}, \citenamefont {Wu}, \citenamefont {Zhang}, \citenamefont {Wang}, \citenamefont {Han}, \citenamefont {Li}, \citenamefont {Xi}, \citenamefont {Pan}, \citenamefont {Wang}, \citenamefont {Watanabe}, \citenamefont {Taniguchi}, \citenamefont {Sacépé}, \citenamefont {Zhang}, \citenamefont {Wang}, \citenamefont {Lu}, \citenamefont {Regnault},\ and\ \citenamefont {Han}}]{zhao2023probing}%
  \BibitemOpen
  \bibfield  {author} {\bibinfo {author} {\bibfnamefont {S.}~\bibnamefont {Zhao}}, \bibinfo {author} {\bibfnamefont {J.}~\bibnamefont {Huang}}, \bibinfo {author} {\bibfnamefont {V.}~\bibnamefont {Crépel}}, \bibinfo {author} {\bibfnamefont {X.}~\bibnamefont {Wu}}, \bibinfo {author} {\bibfnamefont {T.}~\bibnamefont {Zhang}}, \bibinfo {author} {\bibfnamefont {H.}~\bibnamefont {Wang}}, \bibinfo {author} {\bibfnamefont {X.}~\bibnamefont {Han}}, \bibinfo {author} {\bibfnamefont {Z.}~\bibnamefont {Li}}, \bibinfo {author} {\bibfnamefont {C.}~\bibnamefont {Xi}}, \bibinfo {author} {\bibfnamefont {S.}~\bibnamefont {Pan}}, \bibinfo {author} {\bibfnamefont {Z.}~\bibnamefont {Wang}}, \bibinfo {author} {\bibfnamefont {K.}~\bibnamefont {Watanabe}}, \bibinfo {author} {\bibfnamefont {T.}~\bibnamefont {Taniguchi}}, \bibinfo {author} {\bibfnamefont {B.}~\bibnamefont {Sacépé}}, \bibinfo {author} {\bibfnamefont {J.}~\bibnamefont {Zhang}}, \bibinfo {author} {\bibfnamefont {N.}~\bibnamefont {Wang}}, \bibinfo {author}
  {\bibfnamefont {J.}~\bibnamefont {Lu}}, \bibinfo {author} {\bibfnamefont {N.}~\bibnamefont {Regnault}},\ and\ \bibinfo {author} {\bibfnamefont {Z.~V.}\ \bibnamefont {Han}},\ }\href@noop {} {\bibinfo {title} {Probing the fractional quantum hall phases in valley-layer locked bilayer mos$_{2}$}} (\bibinfo {year} {2023}),\ \Eprint {https://arxiv.org/abs/2308.02821} {arXiv:2308.02821 [cond-mat.mes-hall]} \BibitemShut {NoStop}%
\bibitem [{\citenamefont {Wu}\ \emph {et~al.}(2019)\citenamefont {Wu}, \citenamefont {Lovorn}, \citenamefont {Tutuc}, \citenamefont {Martin},\ and\ \citenamefont {MacDonald}}]{Wu2019}%
  \BibitemOpen
  \bibfield  {author} {\bibinfo {author} {\bibfnamefont {F.}~\bibnamefont {Wu}}, \bibinfo {author} {\bibfnamefont {T.}~\bibnamefont {Lovorn}}, \bibinfo {author} {\bibfnamefont {E.}~\bibnamefont {Tutuc}}, \bibinfo {author} {\bibfnamefont {I.}~\bibnamefont {Martin}},\ and\ \bibinfo {author} {\bibfnamefont {A.~H.}\ \bibnamefont {MacDonald}},\ }\bibfield  {title} {\bibinfo {title} {Topological insulators in twisted transition metal dichalcogenide homobilayers},\ }\href {https://doi.org/10.1103/PhysRevLett.122.086402} {\bibfield  {journal} {\bibinfo  {journal} {Phys. Rev. Lett.}\ }\textbf {\bibinfo {volume} {122}},\ \bibinfo {pages} {086402} (\bibinfo {year} {2019})}\BibitemShut {NoStop}%
\bibitem [{\citenamefont {Laturia}\ \emph {et~al.}(2018)\citenamefont {Laturia}, \citenamefont {Van~de Put},\ and\ \citenamefont {Vandenberghe}}]{laturia2018dielectric}%
  \BibitemOpen
  \bibfield  {author} {\bibinfo {author} {\bibfnamefont {A.}~\bibnamefont {Laturia}}, \bibinfo {author} {\bibfnamefont {M.~L.}\ \bibnamefont {Van~de Put}},\ and\ \bibinfo {author} {\bibfnamefont {W.~G.}\ \bibnamefont {Vandenberghe}},\ }\bibfield  {title} {\bibinfo {title} {Dielectric properties of hexagonal boron nitride and transition metal dichalcogenides: from monolayer to bulk},\ }\href@noop {} {\bibfield  {journal} {\bibinfo  {journal} {npj 2D Materials and Applications}\ }\textbf {\bibinfo {volume} {2}},\ \bibinfo {pages} {6} (\bibinfo {year} {2018})}\BibitemShut {NoStop}%
\bibitem [{\citenamefont {L\"auchli}\ \emph {et~al.}(2013)\citenamefont {L\"auchli}, \citenamefont {Liu}, \citenamefont {Bergholtz},\ and\ \citenamefont {Moessner}}]{lauchli2013hierarchy}%
  \BibitemOpen
  \bibfield  {author} {\bibinfo {author} {\bibfnamefont {A.~M.}\ \bibnamefont {L\"auchli}}, \bibinfo {author} {\bibfnamefont {Z.}~\bibnamefont {Liu}}, \bibinfo {author} {\bibfnamefont {E.~J.}\ \bibnamefont {Bergholtz}},\ and\ \bibinfo {author} {\bibfnamefont {R.}~\bibnamefont {Moessner}},\ }\bibfield  {title} {\bibinfo {title} {Hierarchy of fractional chern insulators and competing compressible states},\ }\href {https://doi.org/10.1103/PhysRevLett.111.126802} {\bibfield  {journal} {\bibinfo  {journal} {Phys. Rev. Lett.}\ }\textbf {\bibinfo {volume} {111}},\ \bibinfo {pages} {126802} (\bibinfo {year} {2013})}\BibitemShut {NoStop}%
\bibitem [{\citenamefont {Rezayi}\ and\ \citenamefont {Simon}(2011)}]{Rezayi2011}%
  \BibitemOpen
  \bibfield  {author} {\bibinfo {author} {\bibfnamefont {E.~H.}\ \bibnamefont {Rezayi}}\ and\ \bibinfo {author} {\bibfnamefont {S.~H.}\ \bibnamefont {Simon}},\ }\bibfield  {title} {\bibinfo {title} {Breaking of particle-hole symmetry by landau level mixing in the $\ensuremath{\nu}=5/2$ quantized hall state},\ }\href {https://doi.org/10.1103/PhysRevLett.106.116801} {\bibfield  {journal} {\bibinfo  {journal} {Phys. Rev. Lett.}\ }\textbf {\bibinfo {volume} {106}},\ \bibinfo {pages} {116801} (\bibinfo {year} {2011})}\BibitemShut {NoStop}%
\bibitem [{\citenamefont {Liu}\ \emph {et~al.}(2015)\citenamefont {Liu}, \citenamefont {Bhatt},\ and\ \citenamefont {Regnault}}]{Liu2015characterization}%
  \BibitemOpen
  \bibfield  {author} {\bibinfo {author} {\bibfnamefont {Z.}~\bibnamefont {Liu}}, \bibinfo {author} {\bibfnamefont {R.~N.}\ \bibnamefont {Bhatt}},\ and\ \bibinfo {author} {\bibfnamefont {N.}~\bibnamefont {Regnault}},\ }\bibfield  {title} {\bibinfo {title} {Characterization of quasiholes in fractional chern insulators},\ }\href {https://doi.org/10.1103/PhysRevB.91.045126} {\bibfield  {journal} {\bibinfo  {journal} {Phys. Rev. B}\ }\textbf {\bibinfo {volume} {91}},\ \bibinfo {pages} {045126} (\bibinfo {year} {2015})}\BibitemShut {NoStop}%
\bibitem [{\citenamefont {Roy}(2014)}]{Roy}%
  \BibitemOpen
  \bibfield  {author} {\bibinfo {author} {\bibfnamefont {R.}~\bibnamefont {Roy}},\ }\bibfield  {title} {\bibinfo {title} {Band geometry of fractional topological insulators},\ }\href {https://doi.org/10.1103/PhysRevB.90.165139} {\bibfield  {journal} {\bibinfo  {journal} {Phys. Rev. B}\ }\textbf {\bibinfo {volume} {90}},\ \bibinfo {pages} {165139} (\bibinfo {year} {2014})}\BibitemShut {NoStop}%
\bibitem [{\citenamefont {Wilk}\ \emph {et~al.}(2001)\citenamefont {Wilk}, \citenamefont {Wallace},\ and\ \citenamefont {Anthony}}]{Wilk2001highk}%
  \BibitemOpen
  \bibfield  {author} {\bibinfo {author} {\bibfnamefont {G.~D.}\ \bibnamefont {Wilk}}, \bibinfo {author} {\bibfnamefont {R.~M.}\ \bibnamefont {Wallace}},\ and\ \bibinfo {author} {\bibfnamefont {J.~M.}\ \bibnamefont {Anthony}},\ }\bibfield  {title} {\bibinfo {title} {{High-K gate dielectrics: Current status and materials properties considerations}},\ }\href {https://doi.org/10.1063/1.1361065} {\bibfield  {journal} {\bibinfo  {journal} {Journal of Applied Physics}\ }\textbf {\bibinfo {volume} {89}},\ \bibinfo {pages} {5243} (\bibinfo {year} {2001})},\ \Eprint {https://arxiv.org/abs/https://pubs.aip.org/aip/jap/article-pdf/89/10/5243/19314247/5243\_1\_online.pdf} {https://pubs.aip.org/aip/jap/article-pdf/89/10/5243/19314247/5243\_1\_online.pdf} \BibitemShut {NoStop}%
\bibitem [{\citenamefont {Blason}\ and\ \citenamefont {Fabrizio}(2022)}]{blason2022local}%
  \BibitemOpen
  \bibfield  {author} {\bibinfo {author} {\bibfnamefont {A.}~\bibnamefont {Blason}}\ and\ \bibinfo {author} {\bibfnamefont {M.}~\bibnamefont {Fabrizio}},\ }\bibfield  {title} {\bibinfo {title} {Local kekul\'e distortion turns twisted bilayer graphene into topological mott insulators and superconductors},\ }\href {https://doi.org/10.1103/PhysRevB.106.235112} {\bibfield  {journal} {\bibinfo  {journal} {Phys. Rev. B}\ }\textbf {\bibinfo {volume} {106}},\ \bibinfo {pages} {235112} (\bibinfo {year} {2022})}\BibitemShut {NoStop}%
\bibitem [{\citenamefont {Kwan}\ \emph {et~al.}(2023{\natexlab{b}})\citenamefont {Kwan}, \citenamefont {Wagner}, \citenamefont {Bultinck}, \citenamefont {Simon}, \citenamefont {Berg},\ and\ \citenamefont {Parameswaran}}]{kwan2023electronphonon}%
  \BibitemOpen
  \bibfield  {author} {\bibinfo {author} {\bibfnamefont {Y.~H.}\ \bibnamefont {Kwan}}, \bibinfo {author} {\bibfnamefont {G.}~\bibnamefont {Wagner}}, \bibinfo {author} {\bibfnamefont {N.}~\bibnamefont {Bultinck}}, \bibinfo {author} {\bibfnamefont {S.~H.}\ \bibnamefont {Simon}}, \bibinfo {author} {\bibfnamefont {E.}~\bibnamefont {Berg}},\ and\ \bibinfo {author} {\bibfnamefont {S.~A.}\ \bibnamefont {Parameswaran}},\ }\href@noop {} {\bibinfo {title} {Electron-phonon coupling and competing kekul\'e orders in twisted bilayer graphene}} (\bibinfo {year} {2023}{\natexlab{b}}),\ \Eprint {https://arxiv.org/abs/2303.13602} {arXiv:2303.13602 [cond-mat.str-el]} \BibitemShut {NoStop}%
\bibitem [{\citenamefont {Shi}\ \emph {et~al.}(2024{\natexlab{b}})\citenamefont {Shi}, \citenamefont {Miao},\ and\ \citenamefont {Dai}}]{shi2024moire}%
  \BibitemOpen
  \bibfield  {author} {\bibinfo {author} {\bibfnamefont {H.}~\bibnamefont {Shi}}, \bibinfo {author} {\bibfnamefont {W.}~\bibnamefont {Miao}},\ and\ \bibinfo {author} {\bibfnamefont {X.}~\bibnamefont {Dai}},\ }\href@noop {} {\bibinfo {title} {Moir\'{e} optical phonons dancing with heavy electrons in magic-angle twisted bilayer graphene}} (\bibinfo {year} {2024}{\natexlab{b}}),\ \Eprint {https://arxiv.org/abs/2402.11824} {arXiv:2402.11824 [cond-mat.mes-hall]} \BibitemShut {NoStop}%
\bibitem [{\citenamefont {Zaletel}\ \emph {et~al.}(2015)\citenamefont {Zaletel}, \citenamefont {Mong}, \citenamefont {Pollmann},\ and\ \citenamefont {Rezayi}}]{Zalatel2015}%
  \BibitemOpen
  \bibfield  {author} {\bibinfo {author} {\bibfnamefont {M.~P.}\ \bibnamefont {Zaletel}}, \bibinfo {author} {\bibfnamefont {R.~S.~K.}\ \bibnamefont {Mong}}, \bibinfo {author} {\bibfnamefont {F.}~\bibnamefont {Pollmann}},\ and\ \bibinfo {author} {\bibfnamefont {E.~H.}\ \bibnamefont {Rezayi}},\ }\bibfield  {title} {\bibinfo {title} {Infinite density matrix renormalization group for multicomponent quantum hall systems},\ }\href {https://doi.org/10.1103/PhysRevB.91.045115} {\bibfield  {journal} {\bibinfo  {journal} {Phys. Rev. B}\ }\textbf {\bibinfo {volume} {91}},\ \bibinfo {pages} {045115} (\bibinfo {year} {2015})}\BibitemShut {NoStop}%
\bibitem [{\citenamefont {Herviou}\ and\ \citenamefont {Mila}(2023)}]{Herviou2023}%
  \BibitemOpen
  \bibfield  {author} {\bibinfo {author} {\bibfnamefont {L.}~\bibnamefont {Herviou}}\ and\ \bibinfo {author} {\bibfnamefont {F.}~\bibnamefont {Mila}},\ }\bibfield  {title} {\bibinfo {title} {Possible restoration of particle-hole symmetry in the 5/2-quantized hall state at small magnetic field},\ }\href {https://doi.org/10.1103/PhysRevB.107.115137} {\bibfield  {journal} {\bibinfo  {journal} {Phys. Rev. B}\ }\textbf {\bibinfo {volume} {107}},\ \bibinfo {pages} {115137} (\bibinfo {year} {2023})}\BibitemShut {NoStop}%
\bibitem [{\citenamefont {Simon}\ and\ \citenamefont {Rezayi}(2013)}]{Rezayi2013}%
  \BibitemOpen
  \bibfield  {author} {\bibinfo {author} {\bibfnamefont {S.~H.}\ \bibnamefont {Simon}}\ and\ \bibinfo {author} {\bibfnamefont {E.~H.}\ \bibnamefont {Rezayi}},\ }\bibfield  {title} {\bibinfo {title} {Landau level mixing in the perturbative limit},\ }\href {https://doi.org/10.1103/PhysRevB.87.155426} {\bibfield  {journal} {\bibinfo  {journal} {Phys. Rev. B}\ }\textbf {\bibinfo {volume} {87}},\ \bibinfo {pages} {155426} (\bibinfo {year} {2013})}\BibitemShut {NoStop}%
\bibitem [{\citenamefont {Peterson}\ and\ \citenamefont {Nayak}(2013)}]{LLmix1}%
  \BibitemOpen
  \bibfield  {author} {\bibinfo {author} {\bibfnamefont {M.~R.}\ \bibnamefont {Peterson}}\ and\ \bibinfo {author} {\bibfnamefont {C.}~\bibnamefont {Nayak}},\ }\bibfield  {title} {\bibinfo {title} {More realistic hamiltonians for the fractional quantum hall regime in gaas and graphene},\ }\href {https://doi.org/10.1103/PhysRevB.87.245129} {\bibfield  {journal} {\bibinfo  {journal} {Phys. Rev. B}\ }\textbf {\bibinfo {volume} {87}},\ \bibinfo {pages} {245129} (\bibinfo {year} {2013})}\BibitemShut {NoStop}%
\bibitem [{\citenamefont {Bishara}\ and\ \citenamefont {Nayak}(2009)}]{LLmix2}%
  \BibitemOpen
  \bibfield  {author} {\bibinfo {author} {\bibfnamefont {W.}~\bibnamefont {Bishara}}\ and\ \bibinfo {author} {\bibfnamefont {C.}~\bibnamefont {Nayak}},\ }\bibfield  {title} {\bibinfo {title} {Effect of landau level mixing on the effective interaction between electrons in the fractional quantum hall regime},\ }\href {https://doi.org/10.1103/PhysRevB.80.121302} {\bibfield  {journal} {\bibinfo  {journal} {Phys. Rev. B}\ }\textbf {\bibinfo {volume} {80}},\ \bibinfo {pages} {121302} (\bibinfo {year} {2009})}\BibitemShut {NoStop}%
\bibitem [{\citenamefont {Xu}\ \emph {et~al.}(2024{\natexlab{c}})\citenamefont {Xu}, \citenamefont {Mao}, \citenamefont {Zeng},\ and\ \citenamefont {Zhang}}]{xu2024multiple}%
  \BibitemOpen
  \bibfield  {author} {\bibinfo {author} {\bibfnamefont {C.}~\bibnamefont {Xu}}, \bibinfo {author} {\bibfnamefont {N.}~\bibnamefont {Mao}}, \bibinfo {author} {\bibfnamefont {T.}~\bibnamefont {Zeng}},\ and\ \bibinfo {author} {\bibfnamefont {Y.}~\bibnamefont {Zhang}},\ }\href@noop {} {\bibinfo {title} {Multiple chern bands in twisted mote$_2$ and possible non-abelian states}} (\bibinfo {year} {2024}{\natexlab{c}}),\ \Eprint {https://arxiv.org/abs/2403.17003} {arXiv:2403.17003 [cond-mat.str-el]} \BibitemShut {NoStop}%
\bibitem [{\citenamefont {Ahn}\ \emph {et~al.}(2024)\citenamefont {Ahn}, \citenamefont {Lee}, \citenamefont {Yananose}, \citenamefont {Kim},\ and\ \citenamefont {Cho}}]{ahn2024landau}%
  \BibitemOpen
  \bibfield  {author} {\bibinfo {author} {\bibfnamefont {C.-E.}\ \bibnamefont {Ahn}}, \bibinfo {author} {\bibfnamefont {W.}~\bibnamefont {Lee}}, \bibinfo {author} {\bibfnamefont {K.}~\bibnamefont {Yananose}}, \bibinfo {author} {\bibfnamefont {Y.}~\bibnamefont {Kim}},\ and\ \bibinfo {author} {\bibfnamefont {G.~Y.}\ \bibnamefont {Cho}},\ }\href@noop {} {\bibinfo {title} {First landau level physics in second moir\'e band of $2.1^\circ$ twisted bilayer mote${}_2$}} (\bibinfo {year} {2024}),\ \Eprint {https://arxiv.org/abs/2403.19155} {arXiv:2403.19155 [cond-mat.str-el]} \BibitemShut {NoStop}%
\bibitem [{\citenamefont {Wang}\ \emph {et~al.}(2024)\citenamefont {Wang}, \citenamefont {Zhang}, \citenamefont {Liu}, \citenamefont {Wang}, \citenamefont {Cao},\ and\ \citenamefont {Xiao}}]{wang2024higher}%
  \BibitemOpen
  \bibfield  {author} {\bibinfo {author} {\bibfnamefont {C.}~\bibnamefont {Wang}}, \bibinfo {author} {\bibfnamefont {X.-W.}\ \bibnamefont {Zhang}}, \bibinfo {author} {\bibfnamefont {X.}~\bibnamefont {Liu}}, \bibinfo {author} {\bibfnamefont {J.}~\bibnamefont {Wang}}, \bibinfo {author} {\bibfnamefont {T.}~\bibnamefont {Cao}},\ and\ \bibinfo {author} {\bibfnamefont {D.}~\bibnamefont {Xiao}},\ }\href@noop {} {\bibinfo {title} {Higher landau-level analogues and signatures of non-abelian states in twisted bilayer mote$_2$}} (\bibinfo {year} {2024}),\ \Eprint {https://arxiv.org/abs/2404.05697} {arXiv:2404.05697 [cond-mat.str-el]} \BibitemShut {NoStop}%
\bibitem [{\citenamefont {Ji}\ \emph {et~al.}(2024)\citenamefont {Ji}, \citenamefont {Park}, \citenamefont {Barber}, \citenamefont {Hu}, \citenamefont {Watanabe}, \citenamefont {Taniguchi}, \citenamefont {Chu}, \citenamefont {Xu},\ and\ \citenamefont {xun Shen}}]{ji2024local}%
  \BibitemOpen
  \bibfield  {author} {\bibinfo {author} {\bibfnamefont {Z.}~\bibnamefont {Ji}}, \bibinfo {author} {\bibfnamefont {H.}~\bibnamefont {Park}}, \bibinfo {author} {\bibfnamefont {M.~E.}\ \bibnamefont {Barber}}, \bibinfo {author} {\bibfnamefont {C.}~\bibnamefont {Hu}}, \bibinfo {author} {\bibfnamefont {K.}~\bibnamefont {Watanabe}}, \bibinfo {author} {\bibfnamefont {T.}~\bibnamefont {Taniguchi}}, \bibinfo {author} {\bibfnamefont {J.-H.}\ \bibnamefont {Chu}}, \bibinfo {author} {\bibfnamefont {X.}~\bibnamefont {Xu}},\ and\ \bibinfo {author} {\bibfnamefont {Z.}~\bibnamefont {xun Shen}},\ }\href@noop {} {\bibinfo {title} {Local probe of bulk and edge states in a fractional chern insulator}} (\bibinfo {year} {2024}),\ \Eprint {https://arxiv.org/abs/2404.07157} {arXiv:2404.07157 [cond-mat.str-el]} \BibitemShut {NoStop}%
\bibitem [{\citenamefont {Kang}\ \emph {et~al.}(2024{\natexlab{b}})\citenamefont {Kang}, \citenamefont {Shen}, \citenamefont {Qiu}, \citenamefont {Watanabe}, \citenamefont {Taniguchi}, \citenamefont {Shan},\ and\ \citenamefont {Mak}}]{kang2024observation}%
  \BibitemOpen
  \bibfield  {author} {\bibinfo {author} {\bibfnamefont {K.}~\bibnamefont {Kang}}, \bibinfo {author} {\bibfnamefont {B.}~\bibnamefont {Shen}}, \bibinfo {author} {\bibfnamefont {Y.}~\bibnamefont {Qiu}}, \bibinfo {author} {\bibfnamefont {K.}~\bibnamefont {Watanabe}}, \bibinfo {author} {\bibfnamefont {T.}~\bibnamefont {Taniguchi}}, \bibinfo {author} {\bibfnamefont {J.}~\bibnamefont {Shan}},\ and\ \bibinfo {author} {\bibfnamefont {K.~F.}\ \bibnamefont {Mak}},\ }\href@noop {} {\bibinfo {title} {Observation of the fractional quantum spin hall effect in moir\'e mote2}} (\bibinfo {year} {2024}{\natexlab{b}}),\ \Eprint {https://arxiv.org/abs/2402.03294} {arXiv:2402.03294 [cond-mat.mes-hall]} \BibitemShut {NoStop}%
\bibitem [{\citenamefont {Zhang}(2024{\natexlab{a}})}]{zhang2024vortex}%
  \BibitemOpen
  \bibfield  {author} {\bibinfo {author} {\bibfnamefont {Y.-H.}\ \bibnamefont {Zhang}},\ }\href@noop {} {\bibinfo {title} {Vortex spin liquid with fractional quantum spin hall effect in moir\'e chern bands}} (\bibinfo {year} {2024}{\natexlab{a}}),\ \Eprint {https://arxiv.org/abs/2402.05112} {arXiv:2402.05112 [cond-mat.str-el]} \BibitemShut {NoStop}%
\bibitem [{\citenamefont {May-Mann}\ \emph {et~al.}(2024)\citenamefont {May-Mann}, \citenamefont {Stern},\ and\ \citenamefont {Devakul}}]{maymann2024theory}%
  \BibitemOpen
  \bibfield  {author} {\bibinfo {author} {\bibfnamefont {J.}~\bibnamefont {May-Mann}}, \bibinfo {author} {\bibfnamefont {A.}~\bibnamefont {Stern}},\ and\ \bibinfo {author} {\bibfnamefont {T.}~\bibnamefont {Devakul}},\ }\href@noop {} {\bibinfo {title} {Theory of half-integer fractional quantum spin hall insulator edges}} (\bibinfo {year} {2024}),\ \Eprint {https://arxiv.org/abs/2403.03964} {arXiv:2403.03964 [cond-mat.mes-hall]} \BibitemShut {NoStop}%
\bibitem [{\citenamefont {Jian}\ \emph {et~al.}(2024)\citenamefont {Jian}, \citenamefont {Cheng},\ and\ \citenamefont {Xu}}]{jian2024minimal}%
  \BibitemOpen
  \bibfield  {author} {\bibinfo {author} {\bibfnamefont {C.-M.}\ \bibnamefont {Jian}}, \bibinfo {author} {\bibfnamefont {M.}~\bibnamefont {Cheng}},\ and\ \bibinfo {author} {\bibfnamefont {C.}~\bibnamefont {Xu}},\ }\href@noop {} {\bibinfo {title} {Minimal fractional topological insulator in half-filled conjugate moir\'{e} chern bands}} (\bibinfo {year} {2024}),\ \Eprint {https://arxiv.org/abs/2403.07054} {arXiv:2403.07054 [cond-mat.str-el]} \BibitemShut {NoStop}%
\bibitem [{\citenamefont {Villadiego}(2024)}]{villadiego2024halperin}%
  \BibitemOpen
  \bibfield  {author} {\bibinfo {author} {\bibfnamefont {I.~S.}\ \bibnamefont {Villadiego}},\ }\href@noop {} {\bibinfo {title} {Halperin states of particles and holes in ideal time reversal invariant pairs of chern bands and the fractional quantum spin hall effect in moir\'e mote$_2$}} (\bibinfo {year} {2024}),\ \Eprint {https://arxiv.org/abs/2403.12185} {arXiv:2403.12185 [cond-mat.mes-hall]} \BibitemShut {NoStop}%
\bibitem [{\citenamefont {Zhang}(2024{\natexlab{b}})}]{zhang2024nonabelianvortex}%
  \BibitemOpen
  \bibfield  {author} {\bibinfo {author} {\bibfnamefont {Y.-H.}\ \bibnamefont {Zhang}},\ }\href@noop {} {\bibinfo {title} {Non-abelian and abelian descendants of vortex spin liquid: fractional quantum spin hall effect in twisted mote$_2$}} (\bibinfo {year} {2024}{\natexlab{b}}),\ \Eprint {https://arxiv.org/abs/2403.12126} {arXiv:2403.12126 [cond-mat.str-el]} \BibitemShut {NoStop}%
\bibitem [{\citenamefont {Chou}\ and\ \citenamefont {Sarma}(2024)}]{chou2024composite}%
  \BibitemOpen
  \bibfield  {author} {\bibinfo {author} {\bibfnamefont {Y.-Z.}\ \bibnamefont {Chou}}\ and\ \bibinfo {author} {\bibfnamefont {S.~D.}\ \bibnamefont {Sarma}},\ }\href {https://arxiv.org/abs/2406.06669} {\bibinfo {title} {Composite helical edges from abelian fractional topological insulators}} (\bibinfo {year} {2024}),\ \Eprint {https://arxiv.org/abs/2406.06669} {arXiv:2406.06669 [cond-mat.str-el]} \BibitemShut {NoStop}%
\bibitem [{\citenamefont {Abouelkomsan}\ and\ \citenamefont {Fu}(2024)}]{abouelkomsan2024nonabelian}%
  \BibitemOpen
  \bibfield  {author} {\bibinfo {author} {\bibfnamefont {A.}~\bibnamefont {Abouelkomsan}}\ and\ \bibinfo {author} {\bibfnamefont {L.}~\bibnamefont {Fu}},\ }\href {https://arxiv.org/abs/2406.14617} {\bibinfo {title} {Non-abelian spin hall insulator}} (\bibinfo {year} {2024}),\ \Eprint {https://arxiv.org/abs/2406.14617} {arXiv:2406.14617 [cond-mat.mes-hall]} \BibitemShut {NoStop}%
\bibitem [{\citenamefont {Yu}\ \emph {et~al.}({\natexlab{a}})\citenamefont {Yu} \emph {et~al.}}]{paperYu}%
  \BibitemOpen
  \bibfield  {author} {\bibinfo {author} {\bibfnamefont {J.}~\bibnamefont {Yu}} \emph {et~al.},\ }\bibfield  {title} {\bibinfo {title} {to appear}} ({\natexlab{a}})\BibitemShut {NoStop}%
\bibitem [{\citenamefont {Haldane}(1985)}]{Haldane1985manybody}%
  \BibitemOpen
  \bibfield  {author} {\bibinfo {author} {\bibfnamefont {F.~D.~M.}\ \bibnamefont {Haldane}},\ }\bibfield  {title} {\bibinfo {title} {Many-particle translational symmetries of two-dimensional electrons at rational landau-level filling},\ }\href {https://doi.org/10.1103/PhysRevLett.55.2095} {\bibfield  {journal} {\bibinfo  {journal} {Phys. Rev. Lett.}\ }\textbf {\bibinfo {volume} {55}},\ \bibinfo {pages} {2095} (\bibinfo {year} {1985})}\BibitemShut {NoStop}%
\bibitem [{\citenamefont {Goerbig}\ \emph {et~al.}(2006)\citenamefont {Goerbig}, \citenamefont {Moessner},\ and\ \citenamefont {Dou\ifmmode~\mbox{\c{c}}\else \c{c}\fi{}ot}}]{Goerbig2006}%
  \BibitemOpen
  \bibfield  {author} {\bibinfo {author} {\bibfnamefont {M.~O.}\ \bibnamefont {Goerbig}}, \bibinfo {author} {\bibfnamefont {R.}~\bibnamefont {Moessner}},\ and\ \bibinfo {author} {\bibfnamefont {B.}~\bibnamefont {Dou\ifmmode~\mbox{\c{c}}\else \c{c}\fi{}ot}},\ }\bibfield  {title} {\bibinfo {title} {Electron interactions in graphene in a strong magnetic field},\ }\href {https://doi.org/10.1103/PhysRevB.74.161407} {\bibfield  {journal} {\bibinfo  {journal} {Phys. Rev. B}\ }\textbf {\bibinfo {volume} {74}},\ \bibinfo {pages} {161407} (\bibinfo {year} {2006})}\BibitemShut {NoStop}%
\bibitem [{\citenamefont {Lee}\ \emph {et~al.}(2023)\citenamefont {Lee}, \citenamefont {Ponc{\'e}}, \citenamefont {Bushick}, \citenamefont {Hajinazar}, \citenamefont {Lafuente-Bartolome}, \citenamefont {Leveillee}, \citenamefont {Lian}, \citenamefont {Lihm}, \citenamefont {Macheda}, \citenamefont {Mori} \emph {et~al.}}]{lee2023EPW}%
  \BibitemOpen
  \bibfield  {author} {\bibinfo {author} {\bibfnamefont {H.}~\bibnamefont {Lee}}, \bibinfo {author} {\bibfnamefont {S.}~\bibnamefont {Ponc{\'e}}}, \bibinfo {author} {\bibfnamefont {K.}~\bibnamefont {Bushick}}, \bibinfo {author} {\bibfnamefont {S.}~\bibnamefont {Hajinazar}}, \bibinfo {author} {\bibfnamefont {J.}~\bibnamefont {Lafuente-Bartolome}}, \bibinfo {author} {\bibfnamefont {J.}~\bibnamefont {Leveillee}}, \bibinfo {author} {\bibfnamefont {C.}~\bibnamefont {Lian}}, \bibinfo {author} {\bibfnamefont {J.-M.}\ \bibnamefont {Lihm}}, \bibinfo {author} {\bibfnamefont {F.}~\bibnamefont {Macheda}}, \bibinfo {author} {\bibfnamefont {H.}~\bibnamefont {Mori}}, \emph {et~al.},\ }\bibfield  {title} {\bibinfo {title} {Electron--phonon physics from first principles using the epw code},\ }\href@noop {} {\bibfield  {journal} {\bibinfo  {journal} {npj Computational Materials}\ }\textbf {\bibinfo {volume} {9}},\ \bibinfo {pages} {156} (\bibinfo {year} {2023})}\BibitemShut {NoStop}%
\bibitem [{\citenamefont {Yu}\ \emph {et~al.}({\natexlab{b}})\citenamefont {Yu}, \citenamefont {Jiang}, \citenamefont {Xu},\ and\ \citenamefont {Bernevig}}]{EPCNbSe2}%
  \BibitemOpen
  \bibfield  {author} {\bibinfo {author} {\bibfnamefont {J.}~\bibnamefont {Yu}}, \bibinfo {author} {\bibfnamefont {Y.}~\bibnamefont {Jiang}}, \bibinfo {author} {\bibfnamefont {Y.}~\bibnamefont {Xu}},\ and\ \bibinfo {author} {\bibfnamefont {B.~A.}\ \bibnamefont {Bernevig}},\ }\href@noop {} {\bibfield  {journal} {\bibinfo  {journal} {to appear}\ } ({\natexlab{b}})}\BibitemShut {NoStop}%
\bibitem [{\citenamefont {Miyazaki}\ \emph {et~al.}(2008)\citenamefont {Miyazaki}, \citenamefont {Odaka}, \citenamefont {Sato}, \citenamefont {Tanaka}, \citenamefont {Goto}, \citenamefont {Kanda}, \citenamefont {Tsukagoshi}, \citenamefont {Ootuka},\ and\ \citenamefont {Aoyagi}}]{miyazaki2008inter}%
  \BibitemOpen
  \bibfield  {author} {\bibinfo {author} {\bibfnamefont {H.}~\bibnamefont {Miyazaki}}, \bibinfo {author} {\bibfnamefont {S.}~\bibnamefont {Odaka}}, \bibinfo {author} {\bibfnamefont {T.}~\bibnamefont {Sato}}, \bibinfo {author} {\bibfnamefont {S.}~\bibnamefont {Tanaka}}, \bibinfo {author} {\bibfnamefont {H.}~\bibnamefont {Goto}}, \bibinfo {author} {\bibfnamefont {A.}~\bibnamefont {Kanda}}, \bibinfo {author} {\bibfnamefont {K.}~\bibnamefont {Tsukagoshi}}, \bibinfo {author} {\bibfnamefont {Y.}~\bibnamefont {Ootuka}},\ and\ \bibinfo {author} {\bibfnamefont {Y.}~\bibnamefont {Aoyagi}},\ }\bibfield  {title} {\bibinfo {title} {Inter-layer screening length to electric field in thin graphite film},\ }\href@noop {} {\bibfield  {journal} {\bibinfo  {journal} {Applied Physics Express}\ }\textbf {\bibinfo {volume} {1}},\ \bibinfo {pages} {034007} (\bibinfo {year} {2008})}\BibitemShut {NoStop}%
\bibitem [{\citenamefont {Ma}\ \emph {et~al.}(2022)\citenamefont {Ma}, \citenamefont {Chen}, \citenamefont {Yananose}, \citenamefont {Zhou}, \citenamefont {Wang}, \citenamefont {Li}, \citenamefont {Zhu}, \citenamefont {Wu}, \citenamefont {Xu}, \citenamefont {Yu} \emph {et~al.}}]{ma2022growth}%
  \BibitemOpen
  \bibfield  {author} {\bibinfo {author} {\bibfnamefont {T.}~\bibnamefont {Ma}}, \bibinfo {author} {\bibfnamefont {H.}~\bibnamefont {Chen}}, \bibinfo {author} {\bibfnamefont {K.}~\bibnamefont {Yananose}}, \bibinfo {author} {\bibfnamefont {X.}~\bibnamefont {Zhou}}, \bibinfo {author} {\bibfnamefont {L.}~\bibnamefont {Wang}}, \bibinfo {author} {\bibfnamefont {R.}~\bibnamefont {Li}}, \bibinfo {author} {\bibfnamefont {Z.}~\bibnamefont {Zhu}}, \bibinfo {author} {\bibfnamefont {Z.}~\bibnamefont {Wu}}, \bibinfo {author} {\bibfnamefont {Q.-H.}\ \bibnamefont {Xu}}, \bibinfo {author} {\bibfnamefont {J.}~\bibnamefont {Yu}}, \emph {et~al.},\ }\bibfield  {title} {\bibinfo {title} {Growth of bilayer mote2 single crystals with strong non-linear hall effect},\ }\href@noop {} {\bibfield  {journal} {\bibinfo  {journal} {Nature Communications}\ }\textbf {\bibinfo {volume} {13}},\ \bibinfo {pages} {5465} (\bibinfo {year} {2022})}\BibitemShut {NoStop}%
\bibitem [{\citenamefont {Berkelbach}\ \emph {et~al.}(2013)\citenamefont {Berkelbach}, \citenamefont {Hybertsen},\ and\ \citenamefont {Reichman}}]{berkelbach2013theory}%
  \BibitemOpen
  \bibfield  {author} {\bibinfo {author} {\bibfnamefont {T.~C.}\ \bibnamefont {Berkelbach}}, \bibinfo {author} {\bibfnamefont {M.~S.}\ \bibnamefont {Hybertsen}},\ and\ \bibinfo {author} {\bibfnamefont {D.~R.}\ \bibnamefont {Reichman}},\ }\bibfield  {title} {\bibinfo {title} {Theory of neutral and charged excitons in monolayer transition metal dichalcogenides},\ }\href {https://doi.org/10.1103/PhysRevB.88.045318} {\bibfield  {journal} {\bibinfo  {journal} {Phys. Rev. B}\ }\textbf {\bibinfo {volume} {88}},\ \bibinfo {pages} {045318} (\bibinfo {year} {2013})}\BibitemShut {NoStop}%
\bibitem [{\citenamefont {Chernikov}\ \emph {et~al.}(2014)\citenamefont {Chernikov}, \citenamefont {Berkelbach}, \citenamefont {Hill}, \citenamefont {Rigosi}, \citenamefont {Li}, \citenamefont {Aslan}, \citenamefont {Reichman}, \citenamefont {Hybertsen},\ and\ \citenamefont {Heinz}}]{chernikov2014exciton}%
  \BibitemOpen
  \bibfield  {author} {\bibinfo {author} {\bibfnamefont {A.}~\bibnamefont {Chernikov}}, \bibinfo {author} {\bibfnamefont {T.~C.}\ \bibnamefont {Berkelbach}}, \bibinfo {author} {\bibfnamefont {H.~M.}\ \bibnamefont {Hill}}, \bibinfo {author} {\bibfnamefont {A.}~\bibnamefont {Rigosi}}, \bibinfo {author} {\bibfnamefont {Y.}~\bibnamefont {Li}}, \bibinfo {author} {\bibfnamefont {B.}~\bibnamefont {Aslan}}, \bibinfo {author} {\bibfnamefont {D.~R.}\ \bibnamefont {Reichman}}, \bibinfo {author} {\bibfnamefont {M.~S.}\ \bibnamefont {Hybertsen}},\ and\ \bibinfo {author} {\bibfnamefont {T.~F.}\ \bibnamefont {Heinz}},\ }\bibfield  {title} {\bibinfo {title} {Exciton binding energy and nonhydrogenic rydberg series in monolayer ${\mathrm{ws}}_{2}$},\ }\href {https://doi.org/10.1103/PhysRevLett.113.076802} {\bibfield  {journal} {\bibinfo  {journal} {Phys. Rev. Lett.}\ }\textbf {\bibinfo {volume} {113}},\ \bibinfo {pages} {076802} (\bibinfo {year} {2014})}\BibitemShut {NoStop}%
\bibitem [{\citenamefont {Van~Tuan}\ \emph {et~al.}(2018)\citenamefont {Van~Tuan}, \citenamefont {Yang},\ and\ \citenamefont {Dery}}]{tuan2018coulomb}%
  \BibitemOpen
  \bibfield  {author} {\bibinfo {author} {\bibfnamefont {D.}~\bibnamefont {Van~Tuan}}, \bibinfo {author} {\bibfnamefont {M.}~\bibnamefont {Yang}},\ and\ \bibinfo {author} {\bibfnamefont {H.}~\bibnamefont {Dery}},\ }\bibfield  {title} {\bibinfo {title} {Coulomb interaction in monolayer transition-metal dichalcogenides},\ }\href {https://doi.org/10.1103/PhysRevB.98.125308} {\bibfield  {journal} {\bibinfo  {journal} {Phys. Rev. B}\ }\textbf {\bibinfo {volume} {98}},\ \bibinfo {pages} {125308} (\bibinfo {year} {2018})}\BibitemShut {NoStop}%
\bibitem [{\citenamefont {Santos}\ \emph {et~al.}(2011)\citenamefont {Santos}, \citenamefont {Neupert}, \citenamefont {Ryu}, \citenamefont {Chamon},\ and\ \citenamefont {Mudry}}]{Santos2011TRShierarchy}%
  \BibitemOpen
  \bibfield  {author} {\bibinfo {author} {\bibfnamefont {L.}~\bibnamefont {Santos}}, \bibinfo {author} {\bibfnamefont {T.}~\bibnamefont {Neupert}}, \bibinfo {author} {\bibfnamefont {S.}~\bibnamefont {Ryu}}, \bibinfo {author} {\bibfnamefont {C.}~\bibnamefont {Chamon}},\ and\ \bibinfo {author} {\bibfnamefont {C.}~\bibnamefont {Mudry}},\ }\bibfield  {title} {\bibinfo {title} {Time-reversal symmetric hierarchy of fractional incompressible liquids},\ }\href {https://doi.org/10.1103/PhysRevB.84.165138} {\bibfield  {journal} {\bibinfo  {journal} {Phys. Rev. B}\ }\textbf {\bibinfo {volume} {84}},\ \bibinfo {pages} {165138} (\bibinfo {year} {2011})}\BibitemShut {NoStop}%
\end{thebibliography}
\end{document}